\documentclass[sn-mathphys-num]{sn-jnl}

\usepackage{multirow}
\usepackage{xcolor}
\usepackage{mathtools}
\usepackage{booktabs}%
\usepackage{algorithmic,algorithm}%
\usepackage{tikz}
\usetikzlibrary{arrows.meta, positioning, shapes}
\usetikzlibrary{positioning, fit}
\usepackage{paralist}
\usepackage{amsmath,amsfonts,amsthm,amsmath,amssymb}
\usepackage{listings}

\definecolor{backcolour}{rgb}{0.95,0.95,0.92}
\definecolor{codegreen}{rgb}{0,0.6,0}
\definecolor{codegray}{rgb}{0.5,0.5,0.5}
\definecolor{codepurple}{rgb}{0.58,0,0.82}

\lstdefinelanguage{SMTLIB}{
    keywords={declare-fun,assert,assert-soft, minimize, check-sat, get-model, get-objectives},
    keywordstyle=\color{black},
    identifierstyle=\color{black},
    sensitive=false,
    comment=[l]{;},
    commentstyle=\color{codegreen}\ttfamily,
    stringstyle=\color{magenta}\ttfamily,
    morestring=[b]',
    morestring=[b]",
}

\lstdefinestyle{mystyle}{
    backgroundcolor=\color{backcolour},   
    commentstyle=\color{codegreen},
    keywordstyle=\color{blue}\bfseries,
    numberstyle=\tiny\color{codegray},
    stringstyle=\color{codepurple},
    basicstyle=\ttfamily\scriptsize,
    breakatwhitespace=false,         
    breaklines=true,                 
    captionpos=b,                    
    keepspaces=true,                 
    numbers=left,                    
    numbersep=5pt,                  
    showspaces=false,                
    showstringspaces=false,
    showtabs=false,                  
    tabsize=2
}

\lstdefinelanguage{JavaScript}{
  keywords={typeof, new, true, false, catch, function, return, null, catch, switch, var, if, in, while, do, else, case, break},
  keywordstyle=\color{blue}\bfseries,
  ndkeywords={class, export, boolean, throw, implements, import, this},
  ndkeywordstyle=\color{darkgray}\bfseries,
  identifierstyle=\color{black},
  sensitive=false,
  comment=[l]{//},
  morecomment=[s]{/*}{*/},
  commentstyle=\color{codegreen}\ttfamily,
  stringstyle=\color{magenta}\ttfamily,
  morestring=[b]',
  morestring=[b]"
}
\usepackage{array}                              
\newcolumntype{C}[1]{>{\centering\arraybackslash}m{#1}}

\theoremstyle{thmstyleone}%
\theoremstyle{thmstyletwo}%

\theoremstyle{thmstylethree}%

\begin{document}

\title[Neuro-Symbolic Constrained Optimization for Cloud Application Deployment via Graph Neural Networks and Satisfiability Modulo Theory]{Neuro-Symbolic Constrained Optimization for Cloud Application Deployment via Graph Neural Networks and Satisfiability Modulo Theory}


\author*[1]{\fnm{M\u{a}d\u{a}lina} \sur{Era\c{s}cu}}\email{madalina.erascu@e-uvt.ro}

\affil*[1]{\orgdiv{Faculty of Mathematics and Informatics}, \orgname{West University of Timișoara}, \orgaddress{\street{Blvd. Vasile Pârvan nr. 4}, \city{Timișoara}, \postcode{300223},  \country{Romania}}}


\abstract{This paper proposes a novel hybrid neuro-symbolic framework for the optimal and scalable deployment of component-based applications in the Cloud. The challenge of efficiently mapping application components to virtual machines (VMs) across diverse VM Offers from Cloud Providers is formalized as a constrained optimization problem (COP), considering both general and application-specific constraints. Due to the NP-hard nature and scalability limitations of exact solvers, we introduce a machine learning-enhanced approach where graph neural networks (GNNs) are trained on small-scale deployment instances and their predictions are used as soft constraints within the Z3 SMT solver. The deployment problem is recast as a graph edge classification task over a heterogeneous graph, combining relational embeddings with constraint reasoning. Our framework is validated through several realistic case studies, each highlighting different constraint profiles. Experimental results confirm that incorporating GNN predictions improves solver scalability and often preserves or even improves cost-optimality. This work demonstrates the practical benefits of neuro-symbolic coupling for Cloud infrastructure planning and contributes a reusable methodology for general NP-hard problems.}




\maketitle

\section{Introduction}\label{sec:introduction}
There has been recent interest in moving beyond the traditional and often pessimistic NP-completeness of satisfiability (modulo theories) checking by using machine-learned predictions. On the one hand, the machine learning (ML) community used graph neural networks (GNNs), in particular, directly as solvers or to improve exact solvers~\cite{GNN-opt}. On the other hand, in the formal methods community, existing work focused mainly on predicting the (best) runtime to check the satisfiability of a formula~\cite{scott2021machsmt,pimpalkhare2021medleysolver,leeson2023sibyl}. These approaches are the so-called \emph{learning-augmented algorithms} that aim to utilize ML predictions to improve the performance of classic algorithms.

The extra information assumed in learning-augmented algorithms can be provided in a variety of settings. One such example is that the current input can be similar to past instances, so an existing solution might be reused, or it might help to avoid computing the solution from scratch or to use the existing solution in order to compute the new one.

Such scenarios occur in the task of \emph{automated deployment in the Cloud of component-based applications}: given a set of available computing resources (virtual machines) from various Cloud Providers and a component-based application to be deployed, find an assignment of components to virtual machines (VMs) such that performance objectives are optimized (e.g. cost is minimized) and the interactions between components are satisfied. The problem is related to \emph{bin-packing}, however:
\begin{inparaenum}[\itshape (1)\upshape]
	\item bins (VMs) can have different capacity, which depends on the VM Offers;
	\item the placement of items (components) in bins is limited not only by the capacity constraints, but also by the constraints induced by the components interactions;
	\item the number of items is not known a priori (for component-based applications, several instances of a component can be deployed, depending on specific constraints on the number of instances);
	\item the smallest cost (optimality criteria) is not necessarily obtained by minimizing the number of bins.
\end{inparaenum}

Solving the problem is feasible only if the size of the input (e.g. number of VM offers\footnote{There are hundreds of VMs offers from the main Cloud Providers.} and number of application components) is sufficiently small, since there exists an exponential number of combinations, and consequently large instances are unfeasible within a reasonable time. 
The lack of scalability of naive search techniques is the reason why these kinds of problem were typically approached using suboptimal techniques, such as hand-crafted heuristics and metaheuristics, as well as approximation algorithms (see~\cite{10.1007/s10922-014-9307-7} for a survey). However, an optimal solution, in terms of cost and assignments of components to VMs, is desirable, and thus scalable exact methods are of interest.

As an illustration, Table \ref{tab:Z3-results-raw} presents the computational challenge of solving deployment-related constraint optimization problems (see Section~\ref{sec:CaseStudies}) with the exact SMT solver Z3 \cite{de2008z3}.
When the set of VM Offers is small, the models of at most $60$ variables (\#var) and $400$ constraints (\#con) are solved almost immediately. Increasing the set by an order of magnitude inflates each model to a few thousand constraints and drives runtime into the tens or hundreds of seconds. With the largest number of VM Offers considered, the solver requires about five minutes and several WordPress use case variants time out in the time limit of $40$ minutes. The result \textsc{UNSAT} means that there is no way to assign the application components to existing VM offers that also satisfy the application constraints. The difficulty is further amplified by the strong interactions among components, which translate into intricate logical constraints within the constraint optimization model.
We conclude that the exact methods scale poorly once the set of VM Offers grows or replication demands tighten.
\begin{table}[ht]
\footnotesize
\setlength{\tabcolsep}{4pt}
\begin{tabular}{|C{1.9cm}|C{1.37cm}|*{5}{C{1.53cm}|}}
\hline
\textbf{Problem} &
\shortstack{\textbf{\#comp}\\\textbf{instances}} &
\shortstack{\textbf{\#offers}\\\textbf{= 20}} &
\shortstack{\textbf{\#offers}\\\textbf{= 40}} &
\shortstack{\textbf{\#offers}\\\textbf{= 250}} &
\shortstack{\textbf{\#offers}\\\textbf{= 500}} &
\shortstack{\textbf{\#offers}\\\textbf{= 27}} \\ \hline
Oryx2 & 10 &
\shortstack{\#var=165\\\#con=619\\$t$=7.32} &
\shortstack{\#var=165\\\#con=839\\$t$=5.98} &
\shortstack{\#var=165\\\#con=3149\\$t$=118.01} &
\shortstack{\#var=165\\\#con=3500\\$t$=300.60} &
  \textsc{UNSAT} \\ \hline
\shortstack{Sec.\ Web\\Container} & 5 &
\shortstack{\#var=60\\\#con=279\\$t$=0.30} &
\shortstack{\#var=60\\\#con=399\\$t$=0.62} &  \shortstack{\#var=60\\\#con=659\\$t$=3.15} &
\shortstack{\#var=60\\\#con=3159\\$t$=10.38} &
\shortstack{\#var=60\\\#con=321\\$t$=0.25} \\ \hline
\shortstack{Sec.\ Billing\\Email} & 5 &
\shortstack{\#var=50\\\#con=212\\$t$=0.17} &
  \shortstack{\#var=50\\\#con=312\\$t$=0.30} &
  \shortstack{\#var=50\\\#con=1362\\$t$=1.42} &
  \shortstack{\#var=50\\\#con=2612\\$t$=4.31} &
  \shortstack{\#var=50\\\#con=247\\$t$=0.10} \\ \hline
\shortstack{WordPress\\min\#inst = 3} & 8 &
  \shortstack{\#var=80\\\#con=365\\$t$=1.45} &
  \shortstack{\#var=80\\\#con=525\\$t$=5.05} &
  \shortstack{\#var=80\\\#con=2205\\$t$=28.62} &
  \shortstack{\#var=80\\\#con=4205\\$t$=146.13} &
  \shortstack{\#var=80\\\#con=421\\$t$=0.46} \\ \hline
\shortstack{WordPress\\min\#inst = 4} & 10 &
  \shortstack{\#var=100\\\#con=453\\$t$=10.65} &
  \shortstack{\#var=100\\\#con=653\\$t$=104.10} &
  \shortstack{\#var=100\\\#con=2753\\$t$=--} &
  \shortstack{\#var=100\\\#con=5253\\$t$=--} &
  \shortstack{\#var=100\\\#con=523\\$t$=29.55} \\ \hline
\shortstack{WordPress\\min\#inst = 5} & 12 &
  \shortstack{\#var=120\\\#con=541\\$t$=1106.49} &
  \shortstack{\#var=120\\\#con=781\\$t$=--} &
  \shortstack{\#var=120\\\#con=3301\\$t$=--} &
  \shortstack{\#var=120\\\#con=6300\\$t$=--} &
  \shortstack{\#var=120\\\#con=625\\$t$=--} \\ \hline
\end{tabular}
\caption{Scalability tests for Z3. Time $t$ values are expressed in seconds}
\label{tab:Z3-results-raw}
\end{table}
This motivates the need for novel strategies that can overcome scalability bottlenecks while preserving the optimality of solutions. We advocate a twofold research agenda:
\begin{enumerate}
\item Train GNNs on trivially small deployment instances and transfer the resulting policy to much larger, more complex graphs either zero-shot or after a brief fine-tuning phase. The crux is to uncover inductive biases, architectures, and learning paradigms that allow such models to generalize reliably to unseen application graphs and VM Offer sets. This way we replace costly large-scale training with a learn-once-apply-anywhere strategy.
\item Exploit the trained GNN at inference time by encoding its placement predictions as soft constraints inside the SMT solver Z3. This neuro-symbolic coupling leverages historical knowledge to steer the exact optimizer, cutting the search effort while still guaranteeing formally correct and cost-efficient deployments.
\end{enumerate}

\noindent This work builds on and extends \cite{10650114} in the following ways:
\begin{enumerate}
\item In \cite{10650114}, we devised a method to create a \emph{comprehensive dataset} with the role of historical data. Our strategy revolved around the different VM Offers available from various Cloud Providers. Specifically, we generated a multitude of combinations of subsets from these VM Offers, and each of it, together with the description of the application to be deployed, was processed through an \emph{exact solver} to yield the optimal deployment solution. The corresponding input configurations and output solutions represent samples of the generated dataset which are then transformed into graph data by using supervised learning. The samples of the dataset are generated using our previous work~\cite{ERASCU2021100664}. In this paper, we curate the datasets rawly obtained in order to possess \emph{class balance} and \emph{diverse range of labels} which is crucial for supervised learning to ensure adaptability and generalization (see Section~\ref{sec:datasets}). 
\item In \cite{10650114}, we trained GNNs for predicting the assignment of application components to VMs. This involved constructing a \emph{heterogeneous graph} where components and VMs are represented as nodes, with edges connecting components to other components and to VMs. The task was formulated as an \emph{edge classification problem}, where the GNN predicted whether an edge between a component and a VM was of type \texttt{linked} (indicating the component is placed on the respective VM) or \texttt{unlinked} (indicating the component is not placed on that VM). In this paper, we enhance the training experiments by systematically exploring a wider range of hyperparameters to fine-tune model performance. Furthermore, we investigate various neural network architectures, including different configurations of layers, activation functions, and aggregation strategies, to evaluate their impact on the accuracy and efficiency of the edge classification task (see Section~\ref{sec:ProblemDefinitionGraphEdgeClassification}).
\item In \cite{10650114}, we interpreted the predictions generated by the GNN model as logical constraints, specifically soft constraints, and integrated them with the optimization engine developed in \cite{ERASCU2021100664}. This hybrid approach allowed us to achieve competitive results in terms of scalability and solution quality, using historical data. In this paper, we refine the analysis and extend it to new case studies by answering the following research questions (RQs):
\begin{itemize}
    \item[(RQ 1)] How does scalability of the hybrid approach vary with the number of available VM Offers and with increasing number of component instances which changes the dynamics of component interactions? (see Section~\ref{sec:RQ1})
    \item[(RQ 2)] Is there a correlation between the GNN predictions and the optimal solution? Furthermore, is there a relationship between solution time and the optimal solution? (see Section~\ref{sec:RQ2})
    \item[(RQ 3)] Are there specific GNNs tailored for particular use cases that simultaneously predict well the assignments of components to VMs while minimizing execution time and achieving optimal solution? (see Section~\ref{sec:RQ3})
\end{itemize}
\end{enumerate}

The remainder of the paper is organized as follows. We begin by reviewing the relevant literature in Section~\ref{sec:RelatedWork}, where we position our hybrid neuro-symbolic methodology within the broader context of combinatorial optimization, formal methods, and graph-based machine learning. Section~\ref{sec:Neural-Symbolic-Combinatorial-Optimization-Pipeline} outlines our proposed framework, detailing the theoretical formulation of the Cloud deployment problem, its encoding as a graph edge classification task, and the integration of GNN predictions into the SMT-based solution pipeline. Section~\ref{sec:CaseStudies} introduces case studies that serve as practical validation scenarios, ranging from secure web applications to scalable services, each characterized by diverse architectural and constraint profiles. Section~\ref{sec:ExperimentalSetup} describes the experimental setup, including dataset preparation, GNN architectures, and hyperparameter and SMT solver configuration. Section \ref{sec:results} presents and discusses the experimental results, highlighting the scalability and accuracy of the proposed approach. Finally, Section \ref{sec:ConclusionsAndFuture Work} concludes the paper and outlines directions for future research, particularly focusing on the integration of GNN and SMT through variable initialization and advanced GNN architectures.

\section{Related Work}\label{sec:RelatedWork}
\noindent \textbf{Neural Combinatorial Optimization.} A recent survey by \cite{bengio2021machine} categorizes the integration of machine learning into combinatorial optimization into three primary strategies:
\begin{inparaenum}[\itshape (1)\upshape]
\item learning in conjunction with traditional optimization algorithms \cite{gasse2019exact},
\item learning to configure or adapt such algorithms \cite{wilder2019melding}, and
\item end-to-end learning approaches aimed at directly approximating solutions, commonly referred to as neural combinatorial optimization \cite{bello2016neural}.
\end{inparaenum}
Graph Neural Networks (GNNs) have been explored in solving classical combinatorial problems such as the Traveling Salesman Problem (TSP) \cite{kipf2016semi}, Vehicle Routing \cite{nazari2018reinforcement}, Boolean Satisfiability (SAT) \cite{yolcu2019learning,selsam2018learning}, and Graph Coloring \cite{huang2019coloring}. The results suggest that GNNs can generalize to more complex or larger instances, particularly in non-sequential tasks such as SAT and MaxCut \cite{selsam2018learning}. However, this generalization is less effective in sequential problems such as TSP \cite{joshi2022learning}.

\noindent \textbf{Combinatorial Optimization and GNNs.} Within the realm of graph representation learning, combinatorial problems have served as benchmarks to assess GNN expressiveness \cite{sato2019approximation}. A growing line of work on learning to execute graph algorithms \cite{velivckovic2019neural} has led to more powerful GNN variants \cite{corso2020principal} and a better understanding of their generalization behavior~\cite{xu2019can}. In particular, \cite{joshi2022learning} evaluates how standard GNN architectures generalize in a zero-shot fashion to larger, unseen instances.

\noindent \textbf{Novel Applications.} Recent advances in combinatorial optimization have enabled applications in new domains where traditional solvers are impractical, such as physical sciences \cite{senior2020improved} and computer architecture \cite{mirhoseini2021graph}. For example, autoregressive models show strong inductive biases in tasks such as device placement \cite{mirhoseini2017device}, while non-autoregressive architectures have proven competitive in molecular generation \cite{bresson2019two,jin2018junction}.

The problem of \textbf{automated deployment of component-based applications in the Cloud}, formalized in Section~\ref{sec:ProblemDefinitionConstrainedOptimization}, can be framed as a constraint optimization task. Various exact techniques, such as constraint programming, mathematical programming, and SMT solving~\cite{DBLP:conf/setta/AbrahamCJKM16} have been employed for this purpose. While these methods offer exact solutions, they are computationally intensive. Considering that application deployments typically occur multiple times during the software lifecycle with only minor configuration changes, we argue that ML, particularly GNNs, offers a promising solution.

While neural combinatorial optimization techniques could, in principle, be applied to our constrained optimization problem, especially those targeting optimality, our use of GNNs pursues different objectives:
\begin{inparaenum}[\itshape (1)\upshape]
  \item to train a model of the application deployment scenario that generalizes to slight variations in both the problem instance and the available Cloud offers, and can be extended to previously unseen applications and Cloud configurations. This contrasts with~\cite{GNN-opt}, which integrates GNNs directly into the optimization solver. That work highlights the ability of GNNs to encode combinatorial and relational inputs due to their permutation invariance and awareness of input sparsity, leading to growing interest in using GNNs as either direct solvers or enhancements for exact solvers;
  \item to use the trained GNN to generate predictions, which are subsequently encoded as soft constraints and incorporated into the Z3 solver~\cite{de2008z3}.
\end{inparaenum} By leveraging knowledge from past deployments and combining GNNs with exact solvers, we aim to reduce computational overhead while maintaining feasible and cost-efficient deployments.

The Cloud deployment problem can also be viewed as a more complex variant of the classical \emph{bin-packing problem}. However, to the best of our knowledge, no prior work has framed bin-packing as a \emph{graph analysis or prediction} task suitable for GNNs. For example, \cite{TSP} applies GNNs to the TSP by constructing efficient graph representations and generating tours in a non-autoregressive fashion using parallelized beam search. This inspired our problem modeling (see Section~\ref{sec:ProblemDefinitionGraphEdgeClassification}): initially, all edges between components and virtual machines (VMs) are marked as \texttt{unlinked}, and may be transformed into \texttt{linked} edges during training. Nevertheless, our problem differs from TSP in several key aspects. First, while TSP is modeled as a \emph{homogeneous graph}, our setting involves a \emph{heterogeneous graph} comprising two node types: components and VMs. Second, we face an \emph{edge classification problem}, where the goal is to predict whether a component--VM edge should be linked. This is more challenging than the TSP case, which uses a single edge type with varying weights. Furthermore, \cite{TSP} does not guarantee optimal tour lengths, as GNNs are not combined with exact solvers. Instead, the exact Concorde TSP solver~\cite{applegate2009certification} is used only to generate training data, a strategy we also adopt, as discussed in Section~\ref{sec:datasets}.


\section{Neural-Symbolic Combinatorial Optimization Pipeline}\label{sec:Neural-Symbolic-Combinatorial-Optimization-Pipeline}
\begin{figure}[htbp]
  \centering
  \begin{tikzpicture}[scale=0.8, transform shape, >=Stealth,
      pipelineStep/.style={draw, rectangle, align=center, font=\small},
      lowerStep/.style={pipelineStep},
      outer/.style={draw, rectangle, thick, inner sep=5mm}
  ]
    \node[pipelineStep, fill=blue!20] (P1) {Problem definition\\(as graph edge classification)};
    \node[pipelineStep, fill=green!20, right=of P1] (GE) {Graph\\embedding};
    \node[pipelineStep, fill=yellow!20, right=of GE] (SD) {Solution\\decoding};
    \node[pipelineStep, fill=orange!20, right=of SD] (SS) {Solution\\search};
    \node[pipelineStep, fill=purple!20, right=of SS] (PL) {Policy\\learning};
    
    \node[lowerStep, fill=red!20, below=1cm of P1] (P1b) {Problem definition\\(as constrained optimization)};
    
    \node[lowerStep, fill=magenta!20, right=of P1b] (CGR) {Constraint-Guided\\Solution Refinement};
    
    \draw[->, thick] (P1.east) -- (GE.west);
    \draw[->, thick] (GE.east) -- (SD.west);
    \draw[->, thick] (SD.east) -- (SS.west);
    \draw[->, thick] (SS.east) -- (PL.west);
    
    \draw[->, thick] (P1b.east) -- (CGR.west);
    
    \draw[->, thick] (SS.south) .. controls +(0,-1) and +(-1,1) .. (CGR.north);
    
  \end{tikzpicture}
  \caption{Neural-symbolic optimization pipeline}
  \label{fig:neural-symbolic-pipeline}
\end{figure}
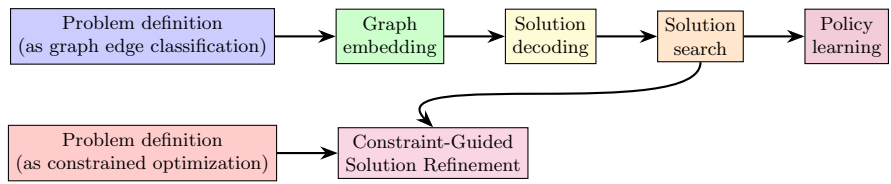

In this work, we take a step toward a systematic investigation of neural-symbolic combinatorial optimization, with a specific focus on the automated deployment of component-based applications in the Cloud. To this end, we propose a hybrid pipeline that integrates a neural end-to-end learning module with a symbolic reasoning module, as illustrated in Fig. \ref{fig:neural-symbolic-pipeline}. The proposed architecture is generic and extensible, making it applicable to a broad class of NP-hard problems beyond Cloud deployment.

\subsection{Problem Definition} \label{sec:probSpec}
Given a component-based application description consisting of:
\begin{inparaenum}[\itshape (i)\upshape]
\item application components together with their hardware/software requirements and the constraints generated by the interactions between them, and 
\item a set of VM Offers from various Cloud Providers (e.g. Amazon Web Services, Microsoft Azure, Google Cloud, Digital Ocean) specifying their hardware/software specifications and costs,
\end{inparaenum} the task is to decide which VM Offers to lease and how to assign the application components to them
in such a way that:
\begin{inparaenum}[\itshape (1)\upshape]
\item the constraints are satisfied, and
\item the leasing price is minimized. 
\end{inparaenum}
These are fulfilled by \emph{formulating and solving the constrained optimization problem (COP)} from Section~\ref{sec:ProblemDefinitionConstrainedOptimization}.

For applications to be deployed in the Cloud, for latency reasons, it is relevant to exploit historic deployment data. This is done by \emph{formulating the problem as a graph analysis problem for predicting graph properties, the latter being solved using GNNs} (see Section~\ref{sec:ProblemDefinitionGraphEdgeClassification}).

\subsubsection{Problem Definition (as Constrained Optimization)}\label{sec:ProblemDefinitionConstrainedOptimization}

A detailed description of this section appeared in~\cite{ERASCU2021100664}. We consider a set of $N$ interacting components, \mbox{$C=\{C_1,\ldots, C_N\}$}, to be assigned to a set of $M$ VMs, $V=\{V_1, \ldots, V_M\}$. Each component $C_i$ is characterized by a set of requirements concerning the hardware resources. Each VM, $V_k$, is characterized by a \emph{type}, which is comprised by hardware/software characteristics and leasing price. There are also \emph{structural constraints} describing the interactions between components. The problem is to find: 
\begin{enumerate}
	\item assignment matrix $a$ with binary entries $a_{ik}\in \{0,1\}$ for $i=\overline{1,N}$, $k=\overline{1,M}$, which are interpreted as follows: 
	\begin{equation*} \label{eq:asignMat}
	a_{ik}=
	\left\{\begin{array}{ll}
	1 & \hbox{if }C_{i} \hbox{ is assigned to } V_k\\  
	0 & \hbox{if }C_{i} \hbox{ is not assigned to } V_k.
	\end{array}
	\right.
	\end{equation*}
	\item the type selection vector $\emph{t}$ with integer entries $\emph{t}_k$ for $k=\overline{1,M}$, representing the type (from a predefined set) of each VM leased.
\end{enumerate}
such that:
\begin{inparaenum}[\itshape (i)\upshape]
	\item the structural constraints,
	\item the hardware requirements  (capacity constraints) of all components are satisfied and 
	\item the purchasing/ leasing price is minimized. 
\end{inparaenum}
For Secure Web Container, a solution is below.
\begin{alignat*}{3}
a=\begin{pmatrix}
0 & 0 & 0 & 0 & 1 & 0\\ 
0 & 0 & 1 & 1 & 0 & 0\\ 
0& 1 & 0 & 0 & 0 & 0\\ 
1& 0 & 0 & 0 & 0 & 0\\ 
0 & 1 & 1 & 1 & 0 & 0
\end{pmatrix} && \qquad && \emph{t} = [12, 12, 13, 13, 15, 0]
\end{alignat*}

The \emph{structural constraints} are \emph{application-specific} and are derived by analyzing various case studies (see Section \ref{sec:CaseStudies}). 

\emph{General constraints} are always considered in the formalization and are related to the following: 
\begin{inparaenum}[\itshape (i)\upshape]
	\item \emph{basic allocation} rules, 
	\item \emph{occupancy} criteria,
	\item hardware \emph{capacity} of the VM Offers,
	\item \emph{link} between the VM Offers and the occupied VMs.
\end{inparaenum}

The problem to solve can be stated as a COP as follows. Let $i,j\!=\!\overline{1, N};$ \ $k\!=\!\overline{1, M}; \ h\!=\!\overline{1, H}; \ o\!=\!\overline{1,O}; \ O,n, n_{ij}, m_{ij}\! \in\! \mathbb{N}^*$.
\[
\begin{tabular}{rl}
\emph{\textbf{Minimize}} & $\sum\limits_{k=1}^{M} v_k \cdot \emph{p}_{k}$ \\
\emph{\textbf{Subject to}} & $a_{ik} \in \{0, 1\} $, $v_{k} \in \{0, 1\} $       \\
                             & $t_{k}\in \text{predefined finite set of natural numbers}$\footnotemark\\
\emph{\textbf{General}}      & 
\textbf{\emph{constraints}}\\
\emph{Basic allocation}& $\sum\limits_{k=1}^M a_{i k} \ge 1$ \\
\emph{Occupancy}& $\sum\limits_{i=1}^N a_{ik} \ge 1 \Rightarrow v_k = 1$  \\
\emph{Capacity}& $\sum\limits_{i=1}^{N}{a_{ik}} \cdot \emph{R}_{i}^{h} \le F^{h}_{t_k}$\\
\emph{Link} & $v_k \!\!=\!\! 1   \Rightarrow \bigwedge\limits_{h=1}^{H} \left(r^{h}_{k} \!\! = \! \!\emph{F}_{o}^{h}\right)  \wedge p_{k}\!\! = \! \!P_{o}$\\
            & $\sum_{i=1}^{N}{a_{ik}} = 0  \Rightarrow \emph{t}_{k} = 0$\\
\emph{\textbf{Application-}}& \emph{\textbf{specific}} \emph{\textbf{constraints}}\\
\emph{Conflicts}& $a_{i k}+a_{j k}\leq 1, \quad \forall(i,j) \ \mathcal{R}_{ij}=1$\\
\emph{Co-location}& $a_{i k} = a_{j k}, \quad \forall(i,j) \ \mathcal{D}_{ij}=1$
\end{tabular}
\]
\[
\begin{tabular}{rl}
\emph{Exclusive }  & \emph{deployment} \\
                   & $\mathcal{H}\left(\sum\limits_{k=1}^M a_{i_{1} k}\right) + ... +\mathcal{H}\left(\sum\limits_{k=1}^M a_{i_{q} k}\right)=1$, \\
                   & \ \ for fixed $q\in\{1, ..., N\}$ and \\
                   & \ \ $\mathcal{H}(u) =  \begin{dcases} 
1 & u > 0 \\[-0.2cm]
0 & u = 0 
\end{dcases}$\\
\emph{Require-}&\emph{Provide} \\
&$n_{ij}\sum\limits_{k=1}^{M}a_{i k} \leq m_{ij}\sum\limits_{k=1}^{M}a_{j k}$, \\
& $\ \ \forall (i,j) \mathcal{Q}_{ij}(n_{ij},m_{ij}) = 1$\\
& $\ \ 0 \le n\sum\limits_{k=1}^{M}a_{j k} - \sum\limits_{k=1}^{M}a_{i k} < n, $\\
\emph{Full deployment} & $\sum\limits_{k=1}^M\left(a_{i k}+\mathcal{H}\left(\sum\limits_{j,\mathcal{R}_{ij}=1}a_{jk}\right)\right)=\sum\limits_{k=1}^{M} v_k$\\
\emph{Deployment with} &\emph{bounded number  of instances} \\
                           &$\sum\limits_{i \in  \overline{C}} \sum\limits_{k=1}^{M}a_{i k} \langle \hbox{op}\rangle n$ with \\
                           & $\ \ |\overline{C}|\le N$, $\langle \hbox{op} \rangle \!\in \! \{=,\leq,\geq\}$
\end{tabular}
\]
\noindent where:
\begin{itemize}
	\item $\mathcal{R}_{ij}=1$ if components $i$ and $j$ are in conflict (can not be placed in the same VM);
	\item $\mathcal{D}_{ij}=1$ if components $i$ and $j$ must be co-located (must be placed in the same VM);
	\item $\mathcal{Q}_{ij}(n,m)\!\!=\!\!1$ if $C_i$ requires at least $n$ instances of $C_j$ and $C_j$ can serve at most $m$ instances of $C_i$;
	\item $R_i^h \in \mathbb{N}^{*}$ is the hardware requirement of type $h$ of the component $i$;
	\item $F_{t_k}^{h} \in \mathbb{N}^{*}$ is the hardware characteristic $h$ of the VM of type $t_k$. $P_{t_k}$ is the price of the VM of type $t_k$.
\end{itemize}
One can observe that the constraints are all linear. What makes the problem challenging is an application with large number of components heavily interacting, to be deployed on a large set of VM Offers.

\paragraph{General Constraints} The \emph{basic allocation rules} specify that each component $C_i$ must be allocated to at least one VM, except those being in \emph{Exclusive Deployment} relation. The \emph{occupancy rules} indicate when a certain VM is occupied or not. \emph{Capacity constraints} describe that the total amount of a certain resource type required by the components hosted on a particular VM does not overpass the corresponding resource type of a VM Offer. In order to have a sound formalization, one also needs to \emph{link} a type of a VM Offer to each of the occupied VMs. 
\paragraph{Application-specific Constraints} We identified two main types of application-specific constraints regarding the components: those concerning the \emph{interactions} between components (conflict, co-location, exclusive deployment) and those concerning the \emph{number of instances} (require-provide, full deployment, deployment with a bounded number of instances).

\noindent\emph{Conflict.} This case corresponds to situations where there are conflictual components which cannot be deployed on the same VM. We consider that all conflicts between components are encoded in a matrix $\mathcal{R}$.

\noindent\emph{Co-location.} This means that the components in the collocation relation should be deployed on the same VM. The co-location relation can be stored in a matrix~$\mathcal{D}$.

\noindent\emph{Exclusive deployment.} This corresponds to the cases when only certain components should be deployed.

\noindent\emph{Require-Provide.} A special case of interaction between components is when one component requires some functionalities offered by other components. Such an interaction induces constraints on the number of instances corresponding to the interacting components.

\noindent\emph{Full deployment.}
There can also be cases where a component must be deployed on all leased VMs (except on those that would induce conflicts on components).

\noindent\emph{Deployment with bounded number of instances.} There are situations where the number of instances corresponding to a set of deployed components should be equal to, greater, or less than some values.

\subsubsection{Problem definition (Graph Edge Classification)}\label{sec:ProblemDefinitionGraphEdgeClassification}

Consider the \emph{heterogeneous graph} $G = (V, E)$ with two types of nodes $V$, components and VMs, and edges \(E\). Suppose that each node and each edge is associated with a feature vector. We defined the \emph{edge classification problem} as follows. Given:
\begin{itemize}
    \item a graph \(G\) with its structural information (connectivity),
    \item its node features and edge features,
    \item the set $E$ of edges between components and between components and VMs, where each edge \(e\) is assigned a label $y_e \in \{\texttt{linked}, \ \texttt{unlinked}, \ \texttt{binding}\}$, 
\end{itemize}
 \noindent learn a model that predicts the label \(y_e\) for every edge between components nodes and VM nodes.

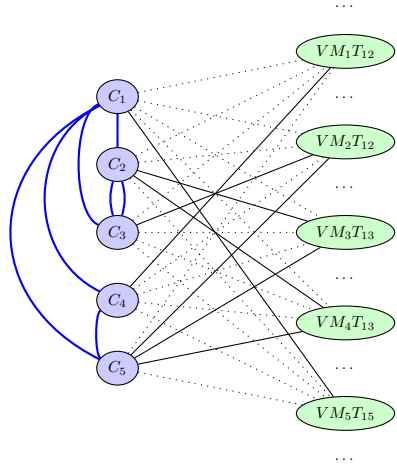
\begin{figure}[htbp]
  \centering
\begin{tikzpicture}[scale=0.6, transform shape, >=Stealth,
    componentNode/.style={draw, ellipse, fill=blue!20},
    vmNode/.style={draw, ellipse, fill=green!20}
]


\node[componentNode] (C1) at (0,-2) {$C_1$};
\node[componentNode] (C2) at (0,-3.5)   {$C_2$};
\node[componentNode] (C3) at (0,-5) {$C_3$};
\node[componentNode] (C4) at (0,-6.5)   {$C_4$};
\node[componentNode] (C5) at (0,-8){$C_5$};

\node at (5,0) {$\dots$};  
\node[vmNode] (VM1T12) at (5,-1)  {$VM_1T_{12}$};
\node at (5,-2) {$\dots$};
\node[vmNode] (VM2T12) at (5,-3)  {$VM_{2}T_{12}$};
\node at (5,-4) {$\dots$};
\node[vmNode] (VM3T13) at (5,-5)  {$VM_3T_{13}$};
\node at (5,-6) {$\dots$};
\node[vmNode] (VM4T13) at (5,-7) {$VM_4T_{13}$};
\node at (5,-8) {$\dots$};
\node[vmNode] (VM5T15) at (5,-9) {$VM_5T_{15}$};
\node at (5,-10) {$\dots$};


\draw[-] (C1) -- (VM5T15);
\draw[-] (C2) -- (VM3T13);
\draw[-] (C2) -- (VM4T13);
\draw[-] (C3) -- (VM2T12);
\draw[-] (C4) -- (VM1T12);
\draw[-] (C5) -- (VM2T12);
\draw[-] (C5) -- (VM3T13);
\draw[-] (C5) -- (VM4T13);


\draw[dotted,-] (C1) -- (VM1T12);
\draw[dotted,-] (C1) -- (VM2T12);
\draw[dotted,-] (C1) -- (VM3T13);
\draw[dotted,-] (C1) -- (VM4T13);

\draw[dotted,-] (C2) -- (VM1T12);
\draw[dotted,-] (C2) -- (VM2T12);
\draw[dotted,-] (C2) -- (VM5T15);

\draw[dotted,-] (C3) -- (VM1T12);
\draw[dotted,-] (C3) -- (VM3T13);
\draw[dotted,-] (C3) -- (VM4T13);
\draw[dotted,-] (C3) -- (VM5T15);

\draw[dotted,-] (C4) -- (VM2T12);
\draw[dotted,-] (C4) -- (VM3T13);
\draw[dotted,-] (C4) -- (VM4T13);
\draw[dotted,-] (C4) -- (VM5T15);

\draw[dotted,-] (C5) -- (VM1T12);
\draw[dotted,-] (C5) -- (VM5T15);


\draw[blue, thick] (C1) -- (C2);

\draw[blue, thick]
      (C1) .. controls +(-1, -0.4) and +(-1, 0.4) .. (C3);

\draw[blue, thick]
      (C1) .. controls +(-2, -1) and +(-2,1) .. (C4);

\draw[blue, thick]
      (C1) .. controls +(-3, -1.5) and +(-3,1.5) .. (C5);

\draw[blue, thick, bend left=15]  (C2) to (C3);
\draw[blue, thick, bend right=15] (C2) to (C3);

\draw[blue, thick]
      (C4) .. controls +(-0.5, -0.3) and +(-0.5, 0.3) .. (C5);

\end{tikzpicture}
\caption{Edge classification problem for Secure Web Container}
  \label{fig:edge-classification-secure-web}
\end{figure}

For example, for the Secure Web Container application, the task is to find a graph similar to the one in Fig.~\ref{fig:edge-classification-secure-web}, where the blue edges are of type \texttt{binding} and represent the application-specific constraints, black ones are of type \texttt{linked} and represent the assignments of component instances to VMs of certain type, dotted black are of type \texttt{unlinked} meaning that the respective component instance is not assigned to the VM of the respective type. The VM type nodes (green) are in total $M \times O$ and were not included in the figure for legibility. Between the missing ones are the nodes of type component are edges of type \texttt{unlinked}.

The first step in solving the edge classification problem is to model the problem as \emph{graph data}, see below.

\paragraph{Component nodes} For each component of the application, there is a corresponding \emph{node of type component} in the graph. For example, in the Secure Web Container application, we have a total of $5$ component nodes. Each component is characterized by the followings: 
\begin{itemize}
    \item the ID of the component
    \item hardware requirements: the CPU requirements (in cores), the memory (RAM) requirements (in MB), the storage requirements (in MB)
    \item application-specific constraints which refer to a single component:
    \begin{itemize}
        \item \emph{Full deployment}. For example, the IDSAgent component should be deployed on all the VMs except those that host the IDSServer and Balancer due to Conflict constraints between IDSServer and Balancer.
        \item \emph{Deployment with bounded number of instances} constraints (upper bound, lower bound, equal bound). For example, the Balancer component has an equal bound of $1$.
    \end{itemize}
\end{itemize}
\noindent These are coded as \emph{node features}, hence each node of type component is described by \emph{$8$ node features}: \emph{(ID, CPU, Mem, Sto, FullDepl, UpperB, LowerB, EqualB).}

The remaining application-specific constraints are encoded using \emph{edges} between component nodes as they involve two or more components. An edge between two components means that there is at least one constraint between them. The actual constraints are specified using \emph{one-hot encoding} on the possible application-specific constraints: \emph{(Conflict, Co-location, RequireProvide, ExclusiveDeployment, UpperB, LowerB, EqualB)}. Note that \emph{Deployment with bounded number of instances} might involve also two or more components, for example, in our case study, there must be at least $3$ instances of Apache and Nginx deployed.

To exemplify the one-hot encoding, the edge between the Balancer and IDSServer components has the features $[1, 0, 0, 0, 0, 0, 0]$ since between the two components there is only one constraint, namely \emph{conflict}.

\paragraph{VM nodes} In order to construct the VM nodes, we have to take into account that the type of VMs leased is not known a priori. Therefore, the total number of  \emph{VM nodes} is $M \cdot O$ (where $M$ is the number of estimated VMs and $O$ is the number of VM Offers). For our case study, $M = 6$, which means that $6$ will be the maximum number of VMs needed to be leased in order to be able to deploy the application. We say maximum because this number can \emph{decrease} if the constraints composition allows it, as it actually is the case, where $5$ VMs are enough to ensure the optimal deployment and the 6th VM remains unassigned. If we have a pool of $20$ VM Offers to choose from, the total number of VM nodes is $120$ as each of the 6 VMs can have one of the $20$ types.

Each node of type VM is characterized by: 
\begin{inparaenum}[\itshape (i)\upshape]
    \item the CPU specifications (in cores)
    \item the memory (RAM) specifications (in MB)
    \item the storage specifications (in MB)
    \item the price of leasing the VM (in USD/hour).
\end{inparaenum}
These specifications are inherited from VM Offer specification and represent the node features of this family of nodes.

Unlike component nodes, there are no interactions between any two VMs so there are \emph{no edges between two VM nodes}.

\subsection{Graph Embedding} \label{sec:GraphEEmbedding}
We denote directed and
labeled multi-graphs as $G = (\mathcal{V}, \mathcal{E}, \mathcal{R})$ with nodes (entities)
$v_i \in \mathcal{V}$ and labeled edges (relations) $(v_i, r, v_j) \in \mathcal{E}$, where
$r \in \mathcal{R}$ is a relation type.

We use relational graph convolutional networks (RGCNs) \cite{schlichtkrull2018modeling} as graph encoder to capture the heterogeneity of the graph and the different types of their edges. The encoder computes \(d\)-dimensional representations for each node in the input graph. In each layer, nodes gather features from their neighbors to represent the local graph structure by recursive message passing \cite{schlichtkrull2018modeling}. Stacking \(L\) layers allows the network to build representations from the \(L\)-hop neighborhood of each node. Let \(h_i^{(\ell)} \in \mathbb{R}^{d}\) denote the \emph{embedding of node} \(i\) at layer \(\ell\). 

In our RGCN implementation, the update at each layer is performed by first applying a relation‐specific graph convolution to aggregate messages from the neighbors of node \(i\) under each relation \(r \in R\) (e.g., \texttt{conflict}, \texttt{linked}, \texttt{unlinked}). Formally, the update rule is given by:

\[
h_i^{(\ell+1)} = \mathrm{ReLU}\left(\sum_{r\in R}\left(\sum_{j\in\mathcal{N}_i^r}\frac{1}{|\mathcal{N}_i^r|}\,W^{(\ell, r)}\, h_j^{(\ell)} + W_0^{(\ell)}\, h_i^{(\ell)}\right)\right),
\]
\noindent where:
\begin{itemize}
\item \(h_i^{(\ell)} \in \mathbb{R}^{d}\) denotes the \(d\)-dimensional embedding for node \(i\) at layer \(\ell\).
\item \(R\) represents the set of relation types in the heterogeneous graph.
\item \(\mathcal{N}_i^r\) denotes the set of neighboring nodes connected to node \(i\) via relation \(r\).
\item \(W^{(\ell, r)} \in \mathbb{R}^{d\times d}\) is a learnable weight matrix for neighbors under relation \(r\) at layer \(\ell\), and \(W_0^{(\ell)} \in \mathbb{R}^{d\times d}\) is a weight matrix for the self-loop (i.e. node \(i\)’s own features).
\item The normalization factor \(\frac{1}{|\mathcal{N}_i^r|}\) implements mean aggregation of the neighboring features.
\item \(\mathrm{ReLU}(x)=\max(0,x)\) is the non-linear activation function.
\end{itemize}
Our implementation relies on standard linear transformations, e.g. matrix multiplication, followed by the ReLU activation. The update was implemented using  \texttt{HeteroGraphConv} module available in DGL \cite{dgl} library, where each relation-specific convolution is applied over the graph’s edges and the results are summed. By stacking two such \texttt{HeteroGraphConv} layers, our network is able to capture local as well as higher-order structural information, and the resulting node embeddings are then fed into a downstream multi-layer predictor (MLP) for tasks such as edge prediction (see Section~\ref{sec:SolutionDecoding}).

\subsection{Solution Decoding} \label{sec:SolutionDecoding}
After the final layer, the edge predictions are computed via a simple linear edge-predictor (MLP) by combining node embeddings of the connected nodes. For an edge connecting nodes \(i\) and \(j\), the classification logits are computed as:
\[
\text{score}_{ij} = W_{\text{edge}}\,\left[h_i^{(\ell+1)}\parallel h_j^{(\ell+1)}\right],
\]
\noindent where:
\begin{itemize}
\item \(W_{\text{edge}}\in\mathbb{R}^{C\times 2d}\) is a learnable parameter, with \(C\) representing the number of edge classes (relation types) and $2d$ is the dimension of the concatenated node embeddings.
\item \(\parallel\) denotes the concatenation operation and combines the two node embeddings \(h_i^{(\ell+1)}\) and \(h_j^{(\ell+1)}\) (each of dimension \(d\)) into a single \(2d\)-dimensional vector. This concatenated vector carries information from both nodes.
\end{itemize}

The output \(\text{score}_{ij}\) is a vector of logits, with length equal to the number of edge classes (e.g., \texttt{conflict}, \texttt{linked}, \texttt{unlinked}).
These logits indicate the likelihood that each edge belongs to a specific class. In our implementation, a softmax operation is applied subsequently to convert logits into class probabilities.

We do not explicitly use edge embeddings, as edge predictions are computed directly from the concatenation of the node embeddings. 

Although predicting edges directly from their explicit embeddings potentially gives richer representation, the approach adopted for predicting edges using only the concatenation of node embeddings is simpler, effective for the problem we are solving in which we categorize which edges between components and VMs are of type \texttt{linked} or \texttt{unlinked}. Explicit edge embeddings introduce additional parameters and computational overhead, which is not necessary since node embeddings already sufficiently capture relational information. In our Python implementation, the existing RGCN-based model computes node embeddings by recursively aggregating messages from neighboring nodes through multiple layers. Consequently, node embeddings implicitly encode structural and relational context, making direct concatenation suitable and computationally efficient for edge prediction tasks. This approach leverages the idea that the representation of a relationship (edge) between two nodes can be accurately inferred from their learned node embeddings alone, without explicitly maintaining separate edge features. 

\subsection{Solution Search} \label{sec:SolutionSearch}
Solution search refers to the mechanism used to navigate through the space of possible solutions. In a first step, a valid assignment based on the predicted scores is constructed, namely the raw edge-level predictions (stored in a tensor) are assembled into a two-dimensional assignment matrix that indicates how components are assigned to VMs. 

In a second step, to evaluate the performance of the trained models, the traditional definition of \emph{accuracy}, being the proportion of correct predictions out of the total number of predictions made by the model, is not useful alone. Given the significantly larger number of \texttt{unlinked} edges compared to \texttt{linked} edges in most applications, a high accuracy could be misleading, possibly indicating an overperformance of models that predominantly predict \texttt{unlinked} edges. To counteract this potential bias, we augment our evaluation with an additional metric focusing specifically on the \texttt{linked} edge type. This metric, which is computed using \emph{predicted True $(\mathbb{T})$ Links} and \emph{predicted False $(\mathbb{F})$ Links}, quantifies the accuracy of \emph{linked} edge predictions exclusively. Note that predicted $\mathbb{T}$ links and $\mathbb{F}$ links are computed by comparing element by element the elements of the assignment matrix $a$ as given by an exact solver, respectively, as predicted by the GNN solution. This is slightly restrictive because the solution given by the exact solver might not be unique.

\subsection{Policy Learning} \label{sec:PolicyLearning}
Policy learning is implemented via an end-to-end supervised training loop, where the model is trained to mimic the optimal solver. The network, comprising a GNN encoder and an edge predictor, is optimized by minimizing the focal loss between predicted and ground-truth edge labels using backpropagation and the Adam optimizer. 

\subsection{Constraint-Guided Solution Refinement} \label{sec:ConstraintGuidedSolutionRefinement}
Although the GNN model provides valuable predictions, it does not guarantee that all application-specific constraints are satisfied. Using these predictions as hard constraints could lead to infeasible solutions, as misclassifications in GNN predictions might incorrectly enforce assignments that violate operational constraints. Instead, we introduce them as soft constraints, allowing the SMT solver to prioritize their satisfaction while still ensuring formal correctness. This approach ensures that GNN guidance accelerates optimization without overriding essential deployment constraints.

The assignment predictions which can be obtained using the GNN model described in Section~\ref{sec:ProblemDefinitionGraphEdgeClassification} can be formalized as a binary 3D tensor $pred$ with $pred_{iko} \in \{0, 1\}$ for $i = \overline{1,N}$, $k = \overline{1,M}$ and $o = \overline{1,O}$ where: 
$$pred_{iko}=\begin{cases}
  1 & \text{ if } C_i \text{ is assigned to } V_j \text{ of type } O_o \\
  0 & \text{ if } C_i \text{ is not assigned to } V_j \text{ of type } O_o
\end{cases}$$
The extra dimension compared to matrix $a$ is induced by the design choice, documented in Section \ref{sec:ProblemDefinitionGraphEdgeClassification}, where when computing the number of VM nodes, we multiply the number $M$ with the total number $O$ of VM Offers. 

Using the 3D tensor $pred$, we can draw conclusions about both the matrix $a$ and the hardware specifications of the VMs (type vector $t$).
\begin{alignat}{3}
& \exists {o \!\in\! \overline{1,O}} \text{ s.t. } pred_{iko} \!= \!1 \!\implies \!a_{ik} \!=\! 1 \!\land\! \bigwedge\limits_{h=1}^{H} \!\!\left(r^{h}_{k} \!\! = \! \!\emph{F}_{o}^{h}\right) \! \wedge \! p_{k}\!\! = \! \!P_{o} \quad \label{eq:pred-a1}\\
& \nexists o \!\in\! \overline{1,O} \text{ s.t. } pred_{iko} \!=\! 1 \!\implies\! a_{ik} \!=\! 0
\label{eq:pred-a0}
\end{alignat}

For Secure Web Container, the $pred$ tensor for the \emph{first component}, generated from running the GNN model prediction on the case study application with 10 VM Offers, looks like: 
$$
\begin{pmatrix}
0 & 0 & 0 & 0 & 0 & 0 & 0 & 0 & 0 & 0 \\
0 & 0 & 0 & 0 & 0 & 0 & 1 & 0 & 0 & 0 \\
0 & 0 & 0 & 0 & 0 & 0 & 1 & 0 & 0 & 0 \\
0 & 0 & 0 & 0 & 1 & 0 & 0 & 0 & 0 & 0 \\
0 & 0 & 0 & 0 & 0 & 0 & 0 & 0 & 0 & 0 \\
0 & 0 & 0 & 0 & 0 & 0 & 0 & 0 & 0 & 0
\end{pmatrix}$$
\noindent where $M = 6$ (number of maximum VMs) rows and $O ~=~ 10$ (number of VM Offers) columns. The corresponding soft constraints are: 
\begin{itemize}
    \item for assignment matrix $a$:
    \begin{small}
    \begin{verbatim}
    (assert-soft (= a11 0))
    (assert-soft (= a12 1))
    (assert-soft (= a13 1))
    (assert-soft (= a14 1))
    (assert-soft (= a15 0))
    (assert-soft (= a16 0))
    \end{verbatim}
    \end{small}
    \item for type vector $t$: 
    \begin{small}
        \begin{verbatim}
(assert-soft (and (= PriceProv2 8.403)...))
(assert-soft (and (= PriceProv3 8.403)...))
(assert-soft (and (= PriceProv4 0.093)...))
\end{verbatim}
\end{small}
where the predictions obtained were $t_2 = 7$, $t_3~=~7$, $t_4 = 5$. We assumed that the VM Offer 7 has the specification $ (CPU, Mem, Sto, Price) = (64, 976000, 1000, 8.403)$ and the VM Offer 5 has the specification $(1, 3750, 1000, 0.093)$.
\end{itemize}

\section{Case Studies}\label{sec:CaseStudies}
To validate our proposed approach, we analyze a set of real-world case studies that highlight the complexities and constraints of \emph{Cloud-based application deployment}. These case studies illustrate the diverse characteristics of Cloud-hosted components and the challenges associated with their \emph{scalability, cost optimization, and constraint satisfaction}.  

Each case study represents a distinct deployment scenario, encompassing security-focused architectures, scalable web services, and complex distributed computing systems. By analyzing these cases, we aim to:
\begin{inparaenum}[\itshape (i)\upshape]
    \item assess the effectiveness of our optimization approach in diverse application settings. 
    \item identify scalability challenges associated with an increasing number of VM Offers. 
    \item examine the accuracy of GNN predictions in guiding optimal deployment strategies.  
\end{inparaenum}
The case studies were chosen to ensure that the findings of our research are practically applicable. In particular, we focus on:  
\begin{inparaenum}[\itshape (i)\upshape]
    \item Secure Web Container, emphasizing security and resilience through distributed intrusion detection and load balancing (see Section~\ref{sec:SWC}).
    \item Secure Billing Email Service, showcasing inter-component dependencies and security constraints in a critical web service (see Section~\ref{sec:SecureBillingEmailService}).  
    \item Oryx2, a large-scale data processing pipeline illustrating the challenges of distributed computing architectures (see Section~\ref{sec:oryx2}). 
    \item WordPress, demonstrating how increasing user demand affects resource allocation and optimization in Cloud environments (see Section~\ref{sec:WordPress}). 
\end{inparaenum}

Note that the description of the case studies is taken from our previous work \cite{ERASCU2021100664}.
\subsection{Secure Web Container}\label{sec:SWC}
Secure Web Container \cite{SWC} provides both resilience and security. It uses redundancy and diversity strategies to fend off attacks and system failures, while also integrating intrusion detection tools for protection from unauthorized access. Resilience is assured by using a set of Web container components (Apache, Nginx) and a Balancer component in order to distribute the requests evenly between the web containers, ensuring load balancing. Intrusion detection is ensured using multiple IDSAgent components that constantly generate intrusion detection reports that are captured by another component, the IDSServer, which then performs detection activities.
\begin{figure}[h!]
    \centering    \includegraphics[scale=0.6]{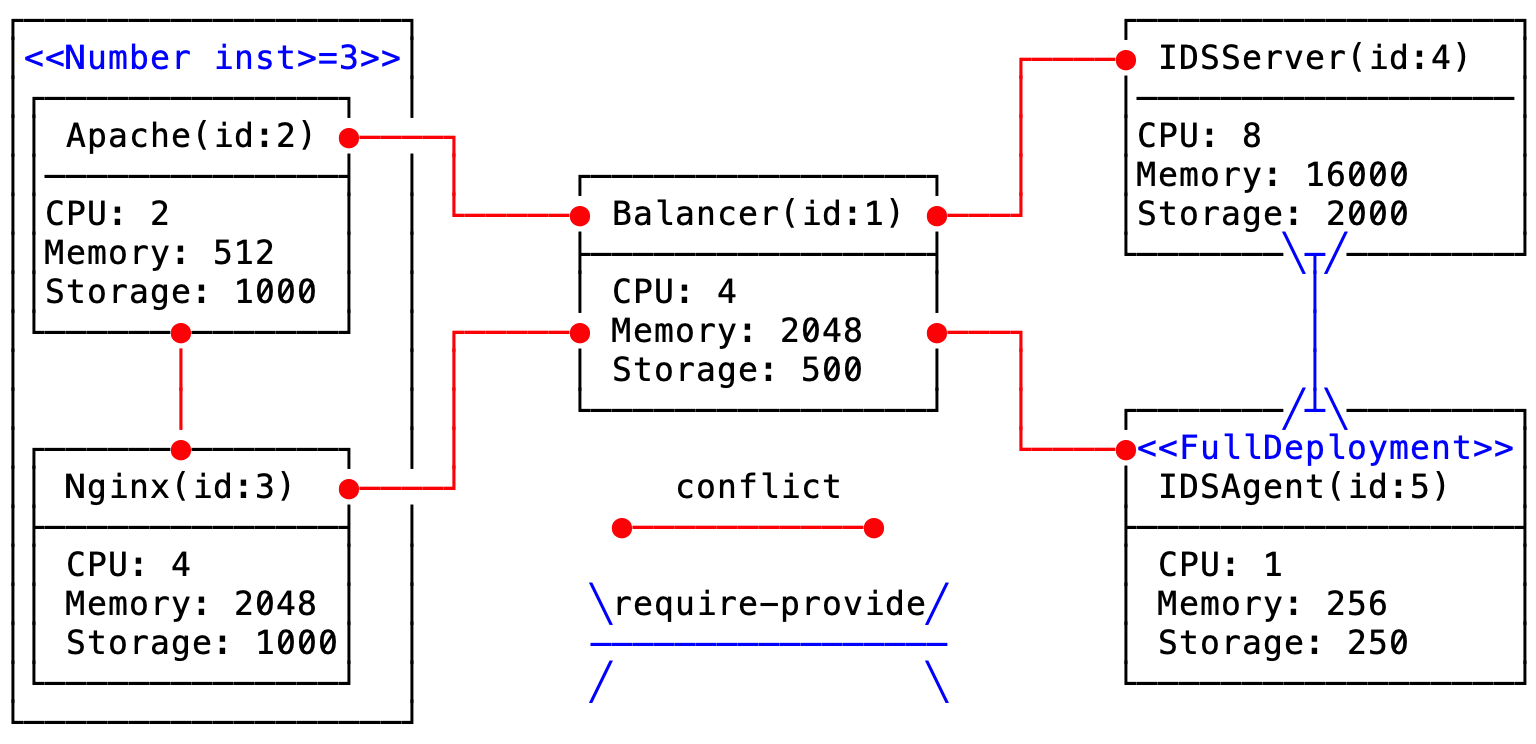}
    \caption{Secure Web Container application ~\cite{ERASCU2021100664}
    }
    \label{fig:SecureWebContainer}
\end{figure}

The \emph{interactions} between the components are as follows.
\begin{inparaenum}[\itshape (1)\upshape]    
\item No two components among the Balancer, Apache and Nginx components can be deployed on the same VM (\emph{Conflict}).
    \item There must be exactly one Balancer component (\emph{Deployment with bounded number of instances constraint}, specifically \emph{equal bound}).
    \item At least 3 instances of the Nginx and Apache components (\emph{Deployment with bounded number of instances constraint}, specifically \emph{lower bound}).
    \item The IDSServer component needs exclusive use of the VM (\emph{Conflict} with all the other components).
    \item For every 10 IDSAgent instances there must be an additional IDSServer instance (\emph{Require-Provide}).
    \item There must be an IDSAgent service instance present on each of the acquired VM, with the exceptions of the machines where the Balancer and IDSServer instances are present (\emph{Full Deployment}).
\end{inparaenum}
\subsection{Secure Billing Email Service}\label{sec:SecureBillingEmailService}
\begin{figure}[h!]
	\centering
	\includegraphics[scale=0.6]{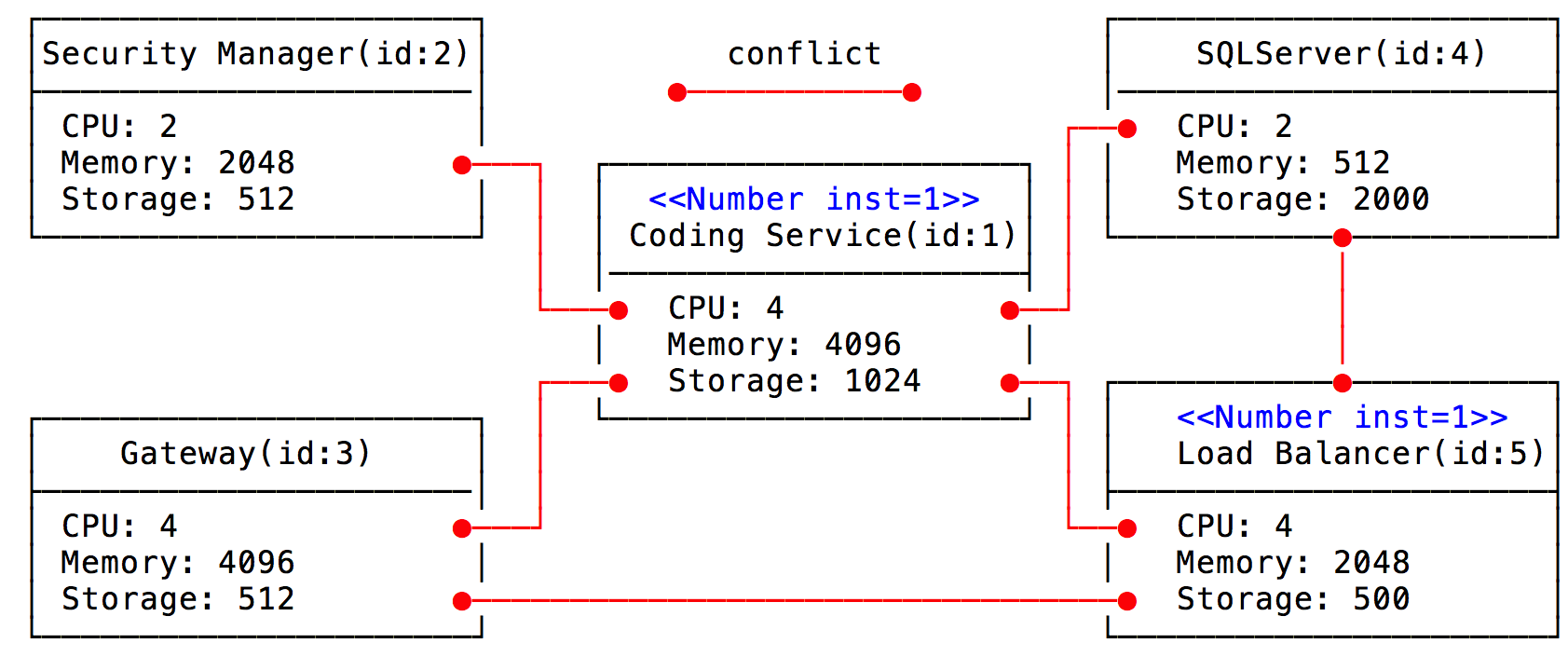}
	\caption{Secure Billing Email Service}
	\label{fig:SecureBillingEmail}
\end{figure}

Secure Billing Email Service (see Figure~\ref{fig:SecureBillingEmail}) has application-specific constraints similar to Secure Web Container. 

In the context of a web application ensuring a secure billing email service (Figure~\ref{fig:SecureBillingEmail}) we consider an architecture consisting of $5$ components: 
\begin{inparaenum}[\itshape (i)\upshape]
	\item a coding service ($C_1$), 
	\item a software manager of the user rights and privileges ($C_2$), 
	\item a gateway component ($C_3$), 
	\item an SQL server ($C_4$) and 
	\item a load balancer ($C_5$).
\end{inparaenum} 
Component $C_1$ should use exclusively a VM, thus it can be considered in \emph{conflict} with all the other components. In such a case the original optimization problem can be decomposed in two subproblems, one corresponding to component $C_1$ and the other one corresponding to the other $4$ components. The first problem is trivial: find the VM with the smallest price which satisfies the hardware requirements of component $C_1$. 

The load balancing component should not be placed on the same machine as the gateway component and the SQL server (\emph{Conflict} constraint). On the other hand, only one instance of components $C_1$ and $C_5$ should be deployed while the other three components could have a larger number of instances placed on different VMs (\emph{Deployment with bounded number of instances} constraint, in particular \emph{equal bound}).
\subsection{Oryx2}\label{sec:oryx2}
\begin{figure}[h!]
	\centering
	\includegraphics[width=0.9\textwidth]{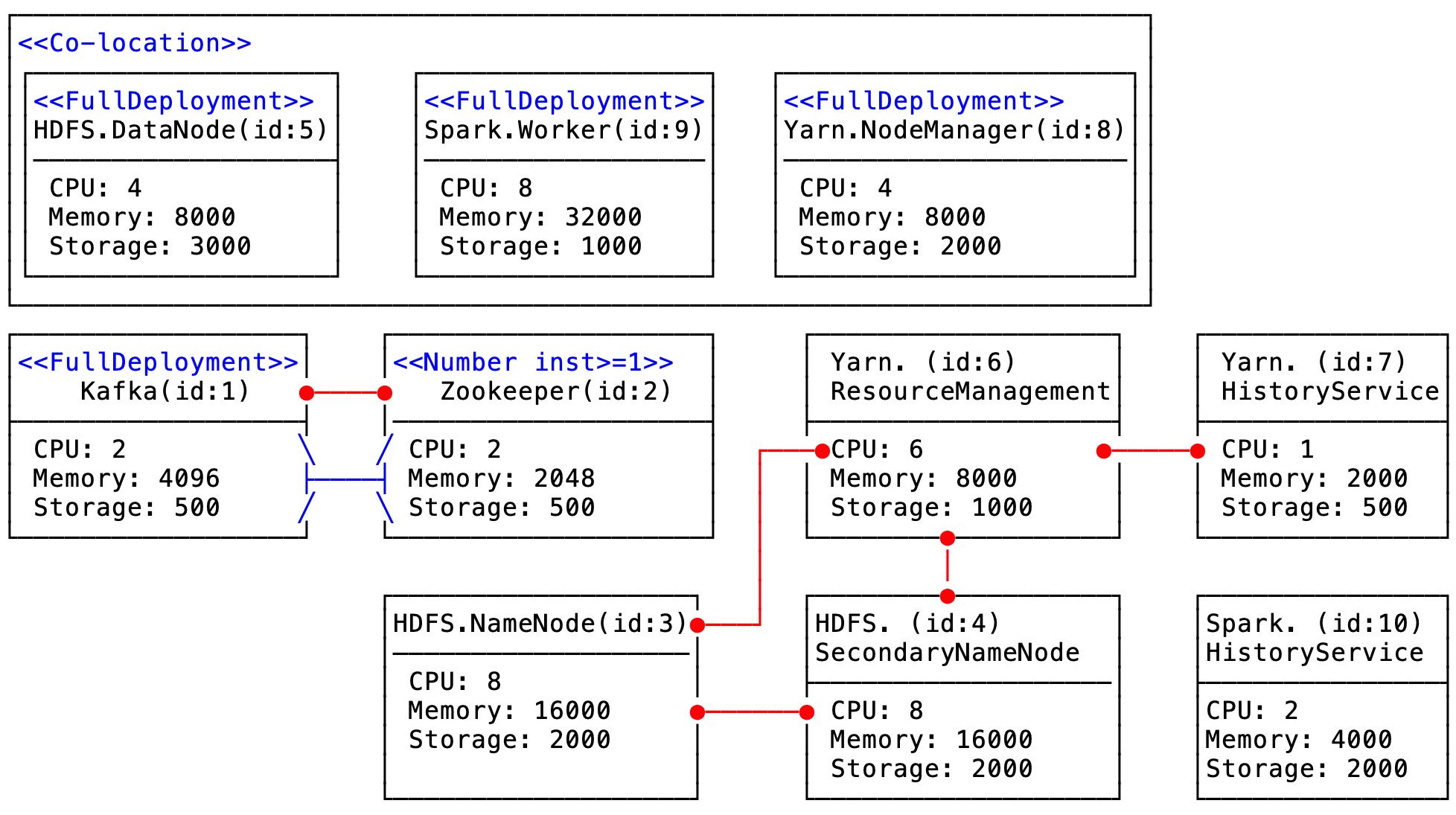}
	\caption{Oryx2 Application}
	\label{fig:Oryx2}
\end{figure}

Oryx2 application (see Figure~\ref{fig:Oryx2}) introduces new types of constraints compared to Secure Web Container

\emph{Oryx2} application (Figure~\ref{fig:Oryx2}) is a realization of the lambda architecture, featuring speed, batch, and serving tiers, with a focus on applying machine learning models in data analysis, and deploys the latest technologies such as Apache Spark\footnote{\url{https://spark.apache.org/}} and Apache Kafka\footnote{\url{https://kafka.apache.org/}}. It has a significant number of components interacting with each other and is highly used in practical applications. It consists of several components which can be distributed over thousands of VMs in the case of a full deployment.
The main goal of Oryx2 is to take incoming data and use them to create and instantiate predictive models for various use-cases, e.g. movie recommendation.
It is comprised of several technologies. Both the batch and serving layer are based on Apache Spark which in turn uses both Apache Yarn\footnote{\url{http://hadoop.apache.org/}} for scheduling and Apache HDFS as a distributed file system. For a processing pipeline Oryx2 uses Apache Kafka with at least two topics; one for incoming data and one for model update. Apache Zookeeper\footnote{\url{https://zookeeper.apache.org/}} is used by Kafka for broker coordination. All of the aforementioned technologies have subservices with a minimum system requirement and recommended deployment as of Figure~\ref{fig:Oryx2}.
The constraints corresponding to the interactions between the components are described in the following.
\begin{inparaenum}[\itshape (i)\upshape]
\item Components HDFS.DataNode and Spark.Worker must the deployed on the same VM  (\emph{Co-location}). In this scenario, we also collocated Yarn.NodeManager because we used Yarn as a scheduler for Spark jobs.  
\item Components Kafka and Zookeeper, HDFS.NameNode and HDFS.SecondaryNameNode, YARN.ResourceManagement and HDFS.NameNode, HDFS.SecondaryNameNode, YARN.Histo\-ryService are, respectively, in \emph{conflict}, that is, they must not be placed on the same VM.
\item Components HDFS.DataNode, YARN.NodeManager and Spark.Worker must be deployed on all VMs except those hosting conflicting components  (\emph{Full Deployment}).
\item In our deployment, we consider that for one instance of Kafka there must be deployed exactly 2 instances of Zookeeper  (\emph{Require-Provide} constraint). There can be situations, however, when more Zookeeper instances are deployed for higher resilience.
\item A single instance of YARN.HistoryService, respectively Spark.HistoryService should be deployed (\emph{Deployment with bounded number of instances} constraint, in particular \emph{equal bound}).
\end{inparaenum}
\subsection{WordPress}\label{sec:WordPress}
\begin{figure}[h!]
	\centering
	\includegraphics[scale=0.6]{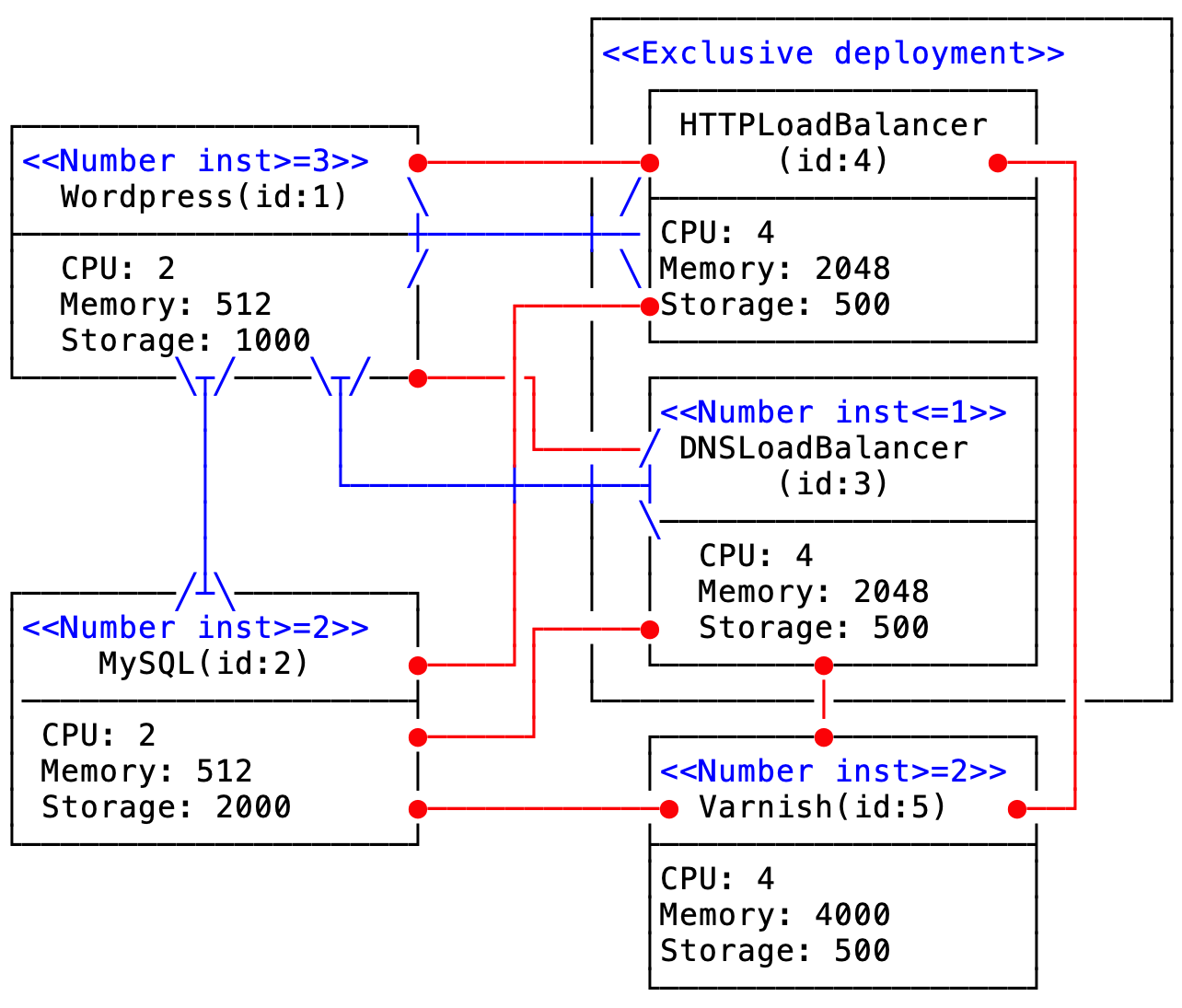}
	\caption{WordPress Application}
	\label{fig:WordPress}
\end{figure}

WordPress application (see Figure~\ref{fig:WordPress}) is especially important because the number of WordPress components can dynamically increase due to user requests, so our approach must also handle increasing number of components. Moreover, this application contains the application-specific constraint \emph{Exclusive deployment} with the semantics that only one component from a given set has to be deployed. This requires the modification of our graph encoding approach as some component nodes should not be considered. This is an ongoing work. Generalization of our approach also for this type of case studies will be studied after the extension of the implementation.

\emph{WordPress} open-source application is frequently used in creating websites, blogs and applications. We chose it in order to compare our approach to Zephyrus and Zephyrus2 deployment tools~\cite{DBLP:conf/kbse/CosmoLTZZEA14,DBLP:conf/setta/AbrahamCJKM16}. In \cite{DBLP:conf/kbse/CosmoLTZZEA14,DBLP:conf/setta/AbrahamCJKM16}, the authors present a high-load and fault tolerant WordPress (Figure~\ref{fig:WordPress}) deployment scenario. The two characteristics are ensured by load balancing. One possibility is to balance load at the DNS level using servers like Bind\footnote{\url{https://www.isc.org/downloads/bind/}}: multiple DNS requests to resolve the website name will result in different IPs from a given pool of machines, on each of which a separate WordPress instance is running. Alternatively one can use as website entry point an HTTP reverse proxy capable of load balancing (and caching, for added benefit) such as Varnish. In both cases, WordPress instances will need to be configured to connect to the same MySQL database, to avoid delivering inconsistent results to users. Also, having redundancy and balancing at the front-end level, one usually expects to have them also at the Database Management System (DBMS) level. One way to achieve that is to use a MySQL cluster, and configure the WordPress instances with multiple entry points to it.
In the deployment scenario considered by us, the following constraints must be fulfilled:
\begin{inparaenum}[\itshape (i)\upshape]
	\item DNSLoadBalancer requires at least one instance of WordPress and DNSLoadBalancer can serve at most 7 WordPress instances (\emph{Require-Provide} constraint).
	\item HTTPLoadBalancer requires at least one WordPress instance and HTTPLoadBalancer can serve at most 3 WordPress instances (\emph{Require-Provide} constraint).
	\item WordPress requires at least 2 instances of MySQL and MySQL can serve at most 3 WordPress (\emph{Require-Provide} constraint).
	\item Only one type of Balancer must be deployed (\emph{Exclusive deployment} constraint).
	\item Since Varnish exhibits load balancing features, it should not be deployed on the same VM as with another type of Balancer (\emph{Conflict} constraint). Moreover, Varnish and MySQL should not be deployed on the same VM because it is best practice to isolate the DBMS level of an application (\emph{Conflict} constraint).
	\item At least 2 instances of Varnish must be deployed too  (\emph{Deployment with bounded number of instances} constraint, in particular \emph{lower bound}).
	\item At least 2 different entry points to the MySQL cluster (\emph{Deployment with bounded number of instance} constraint, in particular \emph{lower bound}).
	\item No more than 1 DNS server deployed in the administrative domain (\emph{Deployment with bounded number of instances} constraint, in particular \emph{upper bound}).
	\item Balancer components must be placed on a single VM, so they are considered to be in \emph{conflict} with all the other components.
\end{inparaenum}
\section{Experimental Setup}\label{sec:ExperimentalSetup}
This section outlines the experimental framework for evaluating our neuro-symbolic optimisation pipeline.
\subsection{Datasets}\label{sec:datasets}
A dataset is used to train, in a supervised manner, models of each of the case studies considered. These models aim to predict as accurately as possible the assignments of components on VM Offers and the leasing price:
\begin{inparaenum}[\itshape (1)\upshape]
\item for applications with similar application-specific constraints, 
\item for applications with similar hardware requirements, and
\item if the list of VMs Offers changes.
\end{inparaenum}

To construct a dataset that accurately represents real-world Cloud deployments, we generate multiple possible VM Offer combinations. We start with a set of 20 VM Offers from various Cloud Providers, each with different hardware configurations and pricing. To ensure diverse scenarios, we generate all possible subsets of 7 VM Offers out of the total 20, leading to approximately 77,000 possible configurations. However, due to the large computational requirements for processing each configuration, we limit the final dataset to a manageable subset.

\begin{figure}[h]
    \centering
\includegraphics[width=0.9\textwidth]{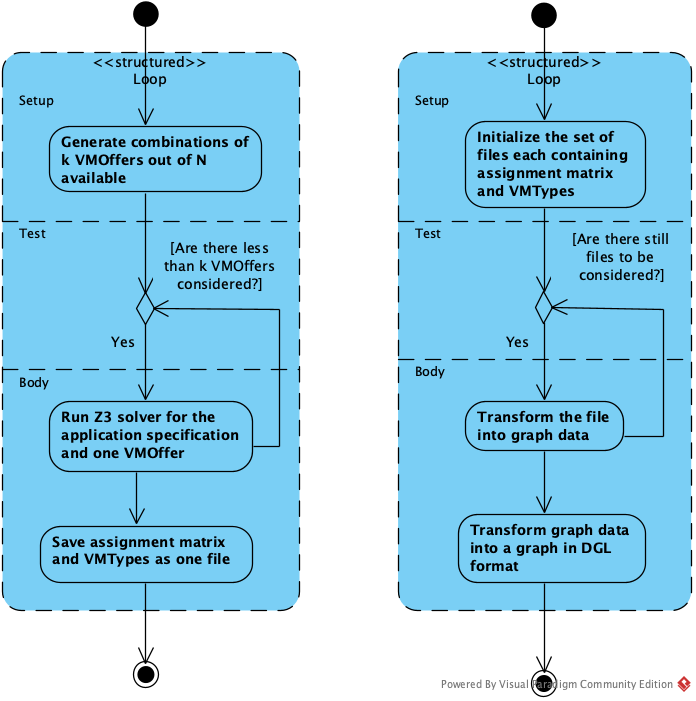}
    \caption{Diagrams showcasing dataset generation process}
    \label{fig:ActDia}
\end{figure}

The dataset generation process is summarized in the two activity diagrams in Fig.~\ref{fig:ActDia}. We decided to break the flow into two separate steps so that the generated output files can be used by any implementation that depends on that dataset, not necessarily a GNN implementation (Figure~\ref{fig:ActDia}-left). Also, the transformation to DGL~\cite{dgl} compliant graph data, the open-source Python library used for deep learning on graph-structured data, is done in two steps (Figure~\ref{fig:ActDia}-right), firstly into generic graph data such that the integration with another graph learning library can be done effortlessly. 
The datasets used for supervised training must show \emph{class balance} and \emph{diverse range of labels} to ensure adaptability and generalization. 
Imbalanced datasets can lead to biased predictions, where the model overfits to majority classes and underperforms on underrepresented ones. To quantify this imbalance, we compute the Gini coefficient, a standard statistical measure used to evaluate class distribution inequality. First, we computed the diversity of labels and the balance of classes for the datasets obtained using the flow from Fig.~\ref{fig:ActDia}-left. The diversity of labels is the number of different minimal prices in the deployment plans. In a classification task, class balance refers to the degree to which samples are distributed evenly in all classes, and this was computed using the Gini coefficient \cite{gini2005measurement} (see Algorithm~\ref{alg:gini}).

\begin{algorithm}
\caption{Compute Gini Coefficient}
\label{alg:gini}
\begin{algorithmic}[1]
\REQUIRE $v$: a list of integers representing the number of occurrences of the $min\_price$
\ENSURE $g$: the Gini coefficinent
\STATE Sort $v$ in ascending order and save the result as $sv$.
\FOR{$i = 0$ \TO $n-1$ /*$n$ is the length of the array $sv$ */ }
\STATE $cv[i] = \sum\limits_{j=0}^{i} sv[j]$  /*compute the cumulative sum of $sv$*/
\ENDFOR
\STATE Compute the sum of elements in $cv$ and assign to $c$.
\STATE Compute the sum of elements in $sv$ and assign to $s$.
\STATE Compute the Gini coefficient:
\(
g = \frac{1}{n} \left( n + 1 - 2\frac{c}{s} \right)
\)
\RETURN $g$
\end{algorithmic}
\end{algorithm}

A Gini coefficient closer to 1 indicates extreme class imbalance, meaning the dataset is heavily skewed toward a few labels. Values above 0.5 suggest significant bias, leading to overfitting on dominant labels, while a Gini of 0.2 – 0.3 ensures a more uniform label distribution. We choose 0.3 as our threshold, as it strikes a balance between maintaining dataset diversity and avoiding model bias.

Given the results of Table \ref{tab:CharacteristicsDatasetsInitialResults}, to mitigate the imbalance of the datasets, we reduced the Gini coefficient to maximum $0.3$ by erasing those labels and the corresponding deployment plans from the dataset. Using Algorithm \ref{alg:reduce_gini1}, 
we ended up with the diversity of labels from Table \ref{tab:CharacteristicsDatasetsFinalResults} (second column). Algorithm~\ref{alg:reduce_gini1} basically removes the element with largest value from the ordered list.

\begin{table}[h!]
\centering
\footnotesize
\renewcommand{\arraystretch}{1.3} 
\setlength{\tabcolsep}{8pt} 
\begin{tabular}{|c|c|c|}
\hline
\textbf{Use Case Name}       & \textbf{\# Distinct Labels} & \textbf{Gini Coefficient} \\ \hline
\textbf{Secure Web Container} & 58                     & 0.6                     \\ \hline
\textbf{Secure Billing Email} & 68                     & 0.66                     \\ \hline
\textbf{Oryx2}                & 1                      & 0                     \\ \hline
\textbf{WordPress}            & 70                     & 0.68                     \\ \hline
\end{tabular}
\caption{Characteristics of the raw datasets} 
\label{tab:CharacteristicsDatasetsInitialResults} 
\end{table}

However, in order to obtain good supervised training results, it is advisable to have a large number of samples of the dataset and the number of samples in each class labeled by $min\_ price$ also large. At this aim, we adopted a manual approach: first we deleted the elements with lowest values from the ordered list computing each time the Gini to become $\le 0.3$. It was also necessary to remove few elements with largest values. In this way, we were left with the distinct labels as in Table \ref{tab:CharacteristicsDatasetsFinalResults} (third column) and the number of samples in the dataset as in \ref{tab:CharacteristicsDatasetsFinalResults} (forth column).

\begin{algorithm}[h!]
\caption{Reduce Gini coefficient}
\label{alg:reduce_gini1}
\begin{algorithmic}[1]
\REQUIRE A list $L$ of positive integers in decreasing order.
\ENSURE Modified list $L$ with Gini coefficient $\leq 0.3$.
\WHILE{Gini coefficient of $L \geq 0.3$}
    \STATE Delete the first element from $L$. /*The first element has largest value*/
    \STATE Compute the Gini coefficient of $L$.
\ENDWHILE
\RETURN $L$.
\end{algorithmic}
\end{algorithm}

\begin{table}[h!]
\centering
\footnotesize
\renewcommand{\arraystretch}{1.3} 
\setlength{\tabcolsep}{4pt} 
\begin{tabular}{|c|c|c|c|}
\hline
\textbf{Use Case} & \textbf{\!\!\! \# Distinct Labels\!\!} & \textbf{\!\!\! \# Distinct Labels \!\!} & \textbf{Dataset} \\ 
\textbf{Name} & \textbf{(Auto)} & \textbf{(Manual)} & \textbf{Size} \\ \hline
\textbf{Secure Web Container} & 12 & 24 & 10582 \\ \hline
\textbf{Secure Billing Email} & 13 & 22 & 7996 \\ \hline
\textbf{Oryx2}                & 1  & 1  & 6925 \\ \hline
\textbf{WordPress}            & 16 & 23 & 5636 \\ \hline
\end{tabular}
\caption{Characteristics of curated datasets} 
\label{tab:CharacteristicsDatasetsFinalResults} 
\end{table}

A notable characteristic of the Oryx2 dataset is that all deployment samples result in the same minimum lease price, making it an interesting case for structured deployment optimization (see Table~\ref{tab:CharacteristicsDatasetsFinalResults}). This is due to the Full Deployment constraint, which ensures that all critical components are deployed across every leased VM. As a result, only VMs with high hardware specifications are feasible, leading to a dataset with a consistent and predictable deployment strategy. While this reduces dataset variability, it also highlights an important aspect of real-world Cloud deployments: when strict operational constraints are in place, deployment decisions become highly structured and optimization becomes more targeted. This consistency provides an opportunity to validate the ability of our model to optimize resource allocation under well-defined conditions, as reflected in our experimental results (see Section \ref{sec:results}).

\subsection{GNNs Architecture and Hyperparameters} The architecture consists of a \emph{Relational Graph Convolutional Network (RGCN)}~\cite{RGCN} layer, used in order to generate the \emph{node representations} for each component and VM node and a \emph{HeteroMLPPredictor} layer, used to generate the \emph{edge class predictions} from the node representations taken as input. The architecture was chosen so because it allows heterogeneity modeling and edge classification. Heterogeneity management is handled by the DGL library.

The RGCN layer consists of \emph{HeteroGraphConv} layers which is a type of layer specially designed for \emph{heterogeneous graphs}. For the Secure Web Container application, we experimented with RGCN layer consisting of:
\begin{enumerate}
    \item two primary \emph{HeteroGraphConv} layers i.e. input and output layer. The hyperparameters of RGCN are:
\begin{inparaenum}[\itshape (1)\upshape]
\item the input features, $8$, representing the component nodes features,
\item hidden features, $10$, and output features, $5$, which were chosen after a preliminary analysis of the architecture performance.
\end{inparaenum}
    \item two primary \emph{HeteroGraphConv} layers and a \emph{HeteroGraphConv} hidden layer;
    \item two primary \emph{HeteroGraphConv} layers and $300$ \emph{HeteroGraphConv} hidden layers. For each of the hidden layers we considered $300$ features.
\end{enumerate}
Two different aggregation functions \texttt{sum} and \texttt{mean} and two different loss functions, cross-entropy and focal were tested. Furthermore, the size of the datasets used for training is as in Table~\ref{tab:CharacteristicsDatasetsFinalResults}, epochs are in $\{100, 1000\}$, batch size in $\{32, 64, 128, 256, 512, 1024\}$.
For the architecture consisting of two primary \emph{HeteroGraphConv} layers, 
For the architecture consisting in two primary \emph{HeteroGraphConv} layers and $300$ \emph{HeteroGraphConv} hidden layers.

Given the preliminary results obtained for the Secure Web Container application with the experimental setup as described above, for the remaining use cases we considered a GNN architecture with the RGCN layer composed of two primary $HeteroGraphConv$ layers and hyperparameters $(8,10,5)$, \texttt{sum} aggregate function, focal loss and parameters as in Table~\ref{tab:Secure-web-container-training}.

For the experiments corresponding to other case studies, we considered a GNN architecture with the RGCN layer composed of two primary $HeteroGraphConv$ layers, \texttt{sum} aggregate function, focal loss. 
The training results are in Tables \ref{tab:secure-billing-email-training}, \ref{tab:oryx-training}, and \ref{tab:WordPress-training} in the Appendix~\ref{sec:Performance-metrics-training-appendix}.

One can observe that the training time \emph{grows approximately linearly} with the size of the dataset in all four use cases when the number of epochs and the size of the batch remain constant. However, the rate of increase varies depending on the dataset characteristics. Secure Web Container exhibits a linear scaling pattern, but the training time is more sensitive to the number of epochs than to the dataset size alone. Secure Billing Email follows a similar trend, where increasing batch size has little impact on reducing training time. In contrast, Oryx2 shows a steeper increase in training time, likely due to more complex constraints among components, making it less scalable than Secure Web Container and Secure Billing Email. WordPress scales more efficiently than Oryx2, as its training time gradually increases with the size of the dataset.

\subsection{SMT Solver}\label{sec:SMTSolver}
The predictions offered by the GNN model are exploited as soft constraints in an optimization solver tailored for the problem we are solving. It results in an optimization problem with the structure as in Listing~\ref{lst:smt-lib-file} in the SMT-LIB format \cite{barrett2010smt}.

\begin{lstlisting}[language=SMTLIB, caption=Template of a constrained optimization problem solved by Z3, label={lst:smt-lib-file}]
; declaration of the variables
(declare-fun C1_VM1 () Int)
...
; declaration of the hard constraints i.e. general and application specific constraints
(assert (Or (= C1_VM4 1) (= C1_VM4 0)))
...
;declaration of the soft constraints i.e. predictions of the GNN model
(assert-soft (= C1_VM4 1) :weight 1)
...
; optimization criteria
(minimize (+ 0 PriceProv1 PriceProv2 PriceProv3 PriceProv4 PriceProv5 PriceProv6))
;verifying the satisfiability
(check-sat)
;obtaining the model
(get-model)
; get the value of the objective functions
(get-objectives)
\end{lstlisting}

The backend of the optimization solver is Z3 \cite{bjorner2015nuz} which handles soft constraints as a MaxSMT problem: the solver tries to satisfy as many soft constraints as possible. Hence, \texttt{assert-soft} constraints are basically forming another optimization function, which leads to a multi-criteria optimization problem: one which maximizes the number of soft constraints, the second minimizing the leasing cost. We used the lexicographic option in Z3 to solve the optimization problem which works as follows: the first objective is optimized, and the values of the decision variables obtained are used in the second objective.

Solving the optimization problem from the Listing \ref{lst:smt-lib-file} lexicographically always ensures that the first objective is optimized. In our case, this means that the application constraints, which are crucial for well-functioning of the application, are always fulfilled. However, the minimal price might not be guaranteed. In order to prioritize price optimality, we could switch the order between the soft constraints and the optimization function; however, we would lose the benefits of the soft constraints that guide the optimization process for time benefits ensuring the scalability of our approach.

\section{Results}\label{sec:results}
\subsection{Scalability of the Hybrid Approach with Varying VM Offers and Component Instance Counts}\label{sec:RQ1}
To assess the scalability of our approach, we systematically evaluated its performance as the number of VM Offers, respectively, the number of component instances application increased. The key objective was to determine whether incorporating GNN predictions as soft constraints accelerates the optimization process. While in most cases, the addition of GNN constraints leads to a \emph{noticeable reduction in computational time}, we observed that in some scenarios, particularly when the GNN predictions were incomplete or inaccurate, the solver was forced to explore a wider search space, counteracting the expected time improvement. This insight is particularly relevant for \emph{fine-tuning GNN training strategies to enhance deployment optimization efficiency}.

We run the experiments with increasing number of VM Offers crawled from various Cloud Providers: 20, 40, 250, 500; each smaller set is included in the larger ones, and the larger ones also contain other offers. A completely new set that counts 27 VM Offers corresponds to the Digital Ocean Cloud Provider \cite{digitalocean}. 

For each case study, we solved 90 optimization problems following the structure provided in Listing~\ref{lst:smt-lib-file}. These problems come from the 18 GNN models we trained for each use case (see Tables~\ref{tab:Secure-web-container-training}--\ref{tab:WordPress-training}), applied to each of the 5 VM Offers sets. The results appear in Figures~\ref{fig:ScalabilityGNNMethodNumberOffersSecureWeb}--\ref{fig:ScalabilityGNNMethodNumberOffersWordPress8}. Along the horizontal axis, each blue dot corresponds to a test ID: for each VM Offer set (separated by a vertical black dotted line), the first test is the optimization problem solved without GNN augmentation (baseline test), while subsequent tests use the GNNs from Tables~\ref{tab:Secure-web-container-training}--\ref{tab:WordPress-training} in sequence. In total, there are 95 reference points on the horizontal axis; the vertical black dotted lines indicate the test without GNN augmentation for offers of 20, 27, 40, 250, and 500, respectively.
Each test was run with a deadline of 40 minutes.

For the Secure Web Container (Figure~\ref{fig:ScalabilityGNNMethodNumberOffersSecureWeb}) and Secure Billing Email (Figure~\ref{fig:ScalabilityGNNMethodNumberOffersSecureBilling}) use cases, all optimization problems were solved within the time frame. However, the improvement in solution time is negligible, and in some instances, it is actually greater than the time required without GNN augmentation. Upon examining the optimization problems that did not see a reduction in time, we found that most of the soft constraints came from GNN models that were unable to predict any VM Offer. Consequently, every possible combination of VM type had to be evaluated, resulting in an increased number of soft constraints.

For the Oryx2 use case (Figure~\ref{fig:ScalabilityGNNMethodNumberOffersOryx2}), no solution was found for the Digital Ocean offers set. This is because the available VMs cannot accommodate the application while meeting its hardware constraints, for example, the
Full Deployment of the three components HDFS.DataNode, Spark.Worker and Yarn.NodeManager.
The results obtained indicate a noticeable improvement in time as the number of offers increases, highlighting the effectiveness of the GNN models.

The WordPress use case exemplifies a realistic scenario where the number of deployed component instances must scale to meet user demand peaks. In our evaluation, we assess the GNN model’s scalability with respect to both the number of available offers and the number of WordPress instances, which introduces slightly modified application-specific constraints as the system scales.

The dataset used to train the GNN model was generated using a configuration of 3 WordPress instances, referred to here as WordPress3. When increasing the number of offers, all 90 optimization problems proved satisfiable, however, the time is not noticeable decreased. In the few problems where runtime did not decrease, we observed that many additional soft constraints had been introduced, often because the GNN model was unable to predict certain VM Offer types. This pattern was similar to what was observed in the Secure Web Container and Secure Billing application, where the addition of numerous soft constraints similarly impacted runtime.

For WordPress4 and WordPress5, additional instances of the WordPress component introduced new application-specific constraints. Reference data were unavailable for an increased number of offer sizes (denoted in the figures by symbol \textcolor{red}{$\times$}), yet in the other cases the solution was provided within the time frame. For the cases where we had reference time, the runtime showed a reduction. Solving the optimization problems GNN-augumented, the phenomenon similar to Secure Web Container and Secure Billing Email appears. Despite this, incorporating soft constraints enabled the solver to find solutions in most of the cases that could not be found by using the exact solver alone.

By the time the system scaled to WordPress7, the exact solver was no longer able to find solutions within the allotted time, illustrating a significant increase in complexity. Nonetheless, the GNN-augmented approach still produced valid solutions, reinforcing its value in handling large-scale, dynamically constrained applications. Given this trend, WordPress8 (and potentially higher-instance deployments) would likewise pose comparable or greater complexity, further highlighting the role of GNN-derived constraints in making the problems tractable within practical time limits.

\subsection{Relationship between GNN Predictions, Solution Time, and the Optimal Solution}\label{sec:RQ2}
As mentioned in Section~\ref{sec:ConstraintGuidedSolutionRefinement}, the GNN model cannot accurately predict the
assignment of components to VMs which is critical for the well-functioning of the application. We solved this issue by prioritizing the fulfillment of the soft-constraints extracted from the GNN model, however, the question remains if the minimal cost can be obtained.

In order to achieve this, from the experiments we performed in Section~\ref{sec:RQ1}, we extracted information about the optimum obtained. The results are in Figures~\ref{fig:price_optimality_secure_web}-- \ref{fig:price_optimality_WordPress8}. Although in the case of GNN scalability, one might infer that an increase in the time taken to obtain a solution indicates that the model did not learn the correct links, the same conclusion does not hold for optimal solutions. For example, in the Secure Web Container scenario, although the time associated with 500 offers is high (see Figure~\ref{fig:ScalabilityGNNMethodNumberOffersSecureWeb}), the optimal price is always correctly identified (see Figure~\ref{fig:price_optimality_secure_web}). For the Oryx2 application, the GNN models help produce the optimal price in every case (see Figure~\ref{fig:Oryx2}).

\subsection{GNNs Tailored for Optimizing Component-to-VM Assignment, Execution Time, and Solution Quality}\label{sec:RQ3}
To identify the most effective GNN configurations for deployment optimization, we analyzed models with different architectures and hyperparameters. The selection of "good models" was based on two primary criteria: 
\begin{inparaenum}[\itshape (1)\upshape]
\item their ability to maintain or improve solver efficiency in terms of runtime, and
\item their capability to generate deployment solutions cost-optimal.
\end{inparaenum}
The tests presented in the following revealed that different architectures yield varying levels of effectiveness across datasets, emphasizing the need for dataset-specific tuning when applying GNN-assisted optimization in real-world Cloud environments.

For Oryx2, all trained GNN models satisfy these criteria. For Secure Billing Email, Secure Web Container, Wordpress3 some models meet these conditions (see Table~\ref{tab:gnn-good-models}).
\begin{table}[h]
\footnotesize
\centering
\begin{tabular}{||c|c|c|c||}
\hline
\textbf{Use Case} & \textbf{Dataset Size} & \textbf{\#Ep} & \textbf{Batch Size} \\
\hline\hline
\multirow{3}{*}{Secure Web Container}  & 1000  & 100  & 32, 64, 128, 256, 512, 1024  \\
                                       & 1000  & 1000 & 512, 1024  \\
                                       & 10582 & 100  & 256, 512, 1024  \\
\hline
Secure Billing Email                   & 1000  & 100  & 1024  \\
\hline
\multirow{3}{*}{WordPress3}           & 1000  & 100  & 512, 1024  \\
                                       & 1000  & 1000 & 64, 128, 256, 512  \\
                                       & 5636  & 100  & 64, 128  \\
\hline
WordPress4                            & 1000  & 100  & 1024  \\
\hline
\multirow{3}{*}{WordPress5}           & 1000  & 100  & 32, 64, 128, 256, 512, 1024  \\
                                       & 1000  & 1000 & 64, 128, 256, 1024  \\
                                       & 5636  & 100  & 32, 64, 256, 512, 1024  \\
\hline
\multirow{3}{*}{WordPress6}           & 1000  & 100  & 32, 64, 128, 256, 512, 1024  \\
                                       & 1000  & 1000 & 64, 128, 256, 512, 1024  \\
                                       & 5636  & 100  & 32, 64, 128, 256, 512, 1024  \\
\hline
\multirow{3}{*}{WordPress7}           & 1000  & 100  & 512, 1024 \\
                                      & 1000  & 1000 & 64, 128, 512  \\
                                       & 5636  & 100  & 64, 128  \\
\hline
\multirow{3}{*}{WordPress8}           & 1000  & 100  & 512  \\
                                       & 1000  & 1000 & 64, 128, 512  \\
                                       & 5636  & 100  & 128  \\
\hline\hline
\end{tabular}
\caption{Top-performing GNN models across Secure Web Container, Secure Billing Email, and WordPress based on time and price criteria}
\label{tab:gnn-good-models}
\end{table}

For WordPress4, due to missing baseline data for offers 250 and 500, comparisons are omitted. However, an optimal model was identified and Figure~\ref{fig:price_optimality_WordPress4} suggests two possible (sub)optimal price solutions for the missing cases. For WordPress5: missing baseline cases for offers 40, 250, and 500 prevent direct comparison. Figure~\ref{fig:price_optimality_WordPress5} provides one or two (sub)optimal price estimates for missing cases. For WordPress6, baseline data is available only for offers 27, so missing cases are excluded. Figure~\ref{fig:price_optimality_WordPress6} identifies one or two likely (sub)optimal price points for the missing cases. For WordPress7 and 8, no baseline cases are available for comparison, but several models offer efficient solutions with potential (sub)optimal price candidates. 

\section{Conclusions and Future Work}\label{sec:ConclusionsAndFuture Work}
In this paper, we investigated how \emph{formal verification} methods can be combined with \emph{machine learning} (in particular, graph neural networks) to improve the scalability and reliability of deploying component-based applications onto Cloud infrastructures. 
By employing a hybrid approach, i.e. training a GNN from synthetically generated configurations to predict potential component-to-VM assignments, we obtained the following key findings: \begin{inparaenum}[\itshape (1)\upshape]
\item \emph{Scalability gains with soft constraints extracted from GNN model}.  In many scenarios, adding GNN-derived soft constraints helped the solver converge faster. However, we observed that inaccuracies in GNN predictions can introduce extra soft constraints that may temporarily slow the solver, underscoring the importance of a well-trained model.
\item \emph{Consistency of optimal solutions.} 
GNN-guided optimization reliably achieves cost-optimal solutions even when predictions are imprecise. While runtime may occasionally increase, solution quality remains unaffected due to the use of soft constraints. This confirms the robustness and effectiveness of the hybrid neuro-symbolic approach.
\item \emph{Use-case specific tuning.} Different Cloud architectures and deployment constraints respond best to different GNN architectures and hyperparameters. This confirms the need for dataset or scenario-specific tuning to maximize both accuracy and solver performance benefits.
\end{inparaenum}

Building on these insights and acknowledging the boundaries of our approach, below are potential directions for further refinement and wider applicability.

\begin{itemize}
\item \emph{Advanced GNN Architectures.} While we used a Relational Graph Convolutional Network (RGCN), alternatives such as Graph Attention Networks or hypergraph-based representations could potentially capture better complex constraints occurring the the application formulation. Future studies could systematically benchmark these architectures to assess their impact on runtime and solution quality.
\item \emph{Integration of GNN and SMT through variable initialization}.
While our current approach adds GNN-based predictions statically, that is, by encoding them as additional constraints in the optimization problem, we plan to explore an alternative that uses the predictions to initialize the solver variables in the Simplex (or other linear programming) algorithm within Z3 which handles the optimization problem. This dynamic guidance would allow the solver to adapt its search path in real time, in contrast to the static approach, where constraints extracted from the GNN predictions can interact badly with those arising from the problem formulation.
\end{itemize}

\section*{Acknowledgements}
We thank Eduard-Brata Laitin for his contribution on the implementation. This work was supported by a grant of the Romanian National Authority for Scientific Research and Innovation, CNCS/CCCDI - UEFISCDI, project number PN-III-P1-1.1-TE-2021-0676, within PNCDI III.

\backmatter





\paragraph{Funding}
This work was supported by a grant of the Romanian National Authority for Scientific Research and Innovation, CNCS/CCCDI - UEFISCDI, project number PN-III-P1-1.1-TE-2021-0676, within PNCDI III.

\paragraph{Data Availability} The source code, benchmarks, and the results of the evaluation can be found at \url{https://github.com/SAGE-Project/SAGE-GNN/tree/Constraints2025}.

\section*{Declarations}

\paragraph{Competing interests} The authors have no relevant financial or non-financial interests to disclose.
\paragraph{Open Access} This article is licensed under a Creative Commons Attribution 4.0 International License, which permits use, sharing, adaptation, distribution and reproduction in any medium or format, as long as you give appropriate credit to the original author(s) and the source, provide a link to the Creative Commons licence, and indicate if changes were made. The images or other third party material in this article are included in the article’s Creative Commons licence, unless indicated otherwise in a credit line to the material. If material is not included in the article’s Creative Commons licence and your intended use is not permitted by statutory
regulation or exceeds the permitted use, you will need to obtain permission directly from the copyright holder.
To view a copy of this licence, visit \url{http://creativecommons.org/licenses/by/4.0/}.







\begin{thebibliography}{39}
\ifx \bisbn   \undefined \def \bisbn  #1{ISBN #1}\fi
\ifx \binits  \undefined \def \binits#1{#1}\fi
\ifx \bauthor  \undefined \def \bauthor#1{#1}\fi
\ifx \batitle  \undefined \def \batitle#1{#1}\fi
\ifx \bjtitle  \undefined \def \bjtitle#1{#1}\fi
\ifx \bvolume  \undefined \def \bvolume#1{\textbf{#1}}\fi
\ifx \byear  \undefined \def \byear#1{#1}\fi
\ifx \bissue  \undefined \def \bissue#1{#1}\fi
\ifx \bfpage  \undefined \def \bfpage#1{#1}\fi
\ifx \blpage  \undefined \def \blpage #1{#1}\fi
\ifx \burl  \undefined \def \burl#1{\textsf{#1}}\fi
\ifx \doiurl  \undefined \def \doiurl#1{\url{https://doi.org/#1}}\fi
\ifx \betal  \undefined \def \betal{\textit{et al.}}\fi
\ifx \binstitute  \undefined \def \binstitute#1{#1}\fi
\ifx \binstitutionaled  \undefined \def \binstitutionaled#1{#1}\fi
\ifx \bctitle  \undefined \def \bctitle#1{#1}\fi
\ifx \beditor  \undefined \def \beditor#1{#1}\fi
\ifx \bpublisher  \undefined \def \bpublisher#1{#1}\fi
\ifx \bbtitle  \undefined \def \bbtitle#1{#1}\fi
\ifx \bedition  \undefined \def \bedition#1{#1}\fi
\ifx \bseriesno  \undefined \def \bseriesno#1{#1}\fi
\ifx \blocation  \undefined \def \blocation#1{#1}\fi
\ifx \bsertitle  \undefined \def \bsertitle#1{#1}\fi
\ifx \bsnm \undefined \def \bsnm#1{#1}\fi
\ifx \bsuffix \undefined \def \bsuffix#1{#1}\fi
\ifx \bparticle \undefined \def \bparticle#1{#1}\fi
\ifx \barticle \undefined \def \barticle#1{#1}\fi
\bibcommenthead
\ifx \bconfdate \undefined \def \bconfdate #1{#1}\fi
\ifx \botherref \undefined \def \botherref #1{#1}\fi
\ifx \url \undefined \def \url#1{\textsf{#1}}\fi
\ifx \bchapter \undefined \def \bchapter#1{#1}\fi
\ifx \bbook \undefined \def \bbook#1{#1}\fi
\ifx \bcomment \undefined \def \bcomment#1{#1}\fi
\ifx \oauthor \undefined \def \oauthor#1{#1}\fi
\ifx \citeauthoryear \undefined \def \citeauthoryear#1{#1}\fi
\ifx \endbibitem  \undefined \def \endbibitem {}\fi
\ifx \bconflocation  \undefined \def \bconflocation#1{#1}\fi
\ifx \arxivurl  \undefined \def \arxivurl#1{\textsf{#1}}\fi
\csname PreBibitemsHook\endcsname

\bibitem[\protect\citeauthoryear{Cappart et~al.}{2021}]{GNN-opt}
\begin{botherref}
\oauthor{\bsnm{Cappart}, \binits{Q.}},
\oauthor{\bsnm{Ch{\'e}telat}, \binits{D.}},
\oauthor{\bsnm{Khalil}, \binits{E.}},
\oauthor{\bsnm{Lodi}, \binits{A.}},
\oauthor{\bsnm{Morris}, \binits{C.}},
\oauthor{\bsnm{Veli{\v{c}}kovi{\'c}}, \binits{P.}}:
{Combinatorial Optimization and Reasoning with Graph Neural Networks}.
arXiv preprint arXiv:2102.09544
(2021)
\end{botherref}
\endbibitem

\bibitem[\protect\citeauthoryear{Scott et~al.}{2021}]{scott2021machsmt}
\begin{bchapter}
\bauthor{\bsnm{Scott}, \binits{J.}},
\bauthor{\bsnm{Niemetz}, \binits{A.}},
\bauthor{\bsnm{Preiner}, \binits{M.}},
\bauthor{\bsnm{Nejati}, \binits{S.}},
\bauthor{\bsnm{Ganesh}, \binits{V.}}:
\bctitle{{{MachSMT}: A Machine Learning-based Algorithm Selector for {SMT} Solvers}}.
In: \bbtitle{International Conference on Tools and Algorithms for the Construction and Analysis of Systems},
pp. \bfpage{303}--\blpage{325}
(\byear{2021}).
\bcomment{Springer}
\end{bchapter}
\endbibitem

\bibitem[\protect\citeauthoryear{Pimpalkhare et~al.}{2021}]{pimpalkhare2021medleysolver}
\begin{bchapter}
\bauthor{\bsnm{Pimpalkhare}, \binits{N.}},
\bauthor{\bsnm{Mora}, \binits{F.}},
\bauthor{\bsnm{Polgreen}, \binits{E.}},
\bauthor{\bsnm{Seshia}, \binits{S.A.}}:
\bctitle{{{MedleySolver}: Online {SMT} Algorithm Selection}}.
In: \bbtitle{Theory and Applications of Satisfiability Testing--SAT 2021: 24th International Conference, Barcelona, Spain, July 5-9, 2021, Proceedings 24},
pp. \bfpage{453}--\blpage{470}
(\byear{2021}).
\bcomment{Springer}
\end{bchapter}
\endbibitem

\bibitem[\protect\citeauthoryear{Leeson et~al.}{2023}]{leeson2023sibyl}
\begin{bchapter}
\bauthor{\bsnm{Leeson}, \binits{W.}},
\bauthor{\bsnm{Dwyer}, \binits{M.B.}},
\bauthor{\bsnm{Filieri}, \binits{A.}}:
\bctitle{{Sibyl: Improving Software Engineering Tools with {SMT} Selection}}.
In: \bbtitle{2023 IEEE/ACM 45th International Conference on Software Engineering (ICSE)},
pp. \bfpage{2185}--\blpage{2197}
(\byear{2023}).
\bcomment{IEEE}
\end{bchapter}
\endbibitem

\bibitem[\protect\citeauthoryear{Jennings and Stadler}{2015}]{10.1007/s10922-014-9307-7}
\begin{barticle}
\bauthor{\bsnm{Jennings}, \binits{B.}},
\bauthor{\bsnm{Stadler}, \binits{R.}}:
\batitle{{Resource Management in Clouds: Survey and Research Challenges}}.
\bjtitle{J. Netw. Syst. Manage.}
\bvolume{23}(\bissue{3}),
\bfpage{567}--\blpage{619}
(\byear{2015})
\end{barticle}
\endbibitem

\bibitem[\protect\citeauthoryear{De~Moura and Bj{\o}rner}{2008}]{de2008z3}
\begin{bchapter}
\bauthor{\bsnm{De~Moura}, \binits{L.}},
\bauthor{\bsnm{Bj{\o}rner}, \binits{N.}}:
\bctitle{{{Z3}: An Efficient {SMT} Solver}}.
In: \bbtitle{Tools and Algorithms for the Construction and Analysis of Systems: 14th International Conference, TACAS 2008, Held as Part of the Joint European Conferences on Theory and Practice of Software, ETAPS 2008, Budapest, Hungary, March 29-April 6, 2008. Proceedings 14},
pp. \bfpage{337}--\blpage{340}
(\byear{2008}).
\bcomment{Springer}
\end{bchapter}
\endbibitem

\bibitem[\protect\citeauthoryear{Eraşcu}{2024}]{10650114}
\begin{bchapter}
\bauthor{\bsnm{Eraşcu}, \binits{M.}}:
\bctitle{{Fast and Exact Synthesis of Application Deployment Plans using Graph Neural Networks and Satisfiability Modulo Theory}}.
In: \bbtitle{2024 International Joint Conference on Neural Networks (IJCNN)},
pp. \bfpage{1}--\blpage{10}
(\byear{2024}).
\doiurl{10.1109/IJCNN60899.2024.10650114}
\end{bchapter}
\endbibitem

\bibitem[\protect\citeauthoryear{Eraşcu et~al.}{2021}]{ERASCU2021100664}
\begin{barticle}
\bauthor{\bsnm{Eraşcu}, \binits{M.}},
\bauthor{\bsnm{Micota}, \binits{F.}},
\bauthor{\bsnm{Zaharie}, \binits{D.}}:
\batitle{{Scalable Optimal Deployment in the Cloud of Component-based Applications using Optimization Modulo Theory, Mathematical Programming and Symmetry Breaking}}.
\bjtitle{Journal of Logical and Algebraic Methods in Programming}
\bvolume{121},
\bfpage{100664}
(\byear{2021})
\end{barticle}
\endbibitem

\bibitem[\protect\citeauthoryear{Bengio et~al.}{2021}]{bengio2021machine}
\begin{barticle}
\bauthor{\bsnm{Bengio}, \binits{Y.}},
\bauthor{\bsnm{Lodi}, \binits{A.}},
\bauthor{\bsnm{Prouvost}, \binits{A.}}:
\batitle{Machine learning for combinatorial optimization: a methodological tour d’horizon}.
\bjtitle{European Journal of Operational Research}
\bvolume{290}(\bissue{2}),
\bfpage{405}--\blpage{421}
(\byear{2021})
\end{barticle}
\endbibitem

\bibitem[\protect\citeauthoryear{Gasse et~al.}{2019}]{gasse2019exact}
\begin{botherref}
\oauthor{\bsnm{Gasse}, \binits{M.}},
\oauthor{\bsnm{Ch{\'e}telat}, \binits{D.}},
\oauthor{\bsnm{Ferroni}, \binits{N.}},
\oauthor{\bsnm{Charlin}, \binits{L.}},
\oauthor{\bsnm{Lodi}, \binits{A.}}:
Exact combinatorial optimization with graph convolutional neural networks.
Advances in neural information processing systems
\textbf{32}
(2019)
\end{botherref}
\endbibitem

\bibitem[\protect\citeauthoryear{Wilder et~al.}{2019}]{wilder2019melding}
\begin{bchapter}
\bauthor{\bsnm{Wilder}, \binits{B.}},
\bauthor{\bsnm{Dilkina}, \binits{B.}},
\bauthor{\bsnm{Tambe}, \binits{M.}}:
\bctitle{Melding the data-decisions pipeline: Decision-focused learning for combinatorial optimization}.
In: \bbtitle{Proceedings of the AAAI Conference on Artificial Intelligence},
vol. \bseriesno{33},
pp. \bfpage{1658}--\blpage{1665}
(\byear{2019})
\end{bchapter}
\endbibitem

\bibitem[\protect\citeauthoryear{Bello et~al.}{2016}]{bello2016neural}
\begin{botherref}
\oauthor{\bsnm{Bello}, \binits{I.}},
\oauthor{\bsnm{Pham}, \binits{H.}},
\oauthor{\bsnm{Le}, \binits{Q.V.}},
\oauthor{\bsnm{Norouzi}, \binits{M.}},
\oauthor{\bsnm{Bengio}, \binits{S.}}:
Neural combinatorial optimization with reinforcement learning.
arXiv preprint arXiv:1611.09940
(2016)
\end{botherref}
\endbibitem

\bibitem[\protect\citeauthoryear{Kipf and Welling}{2016}]{kipf2016semi}
\begin{botherref}
\oauthor{\bsnm{Kipf}, \binits{T.N.}},
\oauthor{\bsnm{Welling}, \binits{M.}}:
Semi-supervised classification with graph convolutional networks.
arXiv preprint arXiv:1609.02907
(2016)
\end{botherref}
\endbibitem

\bibitem[\protect\citeauthoryear{Nazari et~al.}{2018}]{nazari2018reinforcement}
\begin{botherref}
\oauthor{\bsnm{Nazari}, \binits{M.}},
\oauthor{\bsnm{Oroojlooy}, \binits{A.}},
\oauthor{\bsnm{Snyder}, \binits{L.}},
\oauthor{\bsnm{Tak{\'a}c}, \binits{M.}}:
Reinforcement learning for solving the vehicle routing problem.
Advances in neural information processing systems
\textbf{31}
(2018)
\end{botherref}
\endbibitem

\bibitem[\protect\citeauthoryear{Yolcu and P{\'o}czos}{2019}]{yolcu2019learning}
\begin{botherref}
\oauthor{\bsnm{Yolcu}, \binits{E.}},
\oauthor{\bsnm{P{\'o}czos}, \binits{B.}}:
Learning local search heuristics for boolean satisfiability.
Advances in Neural Information Processing Systems
\textbf{32}
(2019)
\end{botherref}
\endbibitem

\bibitem[\protect\citeauthoryear{Selsam et~al.}{2018}]{selsam2018learning}
\begin{botherref}
\oauthor{\bsnm{Selsam}, \binits{D.}},
\oauthor{\bsnm{Lamm}, \binits{M.}},
\oauthor{\bsnm{B{\"u}nz}, \binits{B.}},
\oauthor{\bsnm{Liang}, \binits{P.}},
\oauthor{\bsnm{Moura}, \binits{L.}},
\oauthor{\bsnm{Dill}, \binits{D.L.}}:
Learning a sat solver from single-bit supervision.
arXiv preprint arXiv:1802.03685
(2018)
\end{botherref}
\endbibitem

\bibitem[\protect\citeauthoryear{Huang et~al.}{2019}]{huang2019coloring}
\begin{botherref}
\oauthor{\bsnm{Huang}, \binits{J.}},
\oauthor{\bsnm{Patwary}, \binits{M.}},
\oauthor{\bsnm{Diamos}, \binits{G.}}:
Coloring big graphs with alphagozero.
arXiv preprint arXiv:1902.10162
(2019)
\end{botherref}
\endbibitem

\bibitem[\protect\citeauthoryear{Joshi et~al.}{2022}]{joshi2022learning}
\begin{barticle}
\bauthor{\bsnm{Joshi}, \binits{C.K.}},
\bauthor{\bsnm{Cappart}, \binits{Q.}},
\bauthor{\bsnm{Rousseau}, \binits{L.-M.}},
\bauthor{\bsnm{Laurent}, \binits{T.}}:
\batitle{Learning the travelling salesperson problem requires rethinking generalization}.
\bjtitle{Constraints}
\bvolume{27}(\bissue{1}),
\bfpage{70}--\blpage{98}
(\byear{2022})
\end{barticle}
\endbibitem

\bibitem[\protect\citeauthoryear{Sato et~al.}{2019}]{sato2019approximation}
\begin{botherref}
\oauthor{\bsnm{Sato}, \binits{R.}},
\oauthor{\bsnm{Yamada}, \binits{M.}},
\oauthor{\bsnm{Kashima}, \binits{H.}}:
Approximation ratios of graph neural networks for combinatorial problems.
Advances in Neural Information Processing Systems
\textbf{32}
(2019)
\end{botherref}
\endbibitem

\bibitem[\protect\citeauthoryear{Veli{\v{c}}kovi{\'c} et~al.}{2019}]{velivckovic2019neural}
\begin{botherref}
\oauthor{\bsnm{Veli{\v{c}}kovi{\'c}}, \binits{P.}},
\oauthor{\bsnm{Ying}, \binits{R.}},
\oauthor{\bsnm{Padovano}, \binits{M.}},
\oauthor{\bsnm{Hadsell}, \binits{R.}},
\oauthor{\bsnm{Blundell}, \binits{C.}}:
Neural execution of graph algorithms.
arXiv preprint arXiv:1910.10593
(2019)
\end{botherref}
\endbibitem

\bibitem[\protect\citeauthoryear{Corso et~al.}{2020}]{corso2020principal}
\begin{barticle}
\bauthor{\bsnm{Corso}, \binits{G.}},
\bauthor{\bsnm{Cavalleri}, \binits{L.}},
\bauthor{\bsnm{Beaini}, \binits{D.}},
\bauthor{\bsnm{Li{\`o}}, \binits{P.}},
\bauthor{\bsnm{Veli{\v{c}}kovi{\'c}}, \binits{P.}}:
\batitle{Principal neighbourhood aggregation for graph nets}.
\bjtitle{Advances in neural information processing systems}
\bvolume{33},
\bfpage{13260}--\blpage{13271}
(\byear{2020})
\end{barticle}
\endbibitem

\bibitem[\protect\citeauthoryear{Xu et~al.}{2019}]{xu2019can}
\begin{botherref}
\oauthor{\bsnm{Xu}, \binits{K.}},
\oauthor{\bsnm{Li}, \binits{J.}},
\oauthor{\bsnm{Zhang}, \binits{M.}},
\oauthor{\bsnm{Du}, \binits{S.S.}},
\oauthor{\bsnm{Kawarabayashi}, \binits{K.-i.}},
\oauthor{\bsnm{Jegelka}, \binits{S.}}:
What can neural networks reason about?
arXiv preprint arXiv:1905.13211
(2019)
\end{botherref}
\endbibitem

\bibitem[\protect\citeauthoryear{Senior et~al.}{2020}]{senior2020improved}
\begin{barticle}
\bauthor{\bsnm{Senior}, \binits{A.W.}},
\bauthor{\bsnm{Evans}, \binits{R.}},
\bauthor{\bsnm{Jumper}, \binits{J.}},
\bauthor{\bsnm{Kirkpatrick}, \binits{J.}},
\bauthor{\bsnm{Sifre}, \binits{L.}},
\bauthor{\bsnm{Green}, \binits{T.}},
\bauthor{\bsnm{Qin}, \binits{C.}},
\bauthor{\bsnm{{\v{Z}}{\'\i}dek}, \binits{A.}},
\bauthor{\bsnm{Nelson}, \binits{A.W.}},
\bauthor{\bsnm{Bridgland}, \binits{A.}}, \betal:
\batitle{Improved protein structure prediction using potentials from deep learning}.
\bjtitle{Nature}
\bvolume{577}(\bissue{7792}),
\bfpage{706}--\blpage{710}
(\byear{2020})
\end{barticle}
\endbibitem

\bibitem[\protect\citeauthoryear{Mirhoseini et~al.}{2021}]{mirhoseini2021graph}
\begin{barticle}
\bauthor{\bsnm{Mirhoseini}, \binits{A.}},
\bauthor{\bsnm{Goldie}, \binits{A.}},
\bauthor{\bsnm{Yazgan}, \binits{M.}},
\bauthor{\bsnm{Jiang}, \binits{J.W.}},
\bauthor{\bsnm{Songhori}, \binits{E.}},
\bauthor{\bsnm{Wang}, \binits{S.}},
\bauthor{\bsnm{Lee}, \binits{Y.-J.}},
\bauthor{\bsnm{Johnson}, \binits{E.}},
\bauthor{\bsnm{Pathak}, \binits{O.}},
\bauthor{\bsnm{Nova}, \binits{A.}}, \betal:
\batitle{A graph placement methodology for fast chip design}.
\bjtitle{Nature}
\bvolume{594}(\bissue{7862}),
\bfpage{207}--\blpage{212}
(\byear{2021})
\end{barticle}
\endbibitem

\bibitem[\protect\citeauthoryear{Mirhoseini et~al.}{2017}]{mirhoseini2017device}
\begin{bchapter}
\bauthor{\bsnm{Mirhoseini}, \binits{A.}},
\bauthor{\bsnm{Pham}, \binits{H.}},
\bauthor{\bsnm{Le}, \binits{Q.V.}},
\bauthor{\bsnm{Steiner}, \binits{B.}},
\bauthor{\bsnm{Larsen}, \binits{R.}},
\bauthor{\bsnm{Zhou}, \binits{Y.}},
\bauthor{\bsnm{Kumar}, \binits{N.}},
\bauthor{\bsnm{Norouzi}, \binits{M.}},
\bauthor{\bsnm{Bengio}, \binits{S.}},
\bauthor{\bsnm{Dean}, \binits{J.}}:
\bctitle{Device placement optimization with reinforcement learning}.
In: \bbtitle{International Conference on Machine Learning},
pp. \bfpage{2430}--\blpage{2439}
(\byear{2017}).
\bcomment{PMLR}
\end{bchapter}
\endbibitem

\bibitem[\protect\citeauthoryear{Bresson and Laurent}{2019}]{bresson2019two}
\begin{botherref}
\oauthor{\bsnm{Bresson}, \binits{X.}},
\oauthor{\bsnm{Laurent}, \binits{T.}}:
A two-step graph convolutional decoder for molecule generation.
arXiv preprint arXiv:1906.03412
(2019)
\end{botherref}
\endbibitem

\bibitem[\protect\citeauthoryear{Jin et~al.}{2018}]{jin2018junction}
\begin{bchapter}
\bauthor{\bsnm{Jin}, \binits{W.}},
\bauthor{\bsnm{Barzilay}, \binits{R.}},
\bauthor{\bsnm{Jaakkola}, \binits{T.}}:
\bctitle{Junction tree variational autoencoder for molecular graph generation}.
In: \bbtitle{International Conference on Machine Learning},
pp. \bfpage{2323}--\blpage{2332}
(\byear{2018}).
\bcomment{PMLR}
\end{bchapter}
\endbibitem

\bibitem[\protect\citeauthoryear{{\'{A}}brah{\'{a}}m et~al.}{2016}]{DBLP:conf/setta/AbrahamCJKM16}
\begin{bchapter}
\bauthor{\bsnm{{\'{A}}brah{\'{a}}m}, \binits{E.}},
\bauthor{\bsnm{Corzilius}, \binits{F.}},
\bauthor{\bsnm{Johnsen}, \binits{E.B.}},
\bauthor{\bsnm{Kremer}, \binits{G.}},
\bauthor{\bsnm{Mauro}, \binits{J.}}:
\bctitle{{Zephyrus2: On the Fly Deployment Optimization Using {SMT} and {CP} Technologies}}.
In: \bbtitle{Dependable Software Engineering: Theories, Tools, and Applications - Second International Symposium, {SETTA} 2016, Beijing, China, November 9-11, 2016, Proceedings},
pp. \bfpage{229}--\blpage{245}
(\byear{2016})
\end{bchapter}
\endbibitem

\bibitem[\protect\citeauthoryear{Joshi et~al.}{2019}]{TSP}
\begin{botherref}
\oauthor{\bsnm{Joshi}, \binits{C.K.}},
\oauthor{\bsnm{Laurent}, \binits{T.}},
\oauthor{\bsnm{Bresson}, \binits{X.}}:
{An Efficient Graph Convolutional Network Technique for the Travelling Salesman Problem}.
arXiv preprint arXiv:1906.01227
(2019)
\end{botherref}
\endbibitem

\bibitem[\protect\citeauthoryear{Applegate et~al.}{2009}]{applegate2009certification}
\begin{barticle}
\bauthor{\bsnm{Applegate}, \binits{D.L.}},
\bauthor{\bsnm{Bixby}, \binits{R.E.}},
\bauthor{\bsnm{Chv{\'a}tal}, \binits{V.}},
\bauthor{\bsnm{Cook}, \binits{W.}},
\bauthor{\bsnm{Espinoza}, \binits{D.G.}},
\bauthor{\bsnm{Goycoolea}, \binits{M.}},
\bauthor{\bsnm{Helsgaun}, \binits{K.}}:
\batitle{{Certification of an Optimal TSP Tour Through 85,900 Cities}}.
\bjtitle{Operations Research Letters}
\bvolume{37}(\bissue{1}),
\bfpage{11}--\blpage{15}
(\byear{2009})
\end{barticle}
\endbibitem

\bibitem[\protect\citeauthoryear{Schlichtkrull et~al.}{2018}]{schlichtkrull2018modeling}
\begin{bchapter}
\bauthor{\bsnm{Schlichtkrull}, \binits{M.}},
\bauthor{\bsnm{Kipf}, \binits{T.N.}},
\bauthor{\bsnm{Bloem}, \binits{P.}},
\bauthor{\bsnm{Van Den~Berg}, \binits{R.}},
\bauthor{\bsnm{Titov}, \binits{I.}},
\bauthor{\bsnm{Welling}, \binits{M.}}:
\bctitle{Modeling relational data with graph convolutional networks}.
In: \bbtitle{The Semantic Web: 15th International Conference, ESWC 2018, Heraklion, Crete, Greece, June 3--7, 2018, Proceedings 15},
pp. \bfpage{593}--\blpage{607}
(\byear{2018}).
\bcomment{Springer}
\end{bchapter}
\endbibitem

\bibitem[\protect\citeauthoryear{}{}]{dgl}
\begin{botherref}
{Deep Graph Library}.
\url{https://www.dgl.ai/}.
Accessed: Jan 12th, 2024
\end{botherref}
\endbibitem

\bibitem[\protect\citeauthoryear{Casola et~al.}{2017}]{SWC}
\begin{barticle}
\bauthor{\bsnm{Casola}, \binits{V.}},
\bauthor{\bsnm{De~Benedictis}, \binits{A.}},
\bauthor{\bsnm{Eraşcu}, \binits{M.}},
\bauthor{\bsnm{Modic}, \binits{J.}},
\bauthor{\bsnm{Rak}, \binits{M.}}:
\batitle{{Automatically Enforcing Security {SLAs} in the {Cloud}}}.
\bjtitle{IEEE Transactions on Services Computing}
\bvolume{10}(\bissue{5}),
\bfpage{741}--\blpage{755}
(\byear{2017})
\end{barticle}
\endbibitem

\bibitem[\protect\citeauthoryear{Cosmo et~al.}{2014}]{DBLP:conf/kbse/CosmoLTZZEA14}
\begin{bchapter}
\bauthor{\bsnm{Cosmo}, \binits{R.D.}},
\bauthor{\bsnm{Lienhardt}, \binits{M.}},
\bauthor{\bsnm{Treinen}, \binits{R.}},
\bauthor{\bsnm{Zacchiroli}, \binits{S.}},
\bauthor{\bsnm{Zwolakowski}, \binits{J.}},
\bauthor{\bsnm{Eiche}, \binits{A.}},
\bauthor{\bsnm{Agahi}, \binits{A.}}:
\bctitle{{Automated Synthesis and Deployment of Cloud Applications}}.
In: \bbtitle{{ACM/IEEE} International Conference on Automated Software Engineering, {ASE} '14, Vasteras, Sweden - September 15 - 19, 2014},
pp. \bfpage{211}--\blpage{222}
(\byear{2014})
\end{bchapter}
\endbibitem

\bibitem[\protect\citeauthoryear{Gini}{2005}]{gini2005measurement}
\begin{barticle}
\bauthor{\bsnm{Gini}, \binits{C.}}:
\batitle{{On the Measurement of Concentration and Variability of Characters}}.
\bjtitle{METRON-International Journal of Statistics}
\bvolume{63}(\bissue{1}),
\bfpage{1}--\blpage{38}
(\byear{2005})
\end{barticle}
\endbibitem

\bibitem[\protect\citeauthoryear{Schlichtkrull et~al.}{2018}]{RGCN}
\begin{bchapter}
\bauthor{\bsnm{Schlichtkrull}, \binits{M.}},
\bauthor{\bsnm{Kipf}, \binits{T.N.}},
\bauthor{\bsnm{Bloem}, \binits{P.}},
\bauthor{\bsnm{Berg}, \binits{R.}},
\bauthor{\bsnm{Titov}, \binits{I.}},
\bauthor{\bsnm{Welling}, \binits{M.}}:
\bctitle{{Modeling Relational Data with Graph Convolutional Networks}}.
In: \beditor{\bsnm{Gangemi}, \binits{A.}},
\beditor{\bsnm{Navigli}, \binits{R.}},
\beditor{\bsnm{Vidal}, \binits{M.-E.}},
\beditor{\bsnm{Hitzler}, \binits{P.}},
\beditor{\bsnm{Troncy}, \binits{R.}},
\beditor{\bsnm{Hollink}, \binits{L.}},
\beditor{\bsnm{Tordai}, \binits{A.}},
\beditor{\bsnm{Alam}, \binits{M.}} (eds.)
\bbtitle{The Semantic Web},
pp. \bfpage{593}--\blpage{607}.
\bpublisher{Springer},
\blocation{Cham}
(\byear{2018})
\end{bchapter}
\endbibitem

\bibitem[\protect\citeauthoryear{Barrett et~al.}{2010}]{barrett2010smt}
\begin{bchapter}
\bauthor{\bsnm{Barrett}, \binits{C.}},
\bauthor{\bsnm{Stump}, \binits{A.}},
\bauthor{\bsnm{Tinelli}, \binits{C.}}, \betal:
\bctitle{{The SMT-LIB Standard: Version 2.0}}.
In: \bbtitle{Proceedings of the 8th International Workshop on Satisfiability Modulo Theories (Edinburgh, UK)},
vol. \bseriesno{13},
p. \bfpage{14}
(\byear{2010})
\end{bchapter}
\endbibitem

\bibitem[\protect\citeauthoryear{Bj{\o}rner et~al.}{2015}]{bjorner2015nuz}
\begin{bchapter}
\bauthor{\bsnm{Bj{\o}rner}, \binits{N.}},
\bauthor{\bsnm{Phan}, \binits{A.-D.}},
\bauthor{\bsnm{Fleckenstein}, \binits{L.}}:
\bctitle{{$\nu$Z - An Optimizing SMT Solver}}.
In: \bbtitle{Tools and Algorithms for the Construction and Analysis of Systems: 21st International Conference, TACAS 2015, Held as Part of the European Joint Conferences on Theory and Practice of Software, ETAPS 2015, London, UK, April 11-18, 2015, Proceedings 21},
pp. \bfpage{194}--\blpage{199}
(\byear{2015}).
\bcomment{Springer}
\end{bchapter}
\endbibitem

\bibitem[\protect\citeauthoryear{}{}]{digitalocean}
\begin{botherref}
{Digital Ocean}.
\url{https://www.digitalocean.com/}.
Accessed: Jan 12th, 2024
\end{botherref}
\endbibitem

\end{thebibliography}

\begin{appendices}
\section{Performance metrics for varying dataset and batch sizes, number of epochs}\label{sec:Performance-metrics-training-appendix}
\begin{table}[h!]
  \centering
  \footnotesize
  \caption{Performance metrics for varying dataset sizes, epochs and batch sizes for \emph{Secure Web Container}}
  \label{tab:Secure-web-container-training}
  \begin{tabular}{cccccccc}
    \toprule
        \multirow{2}{*}{\shortstack{\textbf{Dataset}\\ \textbf{Size}}}
      & \multirow{2}{*}{\textbf{\#Ep}} 
      & \multirow{2}{*}{\shortstack{\textbf{Batch}\\ \textbf{Size}}} 
      & \multirow{2}{*}{\shortstack{\textbf{Time}\\\textbf{(in sec.)}}} 
      & \multicolumn{2}{c}{\textbf{Predicted}} 
      & \multirow{2}{*}{\shortstack{\textbf{GT $\mathbb{T}$}\\\textbf{Links}}} 
      & \multirow{2}{*}{\textbf{Acc}} \\
    \cmidrule(lr){5-6}
    & & & & \textbf{$\mathbb{T}$} & \textbf{$\mathbb{F}$} & & \\
    \midrule[\heavyrulewidth]  
    \multirow{12}{*}{1000} 
      & \multirow{6}{*}{100}
      &   32   &  224.07 &         \multirow{4}{*}{8}       & 17 &   \multirow{18}{*}{8}   & 0.91 \\
      \cmidrule(lr){3-4}  \cmidrule(lr){6-6}\cmidrule(lr){8-8}
      &       &   64   &  224.27 &        & 16 &      & \multirow{2}{*}{0.92} \\
      &       &  128   &  221.81 &        & 17 &      &      \\
      \cmidrule(lr){3-4}\cmidrule(lr){6-6}\cmidrule(lr){8-8}
      &       &  256   &  227.83 &        & 14 &      &   \multirow{3}{*}{0.93}   \\
      \cmidrule(lr){3-6} 
      &       &  512   &  234.03 & \multirow{2}{*}{7} & \multirow{2}{*}{13} &      &  \\
      &       & 1024   &  230.51 &        &    &      &      \\
    \cmidrule(lr){2-6}  \cmidrule(lr){8-8}%
      & \multirow{6}{*}{1000}
      &   32   & 2247.25 & \multirow{12}{*}{8} & \multirow{2}{*}{17} &   & \multirow{11}{*}{0.91} \\
      &       &   64   & 2212.59 &                    &  &                   &                   \\
      \cmidrule(lr){3-4}  \cmidrule(lr){6-6}
      &       &  128   & 2464.99 &                    & 18 &                   &                   \\
      \cmidrule(lr){3-4}  \cmidrule(lr){6-6}
      &       &  256   & 2469.77 &                    & \multirow{3}{*}{17} &                   &                   \\
      &       &  512   & 2531.11 &                    &  &                   &                   \\
      &       & 1024   & 2275.88 &                    &    &                   &                   \\
    \cmidrule(lr){1-4}  \cmidrule(lr){6-6}  
    \multirow{6}{*}{10582} 
      & \multirow{6}{*}{100}
      &   32   & 2608.78 &  & \multirow{2}{*}{18} &      &  \\
      &       &   64   & 2631.43 &                    &    &      &                   \\
      \cmidrule(lr){3-4}  \cmidrule(lr){6-6}
      &       &  128   & 2655.73 &                    & \multirow{3}{*}{17} &      &                   \\
      &       &  256   & 2695.72 &                    &    &      &                   \\
      &       &  512   & 2678.12 &                    &    &      &                   \\
      \cmidrule(lr){3-4}  \cmidrule(lr){6-6} \cmidrule(lr){8-8}
      &       & 1024   & 2802.34 &                    & 16 &      &       0.92            \\
    \bottomrule
  \end{tabular}
\end{table}
\begin{table}[h!]
  \centering\footnotesize
  \caption{Performance metrics for varying dataset sizes, epochs and batch sizes for \emph{Secure Billing Email}}
  \label{tab:secure-billing-email-training}
  \begin{tabular}{cccccccc}
    \toprule
        \multirow{2}{*}{\shortstack{\textbf{Dataset}\\ \textbf{Size}}}
      & \multirow{2}{*}{\textbf{\#Ep}} 
      & \multirow{2}{*}{\shortstack{\textbf{Batch}\\ \textbf{Size}}} 
      & \multirow{2}{*}{\shortstack{\textbf{Time}\\\textbf{(s)}}} 
      & \multicolumn{2}{c}{\textbf{Predicted}} 
      & \multirow{2}{*}{\shortstack{\textbf{GT $\mathbb{T}$}\\\textbf{Links}}} 
      & \multirow{2}{*}{\textbf{Acc}} \\
    \cmidrule(lr){5-6}
\cmidrule(lr){5-6} & & & & \textbf{$\mathbb{T}$} & \textbf{$\mathbb{F}$} & & \\
\midrule[\heavyrulewidth]
    \multirow{12}{*}{1000} & \multirow{6}{*}{100}
      & 32   & 252.09  &  \multirow{5}{*}{5} & \multirow{4}{*}{16} & \multirow{18}{*}{8} & \multirow{4}{*}{0.9} \\
    & & 64    & 249.77  &  &  &                     &  \\
    & & 128   & 249.75  &  &  &                     &  \\
    & & 256   & 257.92  &  &  &                     &  \\
    \cmidrule(lr){3-4} \cmidrule(lr){6-6} \cmidrule(lr){8-8}
    & & 512   & 254.82  &  & 10 &                     & 0.94 \\
    \cmidrule(lr){3-6} \cmidrule(lr){8-8}
    & & 1024  & 259.34  & 4 & 11 &                     & 0.93 \\
    \cmidrule(lr){2-6} \cmidrule(lr){8-8}
     & \multirow{6}{*}{1000}
      & 32   & 2468.82 & \multirow{12}{*}{5} & \multirow{2}{*}{16} &                     & \multirow{3}{*}{0.91} \\
    & & 64    & 2452.91 &  &  &                     & \\
    \cmidrule(lr){3-4} \cmidrule(lr){6-6}
    & & 128   & 2476.23 &  & 15 &                     &  \\
    \cmidrule(lr){3-4} \cmidrule(lr){6-6} \cmidrule(lr){8-8}
    & & 256   & 2511.02 &  & \multirow{9}{*}{16} &                     & 0.9 \\
    \cmidrule(lr){3-4}\cmidrule(lr){8-8}
    & & 512   & 2576.07 &  &  &                     & 0.91 \\
    \cmidrule(lr){3-4}\cmidrule(lr){8-8}
    & & 1024  & 2890.11 &  &  &                     & 0.9 \\
    \cmidrule(lr){1-4} \cmidrule(lr){8-8}
    \multirow{6}{*}{7996} & \multirow{6}{*}{100}
      & 32   & 2062.47 &  &  &                     & 0.91 \\
     \cmidrule(lr){3-4}  \cmidrule(lr){8-8}
    & & 64    & 2085.82 &  &  &                     & 0.9 \\
    \cmidrule(lr){3-4} \cmidrule(lr){8-8}
    & & 128   & 2128.80 &  &  &                     & 0.91 \\
    \cmidrule(lr){3-4} \cmidrule(lr){8-8}
    & & 256   & 2126.93 &  &  &                     & \multirow{3}{*}{0.9} \\
    & & 512   & 2132.07 &  &  &                     &  \\
    & & 1024  & 2177.04 &  &  &                     &  \\
    \bottomrule
\end{tabular}
\end{table}
\begin{table}[h!]
  \centering\footnotesize
  \caption{Performance metrics for varying dataset sizes, epochs and batch sizes for \textit{Oryx2}}
  \label{tab:oryx-training}
  \begin{tabular}{cccccccc}
    \toprule
      \multirow{2}{*}{\shortstack{\textbf{Dataset}\\ \textbf{Size}}}
    & \multirow{2}{*}{\textbf{\#Ep}}
    & \multirow{2}{*}{\shortstack{\textbf{Batch}\\ \textbf{Size}}}
    & \multirow{2}{*}{\shortstack{\textbf{Time}\\\textbf{(s)}}}
    & \multicolumn{2}{c}{\textbf{Predicted}}
    & \multirow{2}{*}{\shortstack{\textbf{GT $\mathbb{T}$}\\\textbf{Links}}}
    & \multirow{2}{*}{\textbf{Acc}} \\
    \cmidrule(lr){5-6}
      &  &  &  & \textbf{$\mathbb{T}$} & \textbf{$\mathbb{F}$} &  &  \\
    \midrule[\heavyrulewidth]

    \multirow{12}{*}{1000} & \multirow{6}{*}{100}
      & 32   & 353.15  &  \multirow{5}{*}{25} & \multirow{2}{*}{31} & \multirow{18}{*}{8} & \multirow{8}{*}{0.99} \\
    & & 64   & 368.35  &                     &                     &                     &                      \\
    \cmidrule(lr){6-6}
    & & 128  & 380.30  &                     & 32                  &                     &                      \\
    \cmidrule(lr){6-6}
    & & 256  & 393.93  &                     & 28                  &                     &                      \\
    \cmidrule(lr){6-6}
    & & 512  & 387.28  &                     & 26                  &                     &                      \\
    \cmidrule(lr){3-6}
    & & 1024 & 428.84  & 24                  & 29                  &                     &  \\[2pt]
    \cmidrule(lr){3-6}
    & \multirow{6}{*}{1000}
      & 32   & 6001.85 &  \multirow{6}{*}{25} & 30                  &                     &  \\
    \cmidrule(lr){2-4}\cmidrule(lr){6-6}
    & & 64   & 3339.76 &                     & \multirow{3}{*}{29} &                     &  \\
    \cmidrule(lr){8-8}
    & & 128  & 10274.86&                     &                     &                     & 0.98 \\
    \cmidrule(lr){8-8}
    & & 256  & 9954.72 &                     &                     &                     & \multirow{9}{*}{0.99}\\
    \cmidrule(lr){6-6}
    & & 512  & 13445.49&                     & \multirow{2}{*}{31} &                     &  \\
    & & 1024 & 8621.70 &                     &                     &                     &  \\[2pt]
    \cmidrule(lr){1-4}\cmidrule(lr){6-6}

    \multirow{6}{*}{6925} & \multirow{6}{*}{100}
      & 32   & 9337.18 &                     & \multirow{3}{*}{29} &                     &  \\
    & & 64   & 11610.07&                     &                     &                     &                      \\
    & & 128  & 6357.35 &                     &                     &                     &                      \\
    \cmidrule(lr){6-6}
    & & 256  & 5922.65 &                     & 31                  &                     &                      \\
    \cmidrule(lr){5-6}
    & & 512  & 9110.26 & 24                  & 32                  &                     &                      \\
    \cmidrule(lr){5-6}
    & & 1024 & 10343.80& 25                  & 28                  &                     &                      \\

    \bottomrule
  \end{tabular}
\end{table}
\begin{table}[h!]
  \centering \footnotesize
  \caption{Performance metrics for varying dataset sizes, epochs and batch sizes for \textit{WordPress}}
  \label{tab:WordPress-training}
  \begin{tabular}{cccccccc}
    \toprule
      \multirow{2}{*}{\shortstack{\textbf{Dataset}\\ \textbf{Size}}}
    & \multirow{2}{*}{\textbf{\#Ep}}
    & \multirow{2}{*}{\shortstack{\textbf{Batch}\\ \textbf{Size}}}
    & \multirow{2}{*}{\shortstack{\textbf{Time}\\\textbf{(s)}}}
    & \multicolumn{2}{c}{\textbf{Predicted}}
    & \multirow{2}{*}{\shortstack{\textbf{GT $\mathbb{T}$}\\\textbf{Links}}}
    & \multirow{2}{*}{\textbf{Acc}} \\
    \cmidrule(lr){5-6}
      &  &  &  & \textbf{$\mathbb{T}$} & \textbf{$\mathbb{F}$} &  &  \\
    \midrule[\heavyrulewidth]

    \multirow{12}{*}{1000} & \multirow{6}{*}{100}
      & 32   & 248.47 &  \multirow{4}{*}{8} & \multirow{3}{*}{15} & \multirow{18}{*}{26} & \multirow{4}{*}{0.94}  \\
      
    & & 64   & 235.94 &                     &                     &                      &                      \\
    & & 128  & 271.76 &                     &                     &                      &                      \\
    \cmidrule(lr){6-6}
    & & 256  & 286.53 &                     & 17                  &                      &                      \\
    \cmidrule(lr){5-6}\cmidrule(lr){8-8}
    & & 512  & 266.26 & 7                   & 20                  &                      & 0.92 \\
    \cmidrule(lr){5-6}\cmidrule(lr){8-8}
    & & 1024 & 278.56 & 5                   & 19                  &                      & 0.93 \\[2pt]
    \cmidrule(lr){2-6}\cmidrule(lr){8-8}
    & \multirow{6}{*}{1000}
      & 32   & 3139.90 & \multirow{12}{*}{8} & \multirow{2}{*}{14}                  &                      & 0.94 \\
    \cmidrule(lr){8-8}
    & & 64   & 10239.77&                     &                   &                      & \multirow{2}{*}{0.95} \\
    \cmidrule(lr){6-6}
    & & 128  & 6833.36 &                     & 13                  &                      &                      \\
    \cmidrule(lr){6-6}\cmidrule(lr){8-8}
    & & 256  & 4262.86 &                     & \multirow{3}{*}{15} &                      & \multirow{3}{*}{0.94} \\
    & & 512  & 3089.20 &                     &                     &                      &                      \\
    & & 1024 & 3823.11 &                     &                     &                      &                      \\
    [2pt]\cmidrule(lr){1-4}\cmidrule(lr){6-6}\cmidrule(lr){8-8}
    \multirow{6}{*}{5636} & \multirow{6}{*}{100}
      & 32   & 1370.32 &   & 13                  &                      & \multirow{2}{*}{0.95} \\
    \cmidrule(lr){6-6}
    & & 64   & 2345.47 &                     & 14                  &                      &                      \\
    \cmidrule(lr){6-6}\cmidrule(lr){8-8}
    & & 128  & 2915.26 &                     & \multirow{3}{*}{15} &                      & \multirow{4}{*}{0.94} \\ 
    & & 256  & 2107.45 &                     &                     &                      &                      \\
    & & 512  & 2250.04 &                     &                     &                      &                      \\
    \cmidrule(lr){6-6}
    & & 1024 & 2266.26 &                     & 16                  &                      &                      \\

    \bottomrule
  \end{tabular}
\end{table}

\section{Experimental evaluation of the hybrid approach}\label{sec:Experimental-evaluation-hybrid-approach}
\begin{figure}[h!]
    \centering
\includegraphics[width=0.9\textwidth]{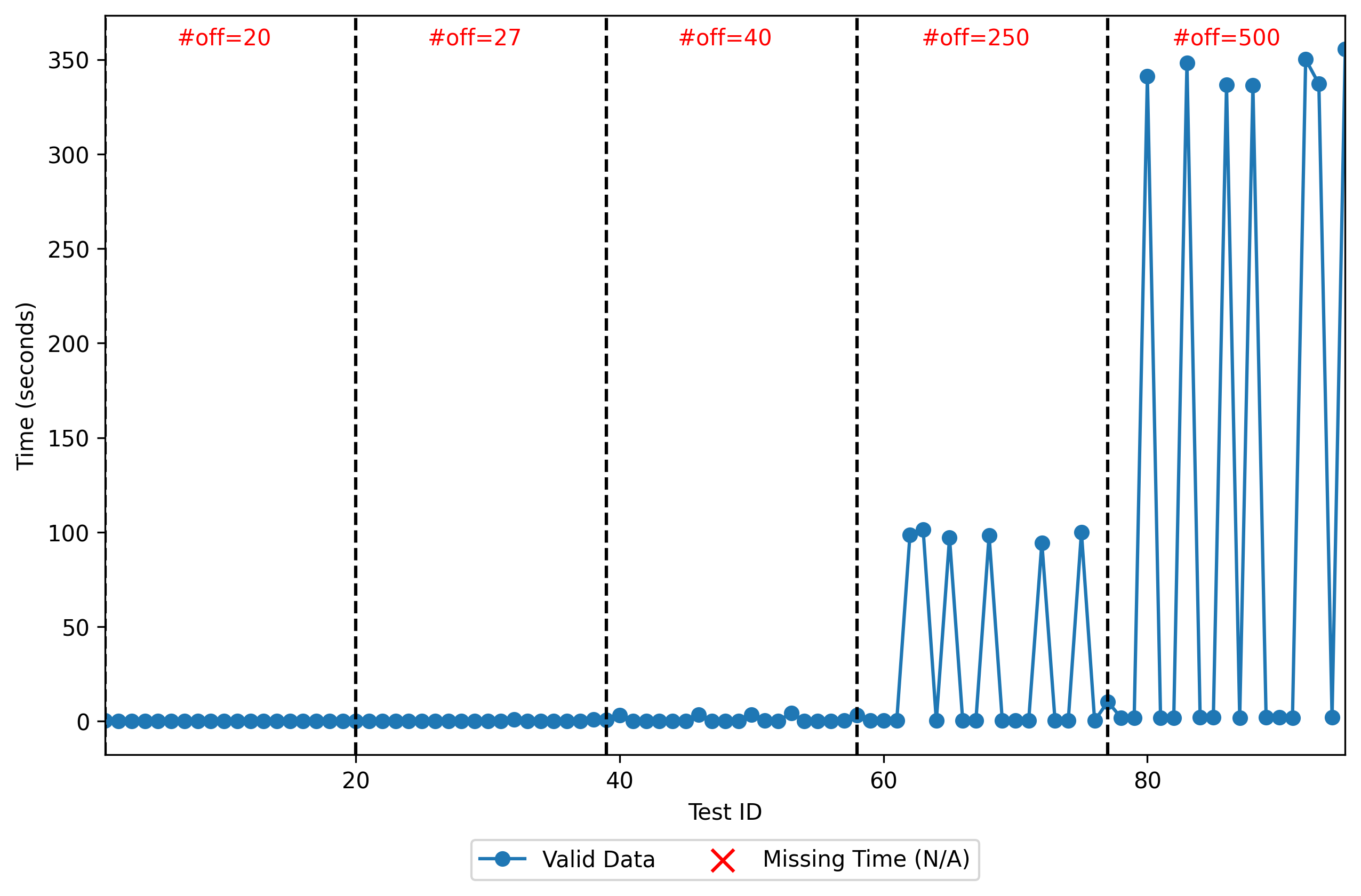}
    \caption{Scalability of the GNN method based on the number of offers (Secure Web Container)}
    \label{fig:ScalabilityGNNMethodNumberOffersSecureWeb}
\end{figure}
\begin{figure}[h!]
    \centering
\includegraphics[width=0.9\textwidth]{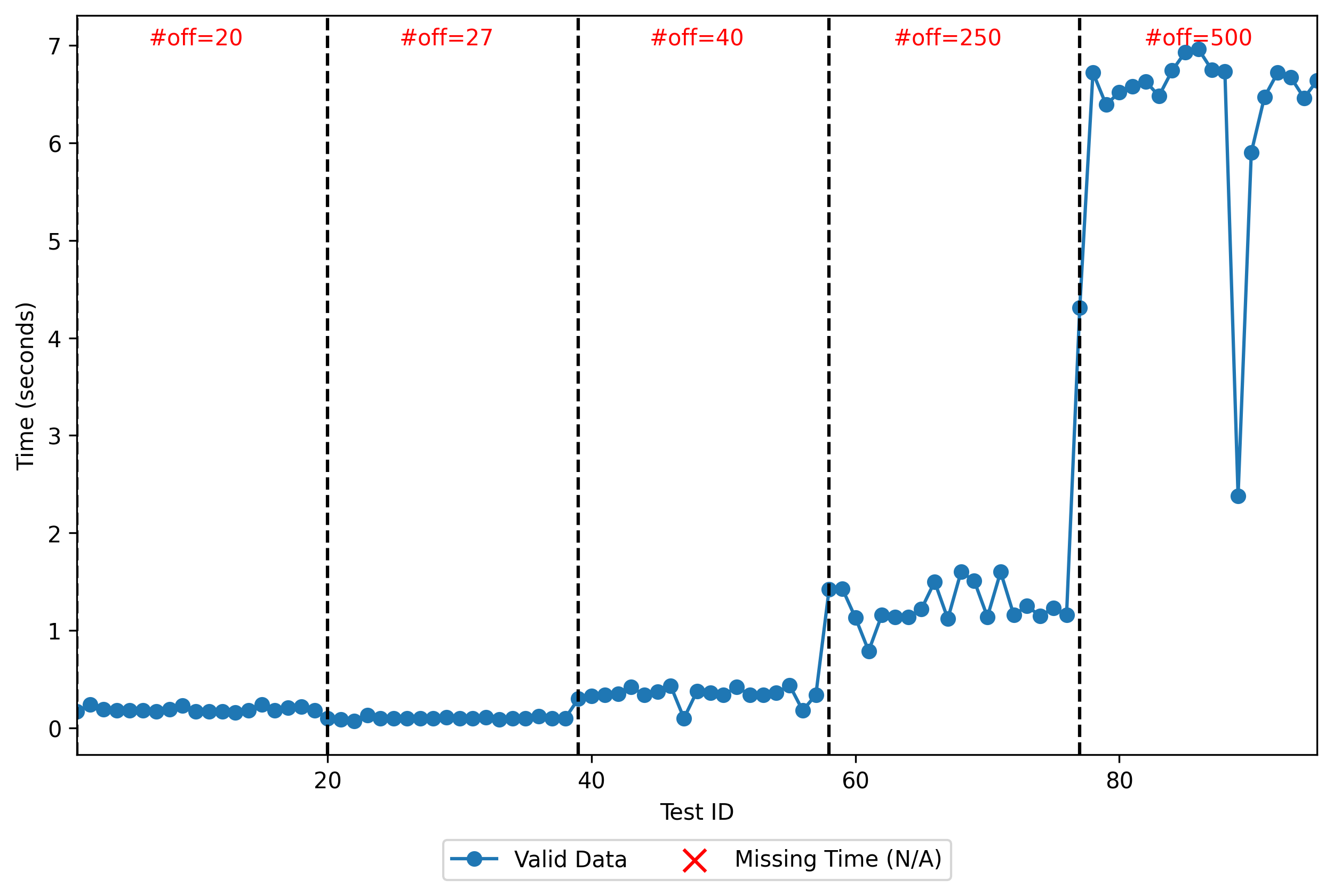}
    \caption{Scalability of the GNN method based on the number of offers (Secure Billing Email)}
\label{fig:ScalabilityGNNMethodNumberOffersSecureBilling}
\end{figure}

\begin{figure}[h!]
    \centering
\includegraphics[width=0.9\textwidth]{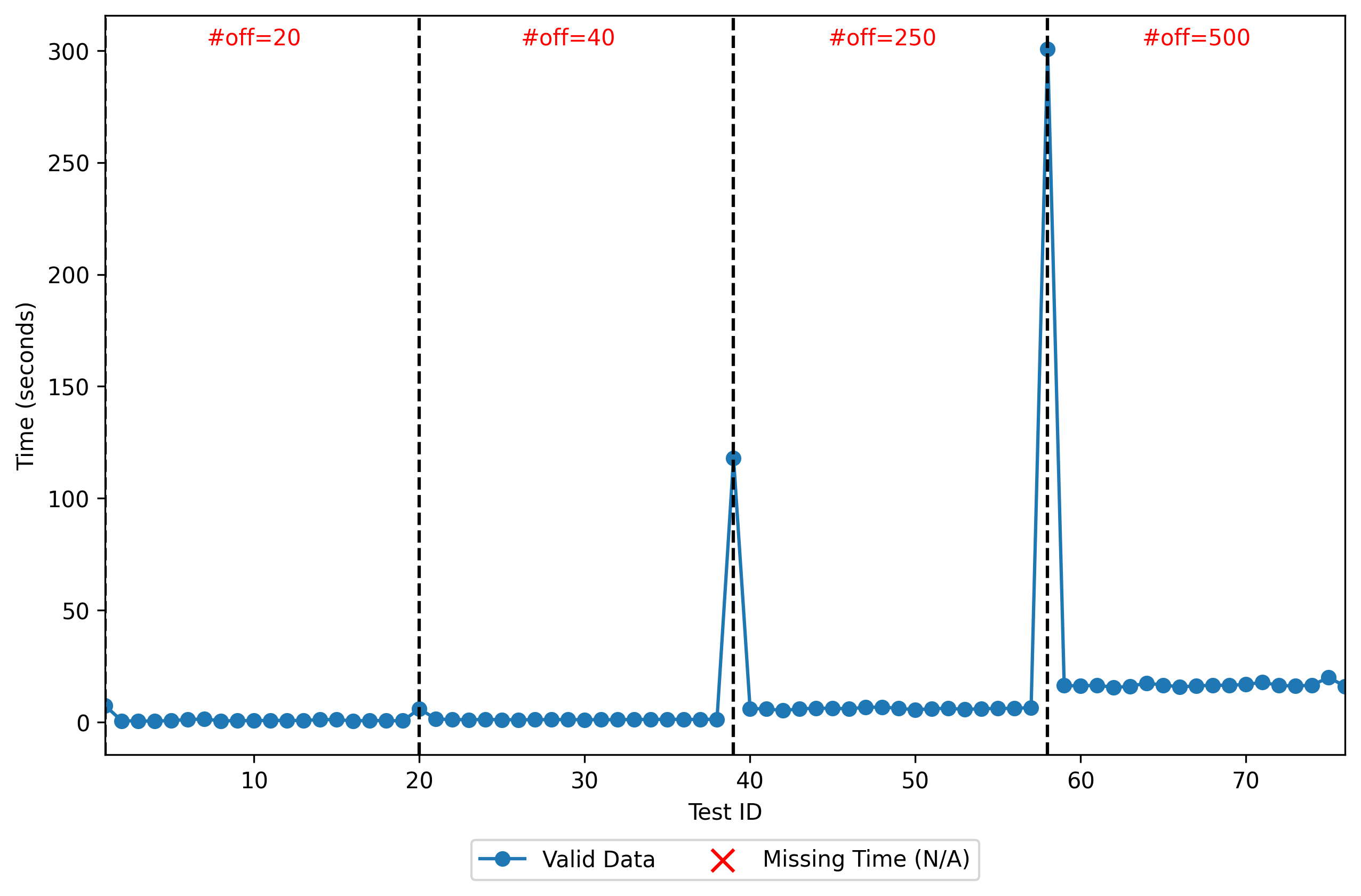}
    \caption{Scalability of the GNN method based on the number of offers (Oryx2)}
\label{fig:ScalabilityGNNMethodNumberOffersOryx2}
\end{figure}

\begin{figure}[h!]
    \centering
\includegraphics[width=0.9\textwidth]{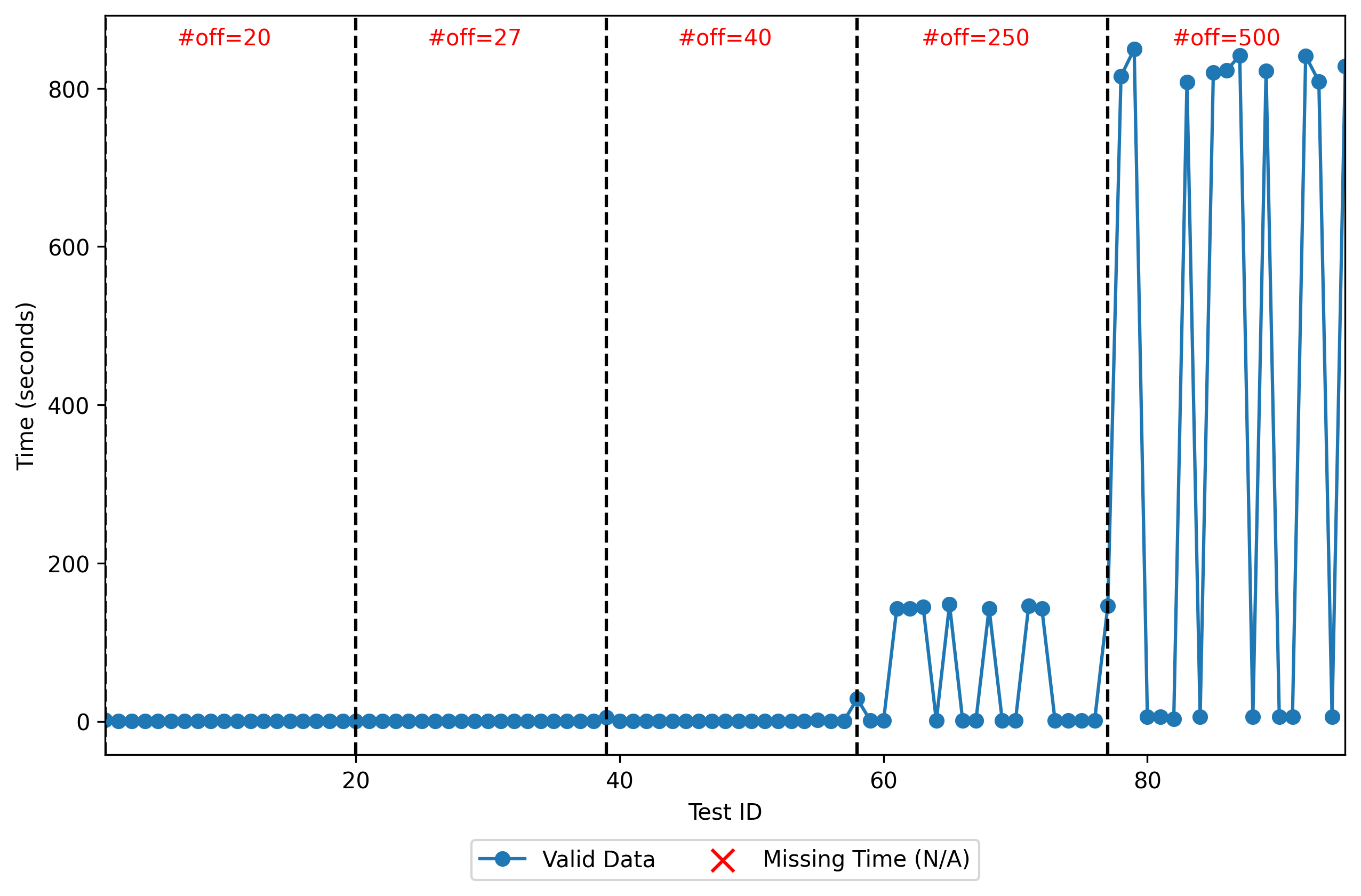}
    \caption{Scalability of the GNN method based on the number of offers (WordPress3)}
\label{fig:ScalabilityGNNMethodNumberOffersWordPress3}
\end{figure}
\begin{figure}[h!]
    \centering
\includegraphics[width=0.9\textwidth]{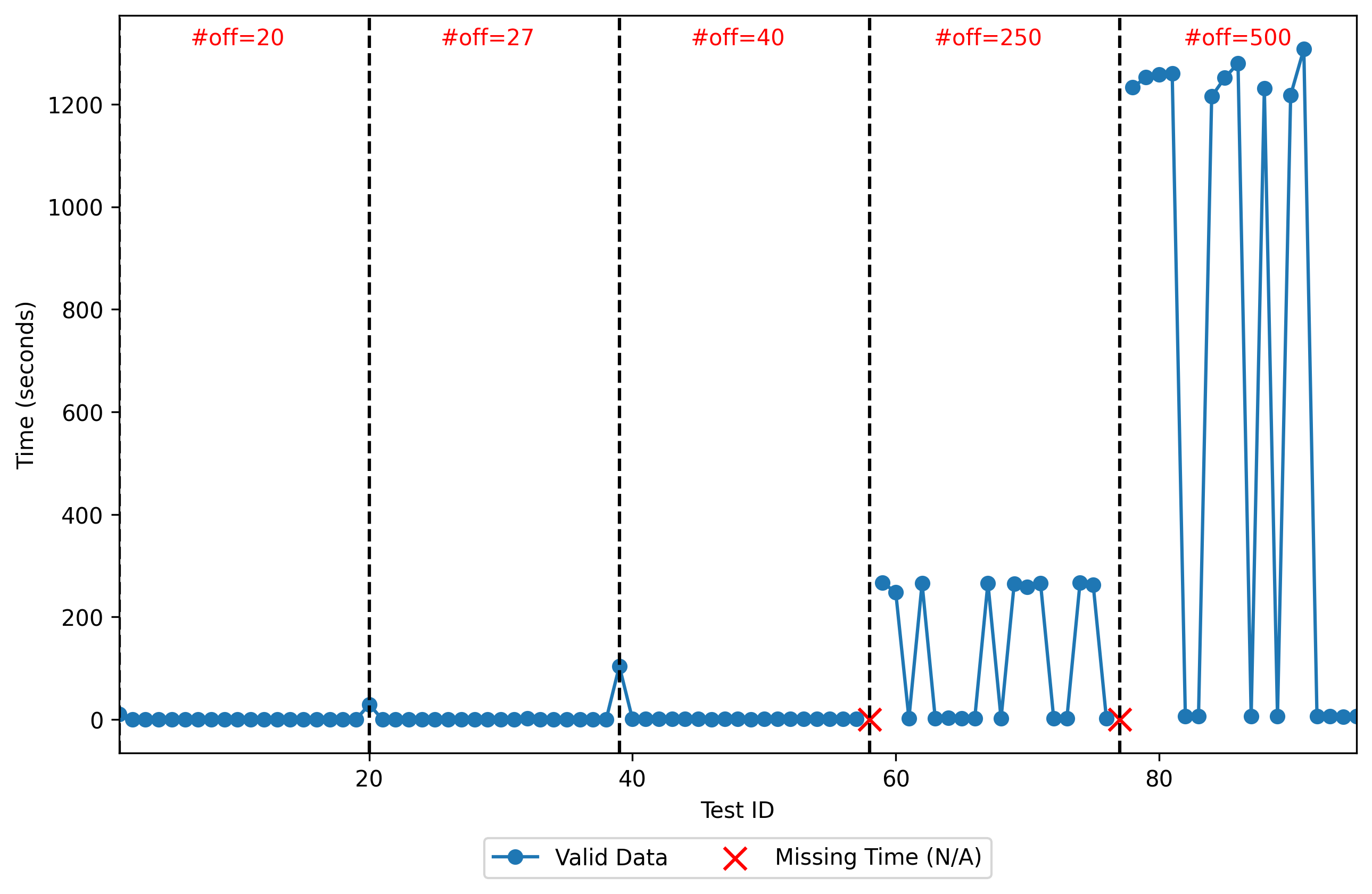}
    \caption{Scalability of the GNN method based on the number of offers (WordPress4)}
\label{fig:ScalabilityGNNMethodNumberOffersWordPress4}
\end{figure}
\begin{figure}[h!]
    \centering
\includegraphics[width=0.9\textwidth]{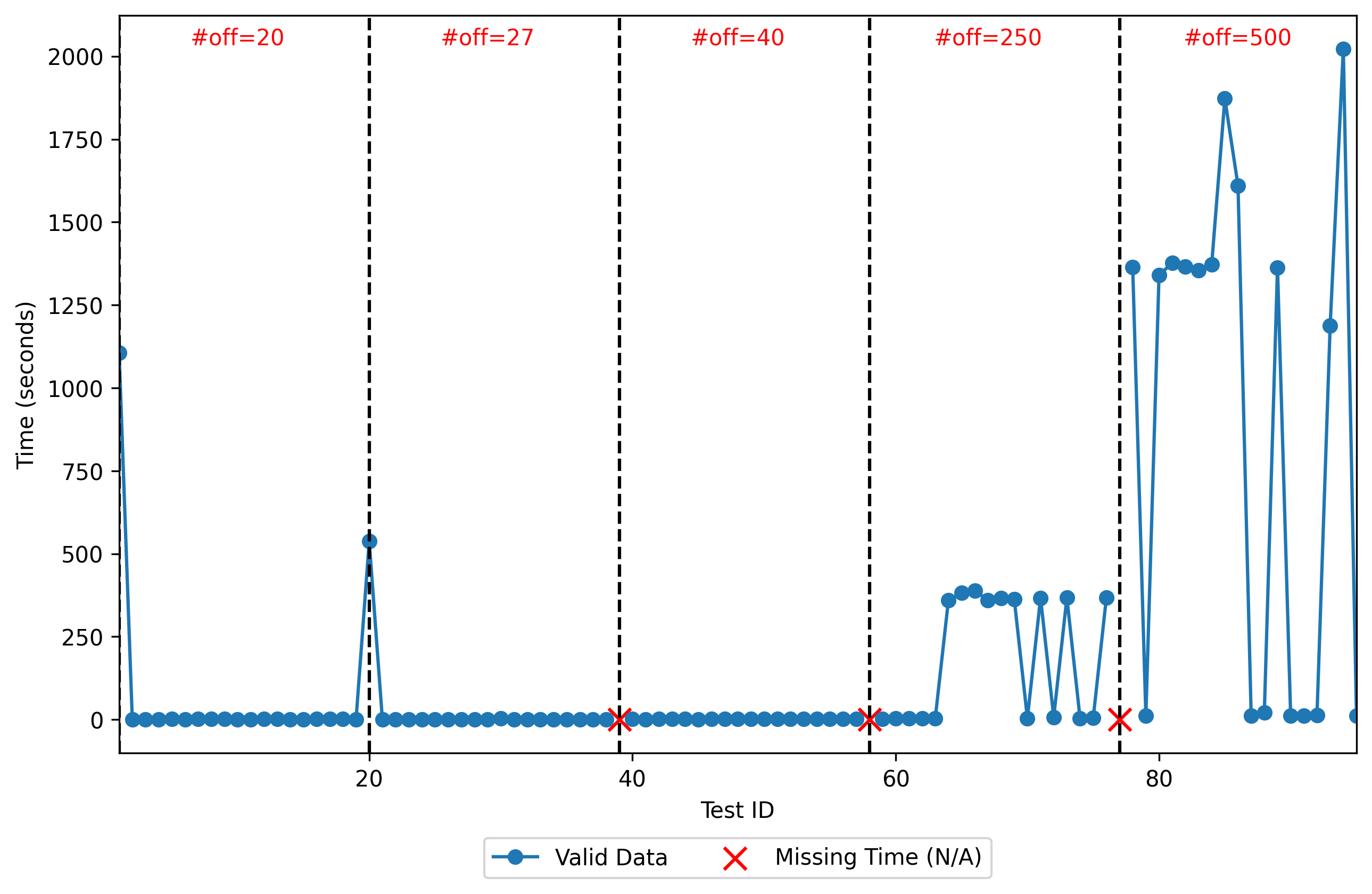}
    \caption{Scalability of the GNN method based on the number of offers (WordPress5)}
\label{fig:ScalabilityGNNMethodNumberOffersWordPress5}
\end{figure}
\begin{figure}[h!]
    \centering
\includegraphics[width=0.9\textwidth]{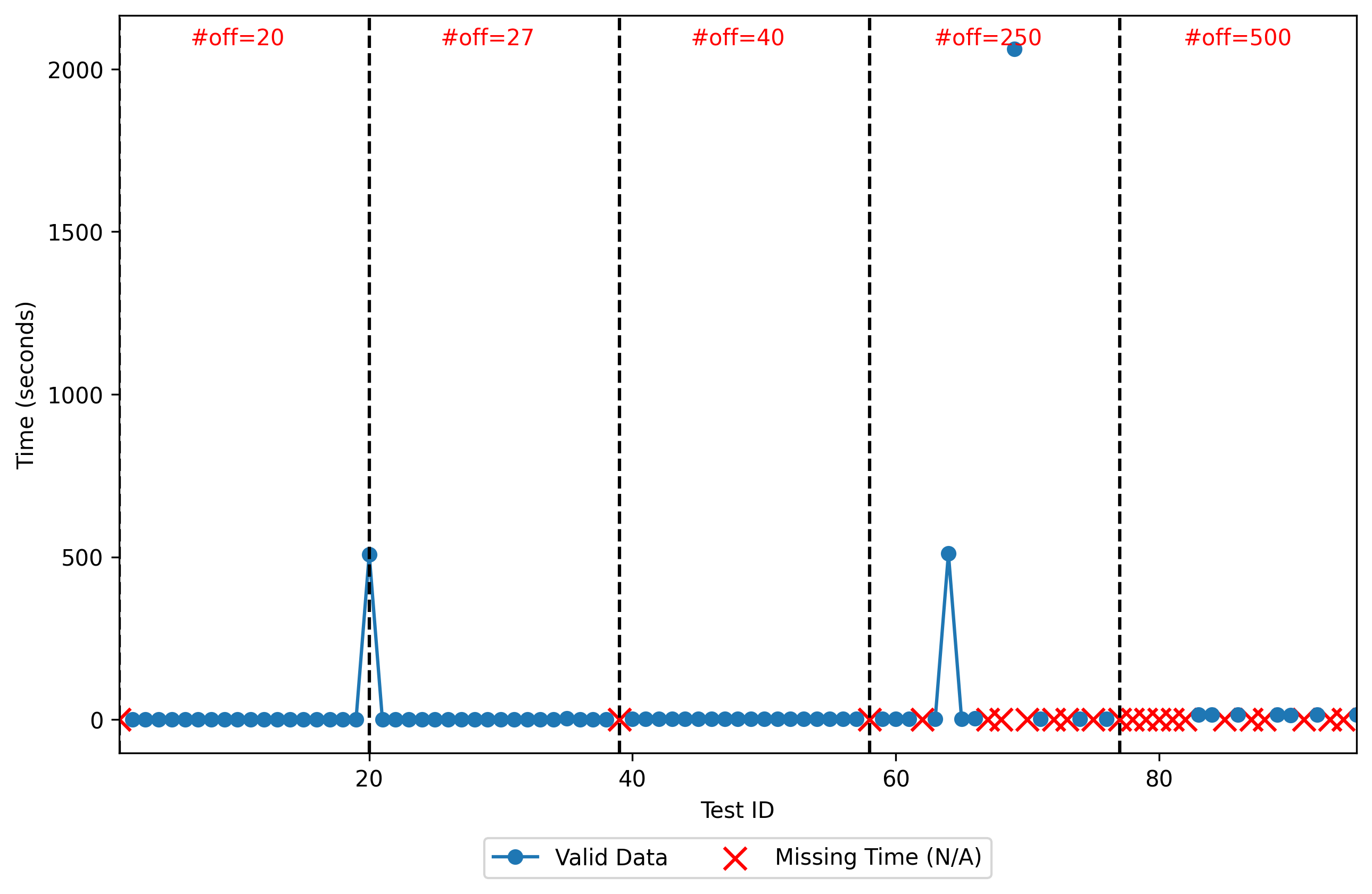}
    \caption{Scalability of the GNN method based on the number of offers (WordPress6)}
\label{fig:ScalabilityGNNMethodNumberOffersWordPress6}
\end{figure}
\begin{figure}[h!]
    \centering
\includegraphics[width=0.9\textwidth]{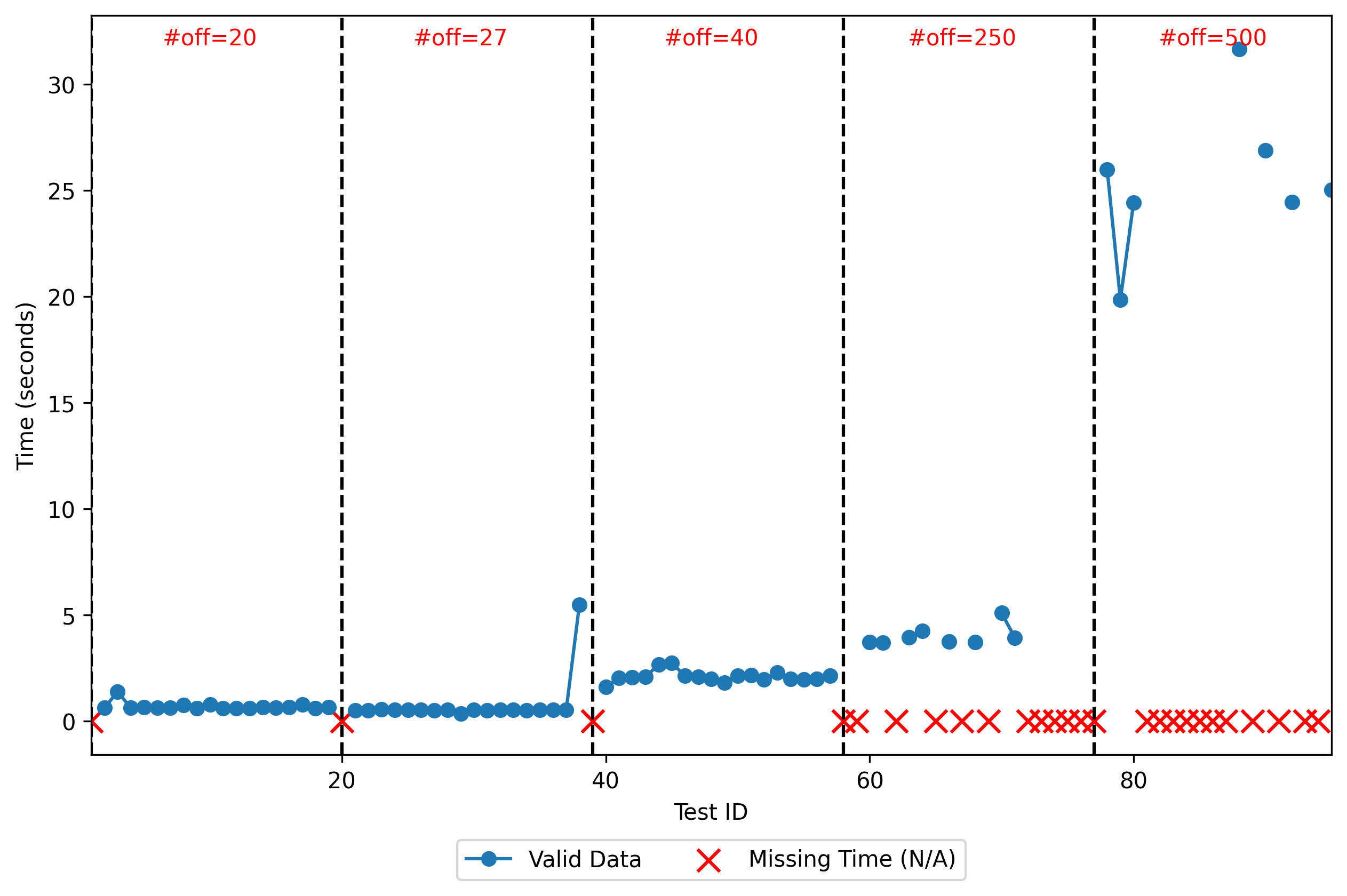}
    \caption{Scalability of the GNN method based on the number of offers (WordPress7)}
\label{fig:ScalabilityGNNMethodNumberOffersWordPress7}
\end{figure}
\begin{figure}[h!]
    \centering
\includegraphics[width=0.9\textwidth]{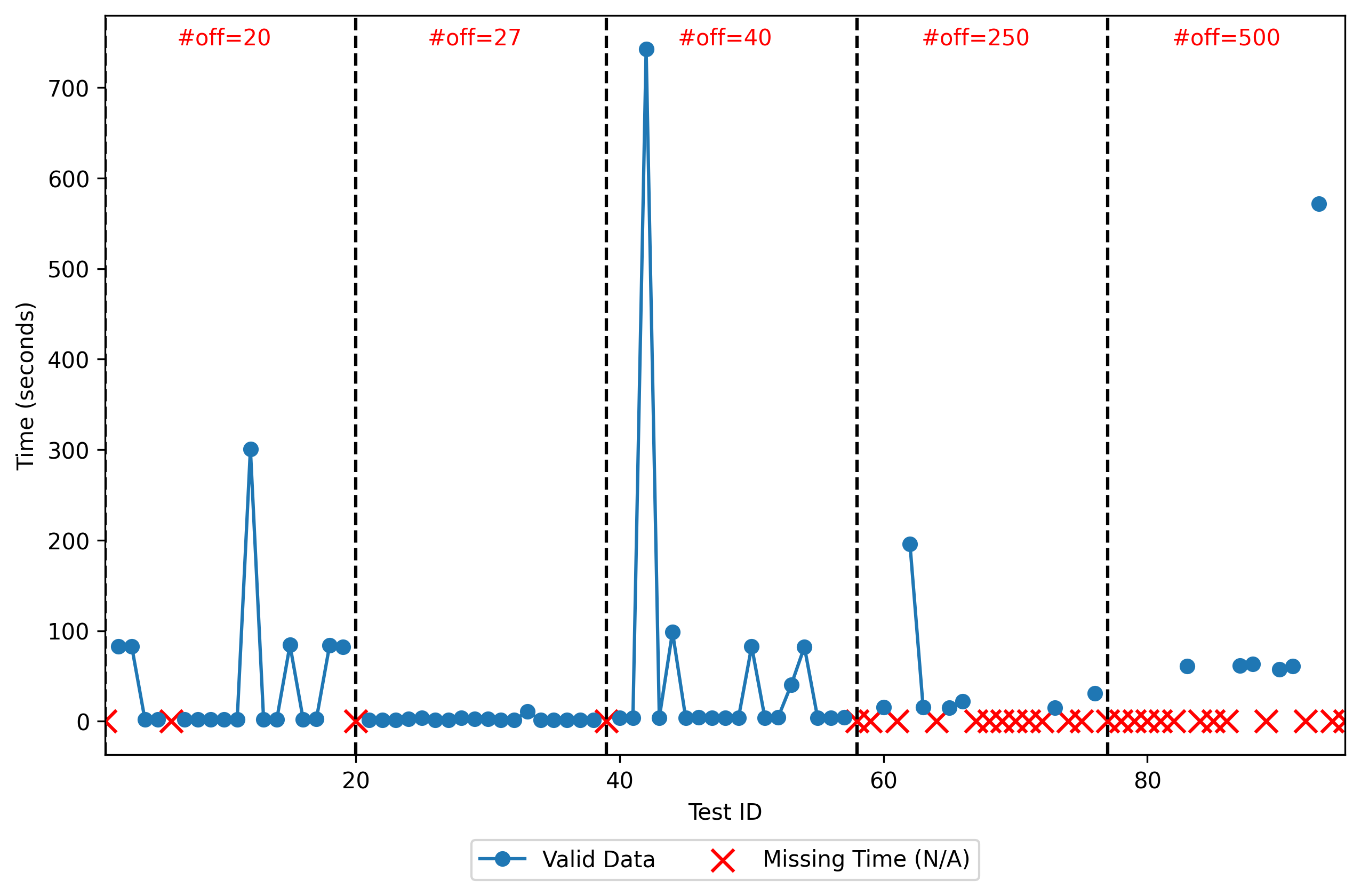}
    \caption{Scalability of the GNN method based on the number of offers (WordPress8)}
\label{fig:ScalabilityGNNMethodNumberOffersWordPress8}
\end{figure}

\begin{figure}[h!]
    \centering
\includegraphics[width=0.9\textwidth]{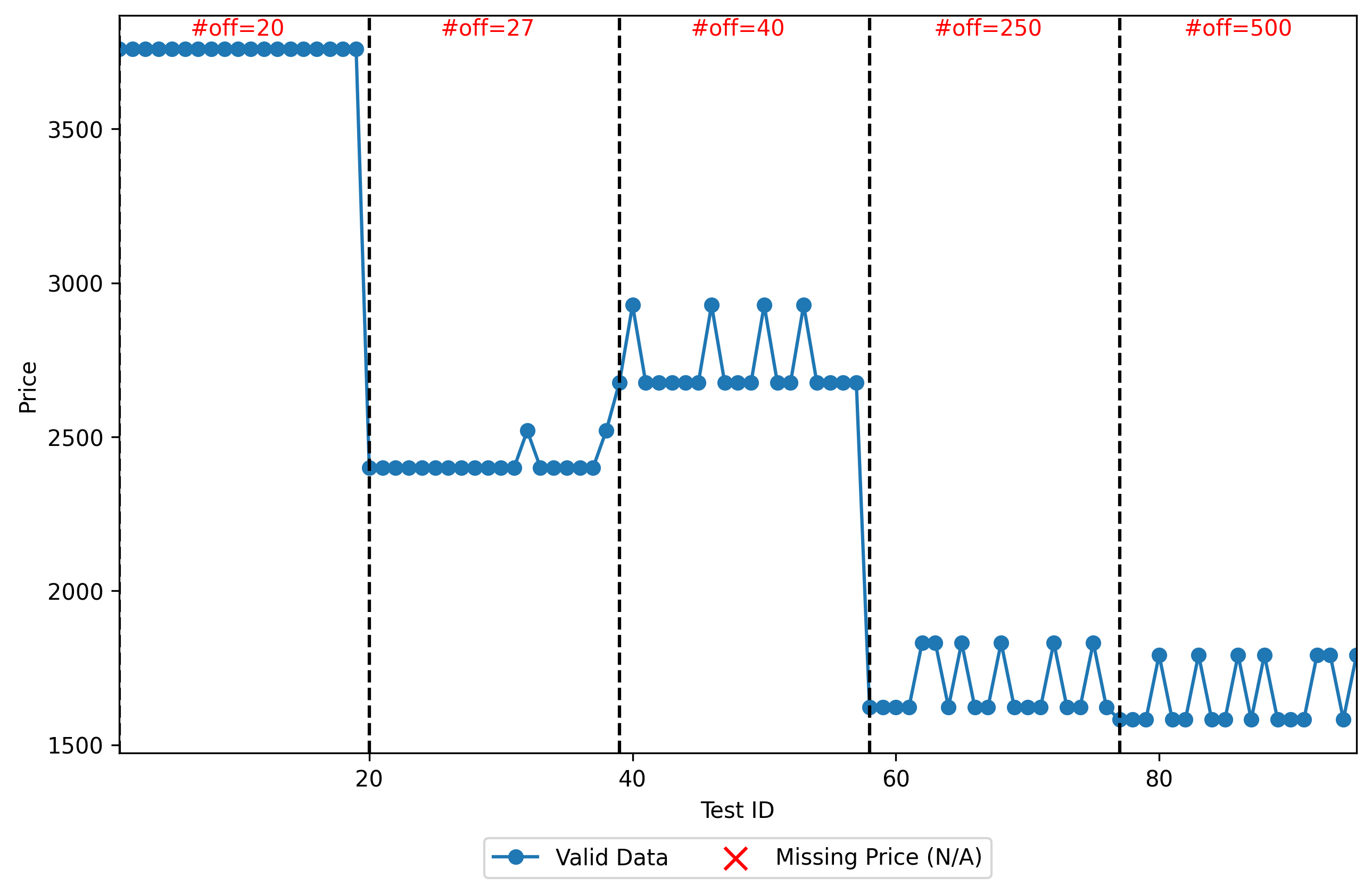}
    \caption{Optimal solution provided by the GNN method based on the number of offers (Secure Web Container)}
\label{fig:price_optimality_secure_web}
\end{figure}
\begin{figure}[h!]
    \centering
\includegraphics[width=0.9\textwidth]{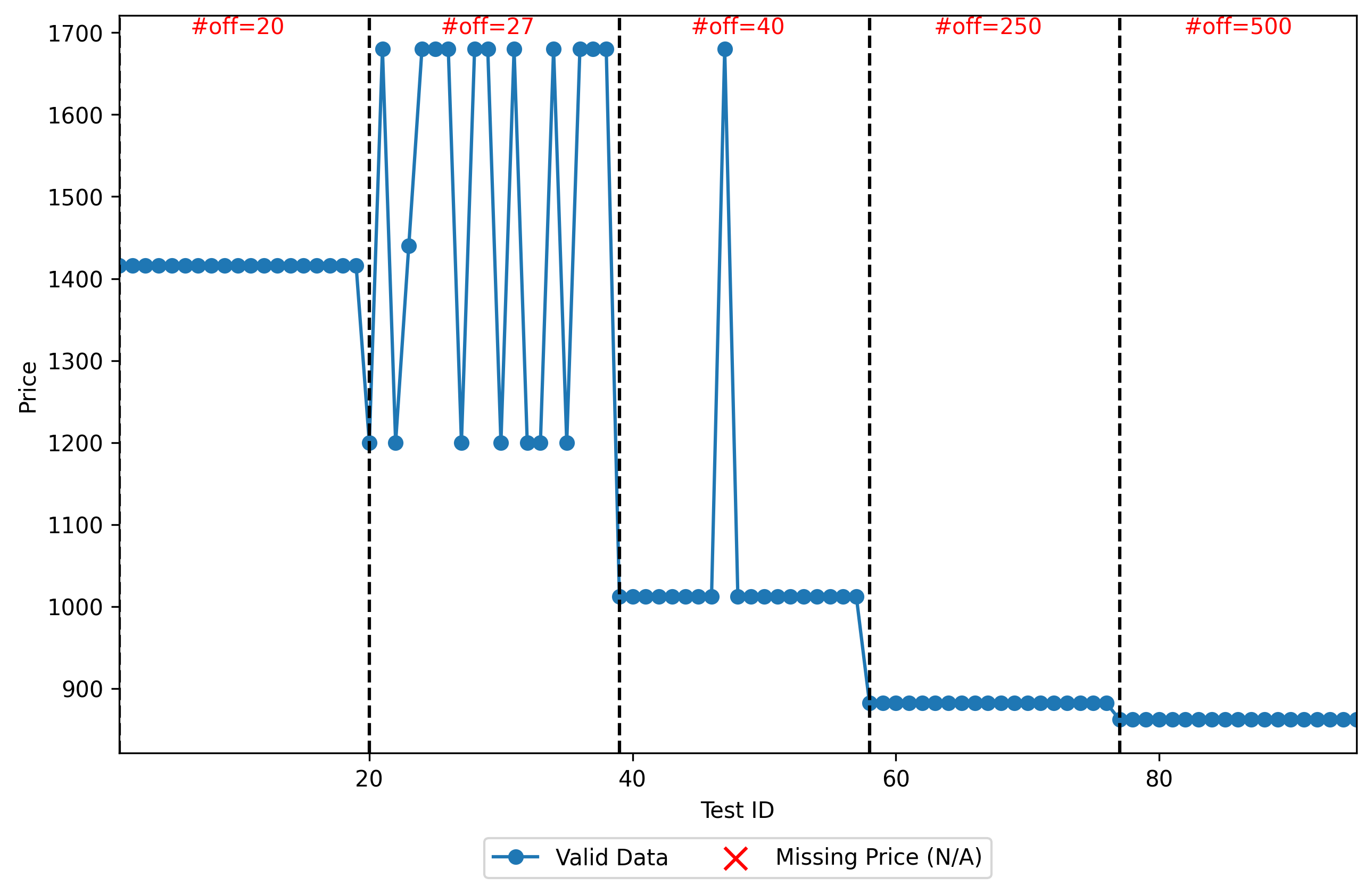}
    \caption{Optimal solution provided by the GNN method based on the number of offers (Secure Billing Email)}
\label{fig:price_optimality_secure_billing}
\end{figure}
\begin{figure}[h!]
    \centering
\includegraphics[width=0.9\textwidth]{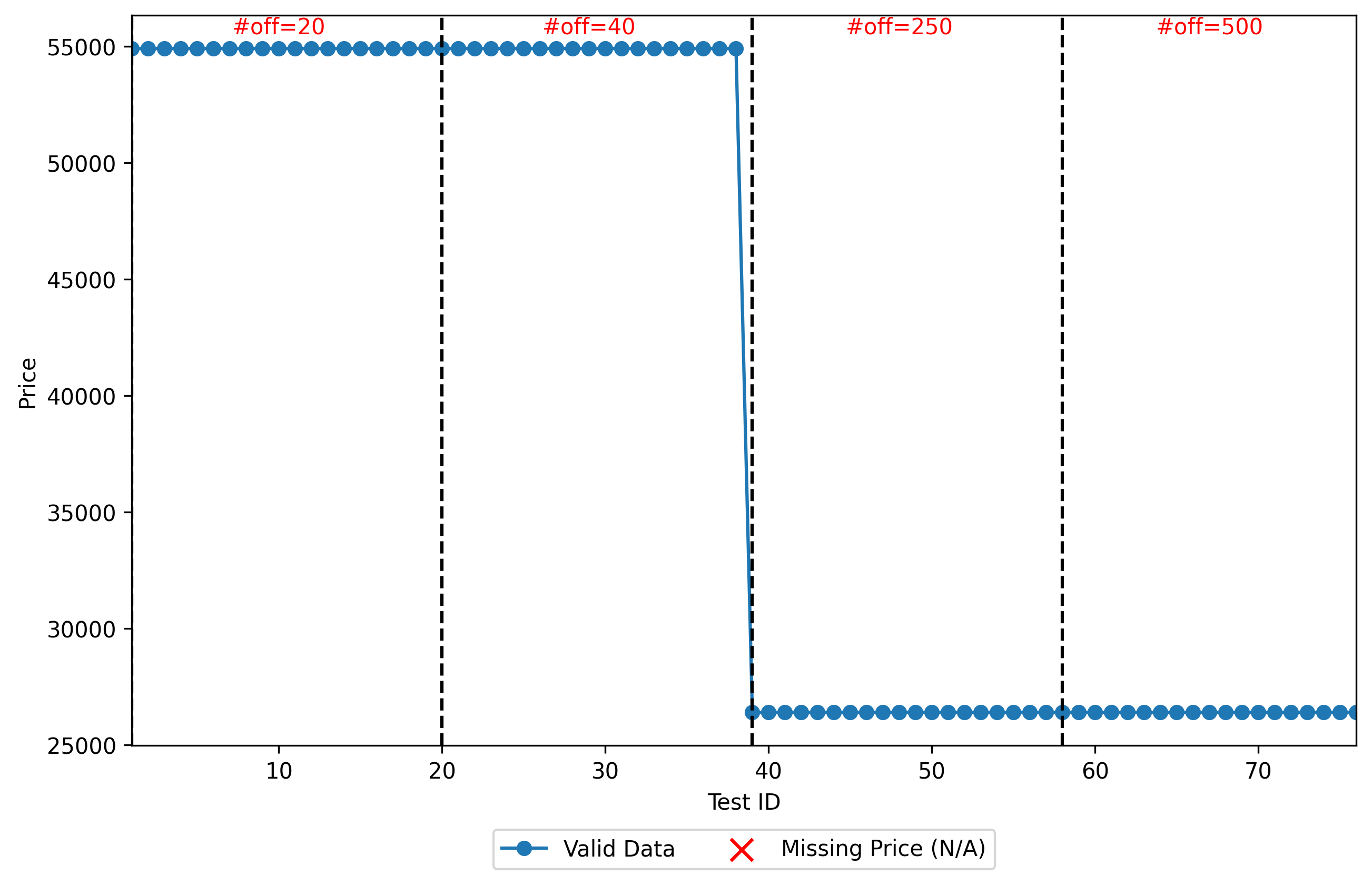}
    \caption{Optimal solution provided by the GNN method based on the number of offers (Oryx 2)}
\label{fig:price_optimality_oryx2}
\end{figure}
\begin{figure}[h!]
    \centering
\includegraphics[width=0.9\textwidth]{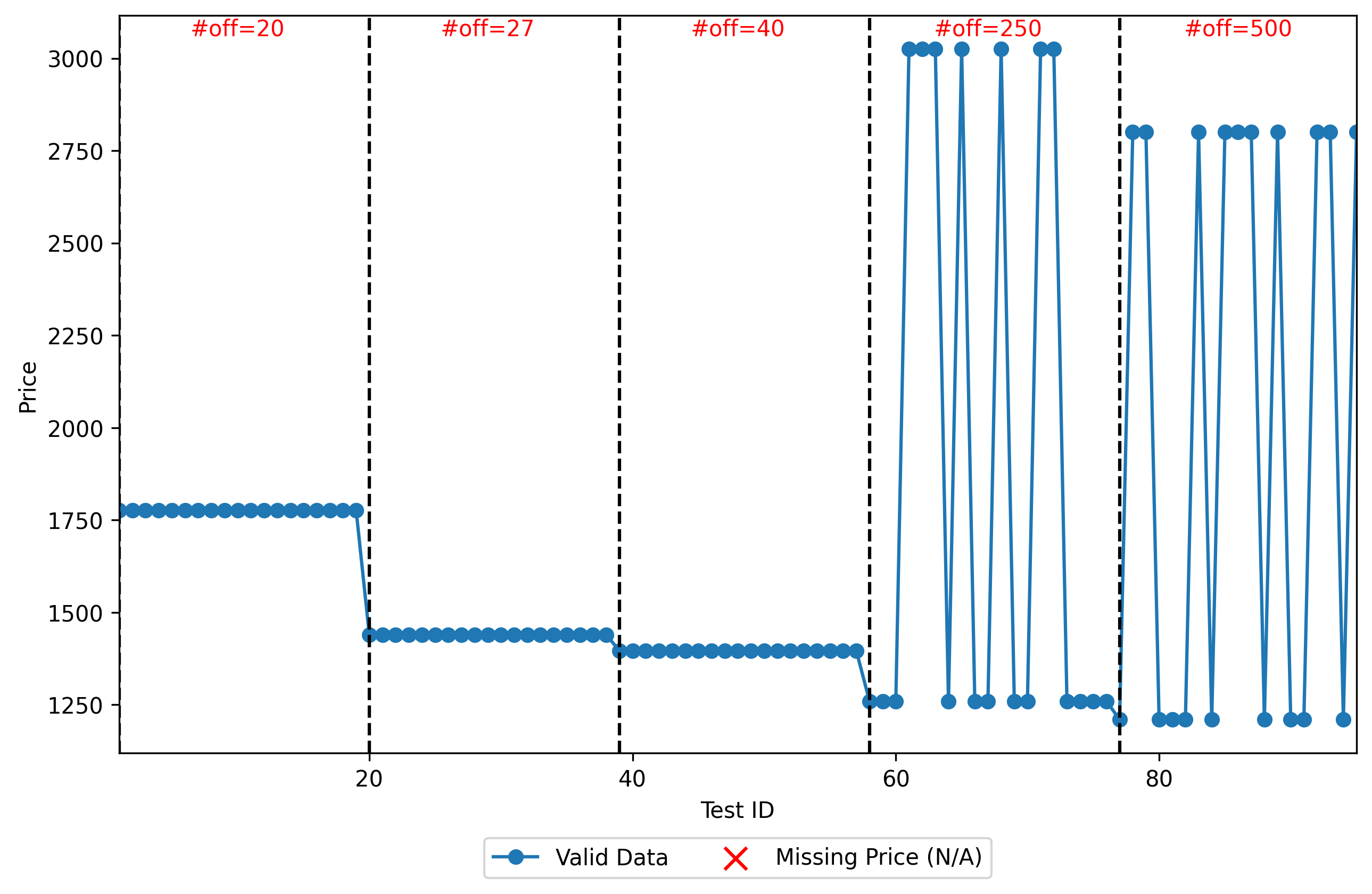}
    \caption{Optimal solution provided by the GNN method based on the number of offers (WordPress3)}
\label{fig:price_optimality_WordPress3}
\end{figure}
\begin{figure}[h!]
    \centering
\includegraphics[width=0.9\textwidth]{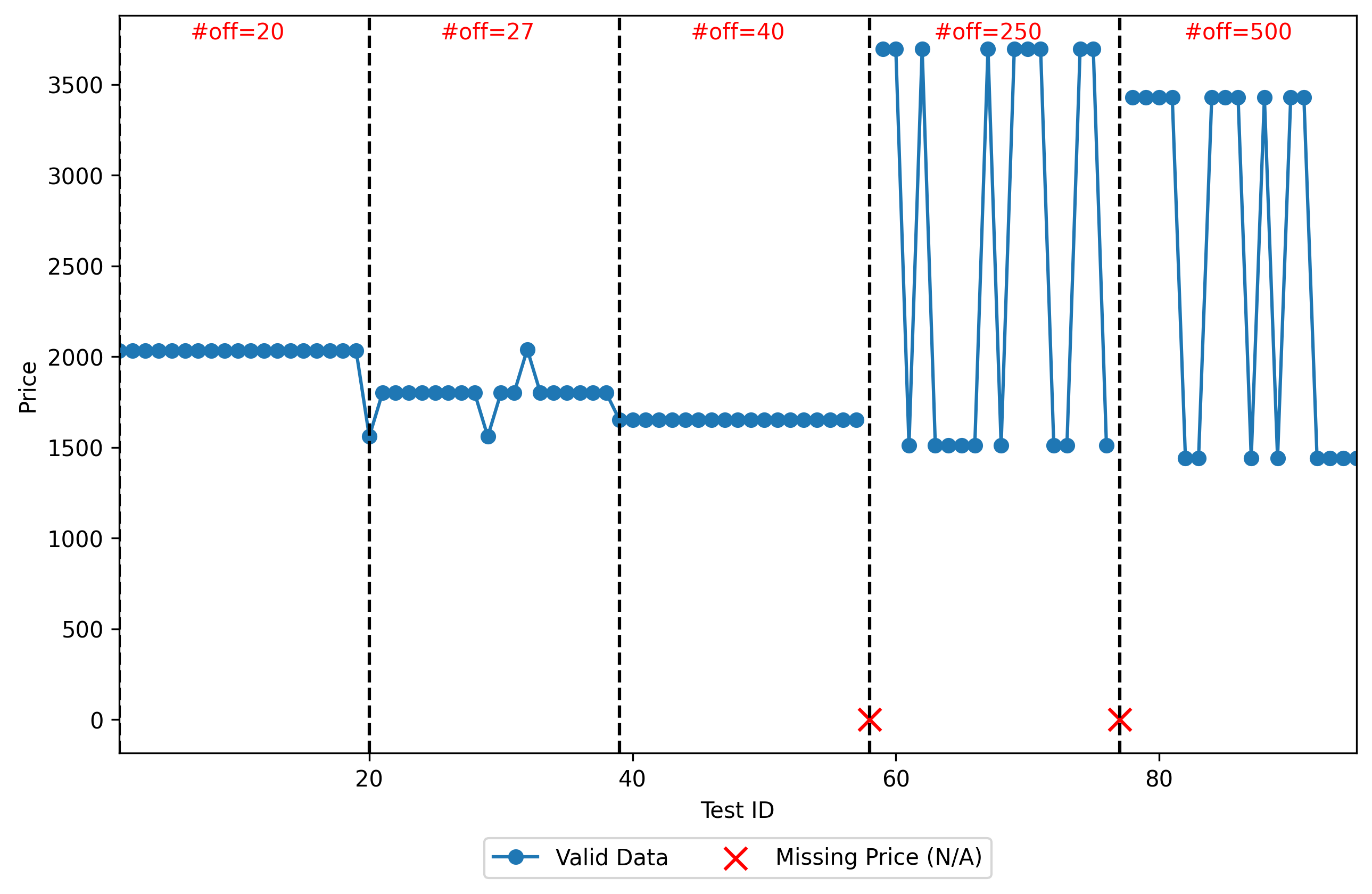}
    \caption{Optimal solution provided by the GNN method based on the number of offers (WordPress4)}
\label{fig:price_optimality_WordPress4}
\end{figure}
\begin{figure}[h!]
    \centering
\includegraphics[width=0.9\textwidth]{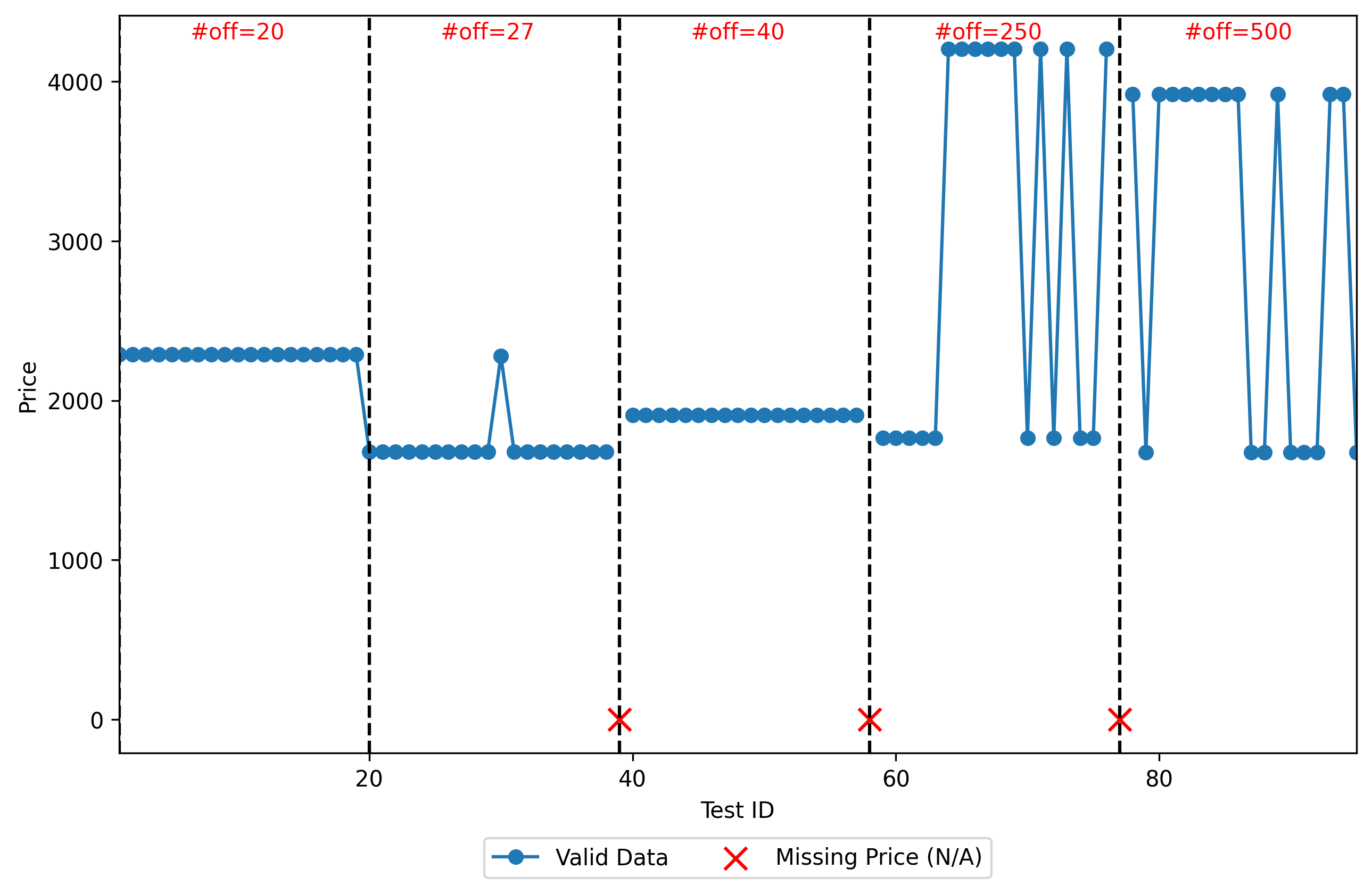}
    \caption{Optimal solution provided by the GNN method based on the number of offers (WordPress5)}
\label{fig:price_optimality_WordPress5}
\end{figure}
\begin{figure}[h!]
    \centering
\includegraphics[width=0.9\textwidth]{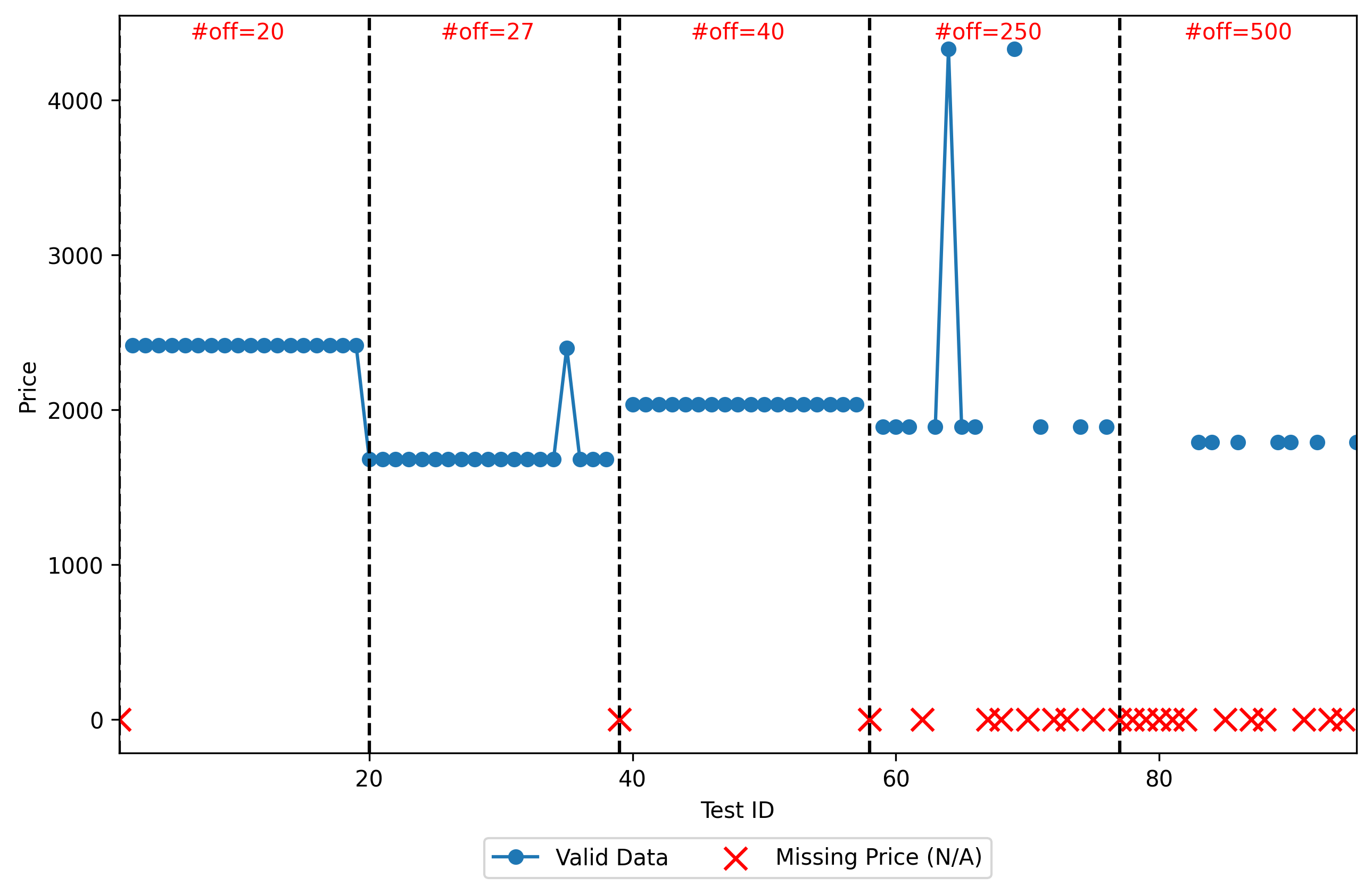}
    \caption{Optimal solution provided by the GNN method based on the number of offers (WordPress6)}
\label{fig:price_optimality_WordPress6}
\end{figure}
\begin{figure}[h!]
    \centering
\includegraphics[width=0.9\textwidth]{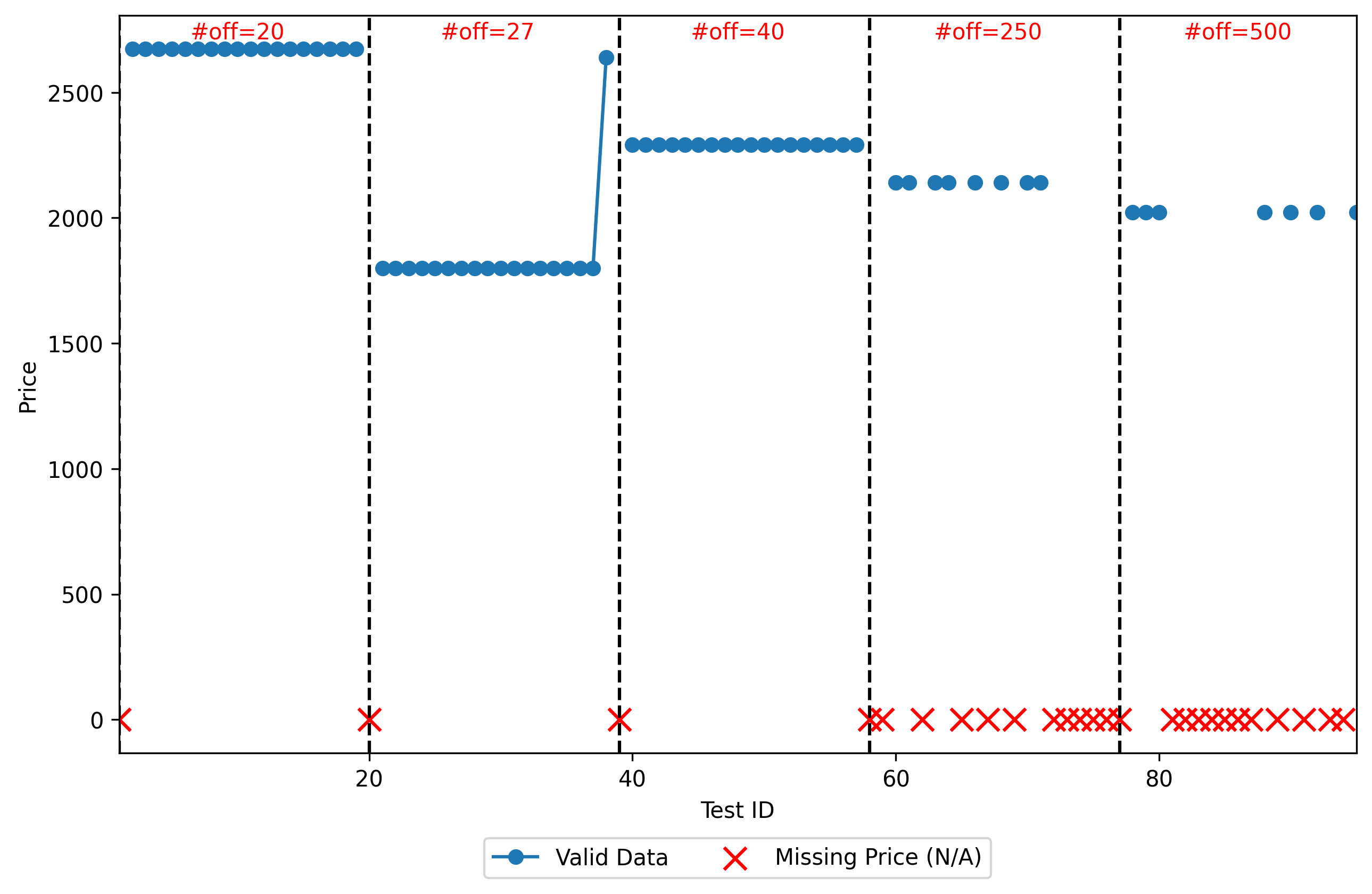}
    \caption{Optimal solution provided by the GNN method based on the number of offers (WordPress7)}
\label{fig:price_optimality_WordPress7}
\end{figure}
\begin{figure}[h!]
    \centering
\includegraphics[width=0.9\textwidth]{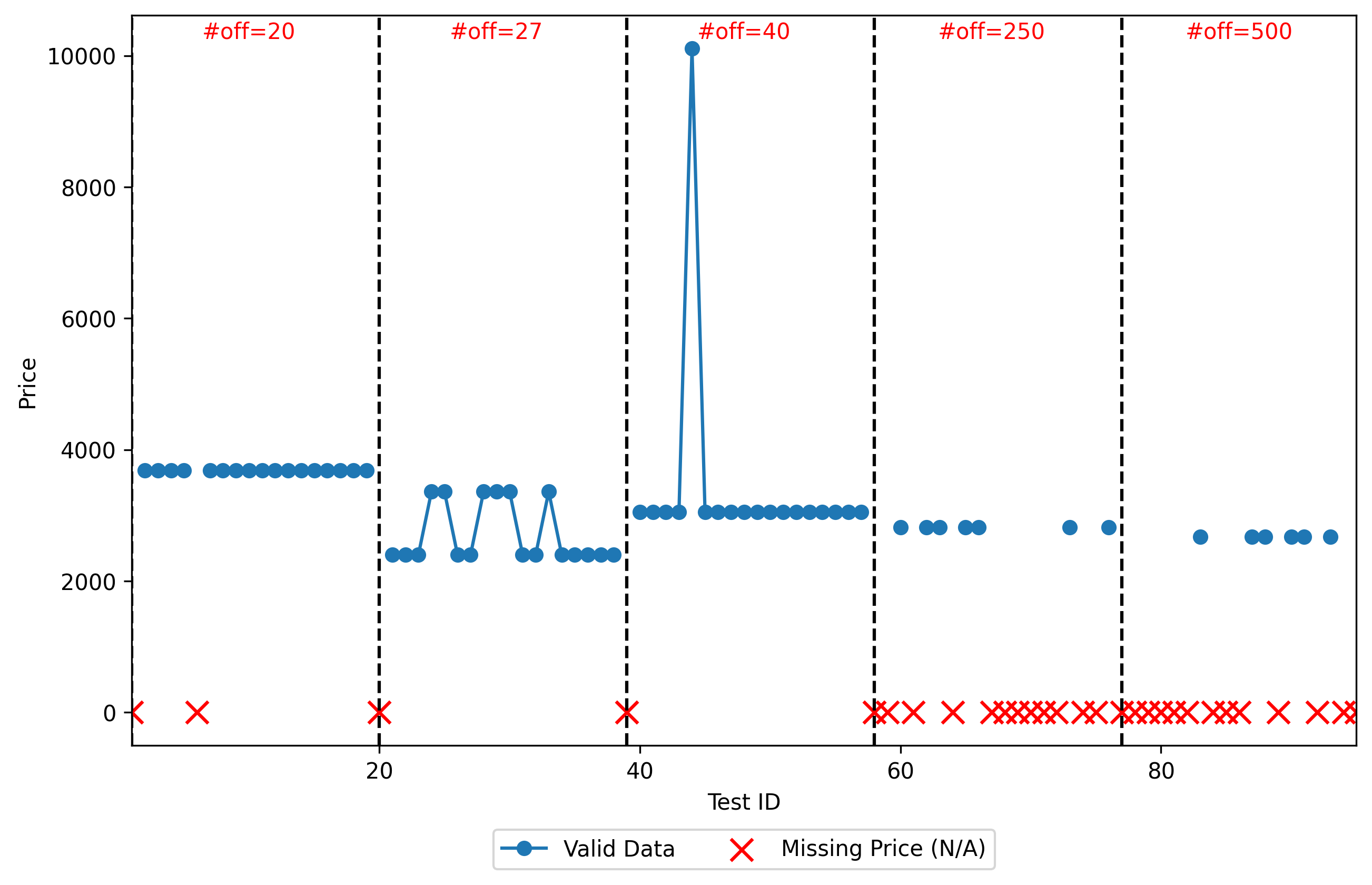}
    \caption{Optimal solution provided by the GNN method based on the number of offers (WordPress8)}
\label{fig:price_optimality_WordPress8}
\end{figure}

\end{appendices}

\end{document}